\documentclass[aps,prd,twocolumn,nofootinbib]{revtex4-1}

\usepackage{soul,xcolor}

\usepackage{graphicx}
\usepackage{epsfig}
\usepackage{amssymb}
\usepackage{color}
\usepackage{epstopdf}
\usepackage{dcolumn}
\usepackage{relsize}
\usepackage{amsmath}
\usepackage{bm}
\usepackage{dcolumn}
\usepackage{txfonts}
\usepackage{physics}
\usepackage{slashed}
\usepackage{float}
\usepackage[mathscr]{euscript}

\newcommand{\mcS}{{\mathcal S}}
\newcommand{\mcL}{{\mathcal L}}

\newcommand{\msbar}{{\overline{\rm MS}}}

\newcommand{\xa}{x}
\newcommand{\xb}{x^{\prime}}
\newcommand{\xc}{x^{\prime\prime}}
\newcommand{\ya}{y}
\newcommand{\yb}{y^{\prime}}
\newcommand{\yc}{y^{\prime\prime}}

\newcommand\bef{\begin{figure*}}
\newcommand\eef[1]{\label{fig:#1}\end{figure*}}
\newcommand\beq{\begin{equation}}
\newcommand\eeq[1]{\label{#1}\end{equation}}
\newcommand\beqa{\begin{eqnarray}}
\newcommand\eeqa[1]{\label{#1}\end{eqnarray}}
\newcommand\bet{\begin{table}}
\newcommand\eet[1]{\label{tb:#1}\end{table}}

\newcommand\fgn[1]{Fig.\ \ref{fig:#1}}
\newcommand\eqn[1]{Eq.\ (\ref{#1})}
\newcommand\scn[1]{Section \ref{sec:#1}}

\begin{document}

\setstcolor{red}

\title{Valence parton distribution function of pion from fine lattice}

\author{
  Taku Izubuchi$^{1,2}$, Luchang Jin$^{2,3}$,
  Christos Kallidonis$^{4}$, Nikhil Karthik$^{1}$, Swagato Mukherjee$^{1}$, Peter Petreczky$^{1}$,
  Charles Shugert$^{1,4}$, Sergey Syritsyn$^{2,4}$
}
\affiliation{
$^1$ Physics Department, Brookhaven National Laboratory, Upton, NY 11973, USA \\
$^2$ RIKEN-BNL Research Center, Brookhaven National Lab, Upton, NY, 11973, USA \\
$^3$ Physics Department, University of Connecticut, Storrs, Connecticut 06269-3046, USA \\
$^4$ Department of Physics and Astronomy, Stony Brook University, Stony Brook, NY 11794, USA
}

\begin{abstract}
    We present a lattice QCD study of the valence parton distribution
inside the pion within the framework of Large Momentum Effective
Theory.  We use a mixed action approach with 1-HYP smeared
valence Wilson clover quarks on 2+1 flavor HISQ sea with the valence quark
mass tuned to 300 MeV pion mass.  We use $48^3 \times 64$ lattice
at a fine lattice spacing $a=0.06$ fm for this computation.  We
renormalize the quasi-PDF matrix element in the non-perturbative RI-MOM
scheme. As a byproduct, we test the validity of 1-loop matching
procedure by comparing the RI-MOM renormalized quasi-PDF matrix
element with off-shell quark external states as computed in the
continuum 1-loop perturbation theory with the lattice results at
$a=0.04$ and 0.06 fm.  By applying the RI-MOM to $\msbar$ one-loop
matching, implemented through a fit to phenomenologically motivated
PDFs, we obtain the valence PDF of pion.
\end{abstract}
\date{\today}
\maketitle

\section{Introduction}
QCD factorization allows us to calculate the cross-section of hard
hadronic processes in terms of the convolution of  partonic
cross-section and parton distribution functions~\cite{Collins:1989gx}.
Parton distribution for a hadron can be defined using hadronic
matrix elements of appropriately chosen gauge invariant operators
separated along the light cone. For example, the quark parton
distribution function (PDF) of a hadron $H$ can be defined in terms
of an operator bilocal in quark field $\psi$
as~\cite{Soper:1976jc,Collins:1989gx}
\begin{equation}
  f(x) = \frac{1}{4\pi}\int d\xi^{-} e^{ixP^{+}\xi^{-}} 
    \bra{H(P)}\overline\psi(\xi^-)\gamma^+W(\xi^-, 0)\psi(0)\ket{H(P)},
\label{soperform}
\end{equation}
where $W(\xi^-, 0) = \mathcal{P} e^{ig\int_0^{\xi^-} d\xi^- A^+}$ is the
path-ordered straight
Wilson Line on the light-cone, and the light-cone coordinates
$\xi^{\pm} = (t \pm z)/\sqrt{2}$.  A straight forward first principle
calculation of PDF is not possible because lattice QCD is formulated
in the Euclidean space-time, and thus, it cannot access quantities defined
on the light-cone.  To circumvent this problem, it has recently
been proposed to calculate the quasi parton distribution function (qPDF),
$\tilde{q}(x,P_z)$, defined in terms of matrix elements of equal
time, but spatially separated, quark bilinears~\cite{Ji:2013dva}
evaluated in a hadron state boosted to a large momentum $P_z$:
\begin{equation}
  \tilde q(x, P_z) = \frac{1}{4\pi}\int dz e^{-ixP^zz}
  \bra{H(P_z)}\overline\psi(z)\Gamma W(z, 0)\psi(0)\ket{H(P_z)},
\label{qpdf}
\end{equation}
where $\Gamma$ is either $\gamma_z$  or $\gamma_t$ for the unpolarized
parton distribution addressed in this paper. Here, 
$W(z,0)$ is a straight spatial Wilson line joining the quark
and anti-quark. For sufficiently boosted hadrons, one can use the Large
Momentum Effective Theory (LaMET)~\cite{Ji:2014gla} to relate the qPDF
to PDF through a convolution with a matching kernel $C$ as
\begin{equation}
\tilde q(x,\mu_L,P_z)=\int_{-1}^{+1}\frac{dy}{|y|}C\left(\frac{x}{y},\frac{y P_z}{\mu},\frac{\mu_L}{y P_z}\right)f(y,\mu).
\label{pdf2qpdf}
\end{equation}
Here $\mu_L$ and $\mu$ are the renormalization scales of the schemes
in which the qPDF and PDF are defined. For the latter, $\msbar$ scheme
is used and $\mu$ is referred to as the factorization scale.  The
matching kernel is perturbative and hence universal for all the hadrons.
Therefore, it is calculated using quark external states in a chosen
gauge.  Such calculations at 1-loop order have been performed using
the cutoff scheme~\cite{Xiong:2013bka} as well as in the $\msbar$
scheme~\cite{Constantinou:2017sej,Liu:2018uuj,Stewart:2017tvs}.
There are also related approaches to calculate the PDF from the
lattice that use similar logic but differ in details, like the
pseudo-PDF approach proposed in
Ref.~\cite{Radyushkin:2017cyf,Orginos:2017kos} and the use of good
lattice cross sections~\cite{Ma:2014jla,Ma:2017pxb}.  The latter
includes the current-current correlators~\cite{Sufian:2019bol}.

Using LaMET and related approaches, various attempts have been made
to calculate the unpolarized and polarized iso-vector quark
distribution of the
nucleon~\cite{Liu:2018uuj,Lin:2018qky,Chen:2018xof,Alexandrou:2018eet,Alexandrou:2018pbm}.
The first studies of the valence quark distribution for the pion
have also been presented~\cite{Sufian:2019bol,Chen:2018fwa}.  One
important issue in the calculation of the PDF from the lattice is
the renormalization and matching.  As indicated above, the PDF and
qPDF are usually defined in different renormalization schemes.  The
qPDF, which is calculated on the lattice, needs a non-perturbative
renormalization scheme because of the self-energy divergence of the
Wilson line~\cite{Ishikawa:2016znu}, and this is usually implemented
using the RI-MOM scheme~\cite{Chen:2017mzz} defined using external
off-shell quark states accessible on the lattice.  Then, one  has
to match the qPDF in this lattice renormalization scheme to the PDF
in the $\msbar$ scheme through \eqn{pdf2qpdf}.  This is achieved through
the convolution using the matching kernel between the RI-MOM and $\msbar$
schemes that is perturbatively calculated in the continuum theory
using dimensional regularization~\cite{Stewart:2017tvs}.  One could
also define the qPDF operator in the $\msbar$ scheme and then perform
the matching between PDF and
qPDF~\cite{Constantinou:2017sej,Alexandrou:2017huk}.  The current
status of this field, including the comparison with the phenomenological
PDF and the issue of renormalization,  is reviewed in
Refs.~\cite{Zhao:2018fyu,Cichy:2018mum,Monahan:2018euv}.

In principle, \eqn{pdf2qpdf} offers a way to calculate PDF from
the lattice, but it is unclear as to what extent this is actually
feasible given the various assumptions that go along with the
equation implicitly.  For example, at any finite hadron momentum
$P_z$, the \eqn{pdf2qpdf} suffers from ${\cal O}\left(\Lambda^2_{\rm
QCD}/(x^2 P^2_z)\right)$ higher twist corrections. This is closely related
to the assumption that the perturbative calculation, currently truncated
at 1-loop order, is able to capture the renormalization as well
as the matching of the qPDF matrix element over a range of
quark-antiquark separations, $z$ -- to be in the perturbative regime, one
would expect $z$ to be smaller than or about $\mathcal{O}(1)$ fm.
It is also important to ensure that $a P_z < 1$ to make sure we are
not overcome with lattice artifacts~\cite{Xiong:2017jtn}.  Therefore,
a closer look at this new methodology is warranted and is actively
being studied~\cite{Alexandrou:2019lfo,Liu:2018uuj}.  The aim of this
paper is to explore these issues further by using finer lattices 
than what is being used in the qPDF literature, and use
pion as a case study.  The smaller mass of the pion makes it easier
to achieve a large boost, the numerical calculations are expected
to be less expensive and it also helps suppress the target mass
correction by ensuring $m_\pi \ll P_z$. We focus on the valence PDF
of pion since it can be accessed using the isotriplet $u-d$ PDF, and
thereby, avoid mixing with the gluon sector.  In our study, we will
use the renormalization and matching strategy outlined
in~\cite{Chen:2017mzz,Liu:2018uuj}.  The pion valence PDF has been
determined through a leading order and next to leading order analyses
of the experimental
data~\cite{Badier:1983mj,Betev:1985pf,Conway:1989fs,Owens:1984zj,Sutton:1991ay,
Gluck:1991ey, Gluck:1999xe, Wijesooriya:2005ir,Aicher:2010cb}, but
it is much less constrained than the nucleon PDF and therefore,
the lattice calculations may have more impact in this case, especially
in constraining the $x\to 1$ limit which is not yet well established.

The paper is organized as follows. In section~\ref{sec:setup}, we
discuss our lattices setup. In section~\ref{sec:2pt}, we present
the calculations of the two point function of the boosted pion and
check how reliable the extractions of the ground state and the first
excited state are. In Section~\ref{sec:3pt}, we present our results
for the pion three point function that defines the qPDF. Here, we
also discuss the problem of excited state contamination. In
section~\ref{sec:pert}, we discuss the non-perturbative renormalization
as well as the validity of 1-loop matching. Our results on the
renormalized pion qPDF and the matching to PDF are presented in
section~\ref{sec:PDF}. Some technical aspects of the calculations
are discussed in the Appendices. Preliminary results on this work
have been reported in conference
proceedings~\cite{Petreczky:2018jqc,Karthik:2018wmj,Shugert:2018pwi}.

\section{Lattice setup}
\label{sec:setup}
We performed the calculations of the pion two-point and
three-point functions needed to obtain the qPDF using the Wilson-Clover
action for valence quarks on 1-HYP smeared gauge
configurations~\cite{Hasenfratz:2001hp} and the Highly Improved Staggered
Quark (HISQ) action~\cite{Follana:2006rc} in the sea. We used the
2+1 flavor gauge configurations corresponding to lattice size
$48^3\times64$ and the lattice spacing of $a=0.06$ fm generated by
the HotQCD collaboration~\cite{Bazavov:2014pvz}.  In addition to this
ensemble, we also used $64^4$ HISQ lattices~\cite{Bazavov:2014pvz}
with the lattice spacing $a=0.04$ fm for the study of the
non-perturbative renormalization (NPR).  In both the ensembles, the
sea quark mass was tuned to a pion mass of 160 MeV.  A similar setup
was used by the PNDME collaboration albeit for 2+1+1 flavor MILC
configurations (c.f., Ref.~\cite{Liu:2018uuj}).  For the valence
quark masses, we used the values $a m=-0.0388$ (i.e., $\kappa=0.12623$)
for the $a=0.06$ fm ensemble and $a m=-0.033$ (i.e., $\kappa=0.12604$)
for the $a=0.04$ fm ensemble, which are tuned such that the pion
mass, $m_\pi$, is 300 MeV. We did not see any exceptional configurations
for these valence quark masses in our calculations.

We used higher statistics at smaller quark-antiquark separations $z$
than at larger ones; to be exact, we used $216$, $100$ and $48$
gauge configurations for $|z|/a\in [0,8]$, $(8,16]$ and $(16,24]$
respectively.  
We further improved the statistics by using the
All-Mode Averaging (AMA)~\cite{Shintani:2014vja} technique in the computations of
the two- and three-point functions, with 32
sloppy calculations to one exact solve for each configuration.  
For the exact and sloppy inversions, we used the stopping criterion
of $10^{-10}$ and $10^{-4}$, respectively.
In our study, we will consider the valence quark
distribution, which in turn is related to the iso-vector $u-d$ quark
distribution in the pion, and thus we do not compute the quark line
disconnected diagrams.

For a reliable extraction of qPDF, a good overlap of the source
operator with the pion state is necessary so as to project out the
ground state at as small source-sink separation as possible. The
quark sources with Gaussian profile, typically implemented through
a gauge covariant Wuppertal smearing~\cite{Gusken:1989ad}, are used
for this purpose when the hadron is at rest.  However, for the fast
moving hadrons that are required in the qPDF framework, the
use of the Gaussian sources is no longer sufficient and this
necessitates the usage of the boosted Gaussian sources~\cite{Bali:2016lva}
instead. Since we are interested in the calculation of the pion two
and three point function at several values of the pion momenta and
several source-sink separations, we found it more practical to
implement the Gaussian sources by using the Coulomb gauge instead of
implementing the Wuppertal smearing. We found the optimal size of
the Gaussian profile to be about 0.3 fm, which roughly corresponds
to 90 steps of Wuppertal smearing. We checked that in terms of
the signal-to-noise ratio, the Wuppertal and Coulomb-gauge Gaussian
sources are similar (see Appendix~\ref{app_2pt}). In the next
section, we discuss the boosted sources in detail and the energy
levels of the boosted pion. In Appendix~\ref{app_momsmear}, we have
explained the construction of boosted sources in detail.

Out of 216 gauge configurations used in our calculations, 24 gauge
configurations were analyzed using the GPU cluster in BNL to calculate
two-point and three-point correlation functions. These calculations
were performed entirely on GPU using the QUDA suite~\cite{Clark:2009wm,
Babich:2011np, Clark:2016rdz}, including the inversion of the fermion
operator with multigrid algorithm, communication between GPU devices
to perform covariant shifts, and the necessary spin-color matrix
multiplications.  In QUDA, the communications between GPUs on the
same node are implemented through MPI or as direct peer-to-peer
communications between the GPU devices. We have found that on rare
occasion the QUDA peer-to-peer communications did not finish by the
time the computations started. These rare glitches happened randomly.
We checked, however, that these glitches did not affect our results
noticeably compared to other errors.

\section{Two point function of the boosted pion}
\label{sec:2pt}

\begin{figure}
    \centering
\includegraphics[scale=0.85]{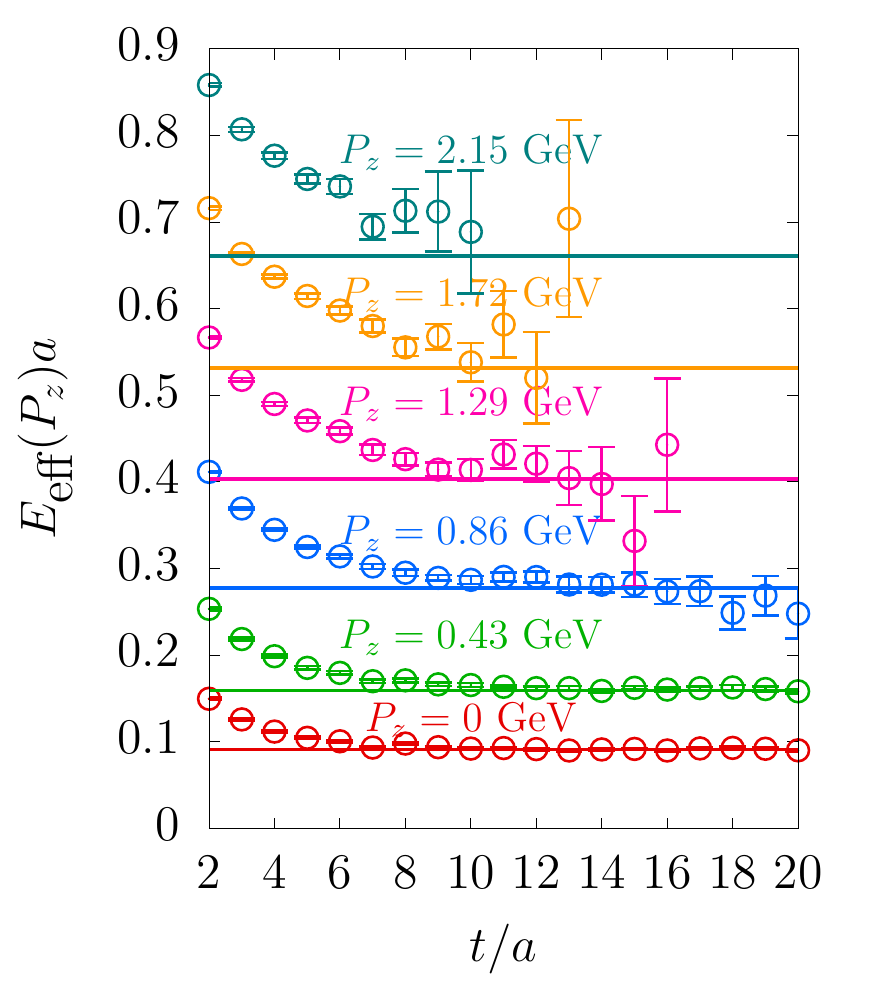}
    \caption{The effective masses $E_{\rm eff}$ from the pion two
    point functions with the boosted Coulomb gauge Gaussian source and sink for
    different momenta as a function of the source-sink separation $t$.
    The horizontal lines are the energy levels from the continuum
    dispersion relation with $m_\pi=300$ MeV.}
\label{fig:effm}
\end{figure}

We calculated the two point functions of the positively charged
pion ($\pi^+=\overline{d}u$),
\begin{equation}
    C^{s s'}_{\rm 2pt}(t,P_z)=\left\langle \left[\pi^+_s(t,\vec{P}\right]\left[ \pi^+_{s'}(0,\vec{P})\right]^\dagger \right\rangle,
\end{equation}
for a spatial pion momentum $\vec{P}=(0,0,P_z)$ which is non-zero
only along the $z$-direction, using the pion source and sink
$\pi^+_{s}(0,\vec{P})$ and  $\pi^+_{s'}(t,\vec{P})$, respectively.
The values of momenta in lattice units are $a P_z=\pm 2\pi n_z/48$ for
$n_z$ ranging from 0 to 5, which in physical units correspond to
$P_z=0,0.43,0.86,1.29,1.72$ and $2.15$ GeV, respectively.  We
always used the Coulomb gauge Gaussian smeared-source ($s=S$), and
either a smeared sink ($s'=S$) or point sink ($s'=P$).  In the rest
of the paper, we will refer to the smeared-source and smeared-sink
set-up to be \emph{SS}, and we will refer to the smeared-source
point-sink set-up as \emph{SP}. 

For the lowest two momenta, we used
the usual Gaussian sources.  To improve the signal for the higher
momenta, we followed Ref.~\cite{Bali:2016lva} and used boosted
sources in which the valence quarks are boosted to a momentum $k_z =
\zeta P_z$, with $\zeta$ being a tunable parameter.  Naively,
one might expect that the optimal choice to be $\zeta=0.5$.  However,
we found that the optimal choice of $\zeta$ for the pion in terms
of the signal-to-noise ratio is between $0.6-0.75$. For $P_z=0.86$
GeV the signal-to-noise ratio is not very sensitive to the value
of $\zeta$.
These findings are in agreement with Ref.~\cite{Bali:2016lva}.  
We discuss the optimization of boosted sources 
further in Appendix \ref{app_2pt}.
Since,
we need to create a source for each value of $\zeta$, we used
$k_z=2(2\pi/48)$ for $n_z=2,3$ and $k_z=3(2\pi/48)$
for $n_z=4,5$, corresponding to the choices of the parameter $\zeta=1,2/3,3/4$
and $3/5$ for $n_z=2,3,4$ and $5$ respectively. We have shown the
corresponding effective masses for the SS two point functions in
\fgn{effm}. By using the boosted smeared sources, one can see that a
reasonable signal for the two point correlation function can be
obtained up to source-sink separations $t=12a$ for all momenta
except for the highest momentum $P_z=2.15$ GeV.  Simply from the
data points in \fgn{effm}, we see that the effective masses approach
a plateau corresponding to the continuum dispersion relation
$E_{\pi}(P_z)=\sqrt{P_z^2+m_{\pi}^2}$, shown as the horizontal lines. 
The effective mass approaches the plateau region at larger
source-sink separations when the momentum is increased, as one would
expect from the shrinking gap between the ground and excited states
as the pion is boosted.

\begin{figure}

\includegraphics[scale=0.53]{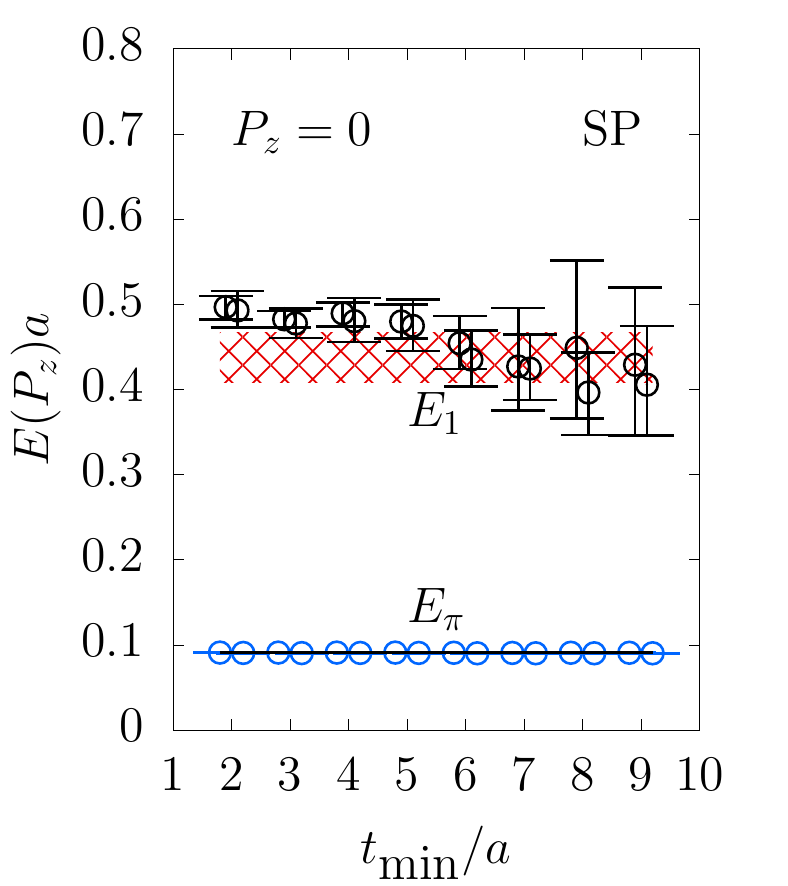}
\includegraphics[scale=0.53]{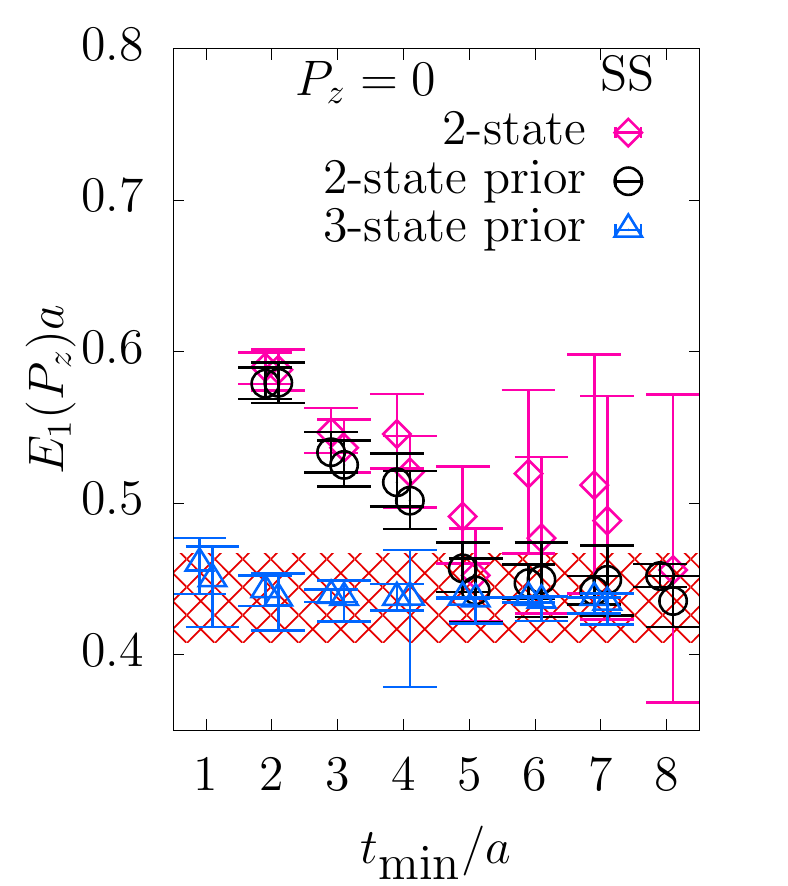}

\includegraphics[scale=0.53]{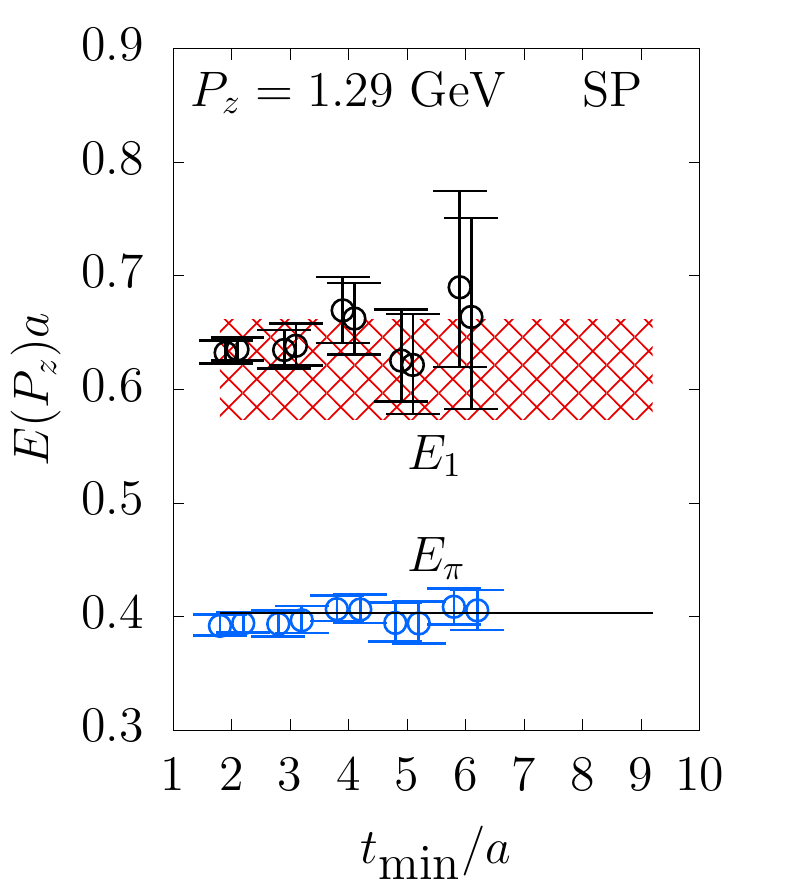}
\includegraphics[scale=0.53]{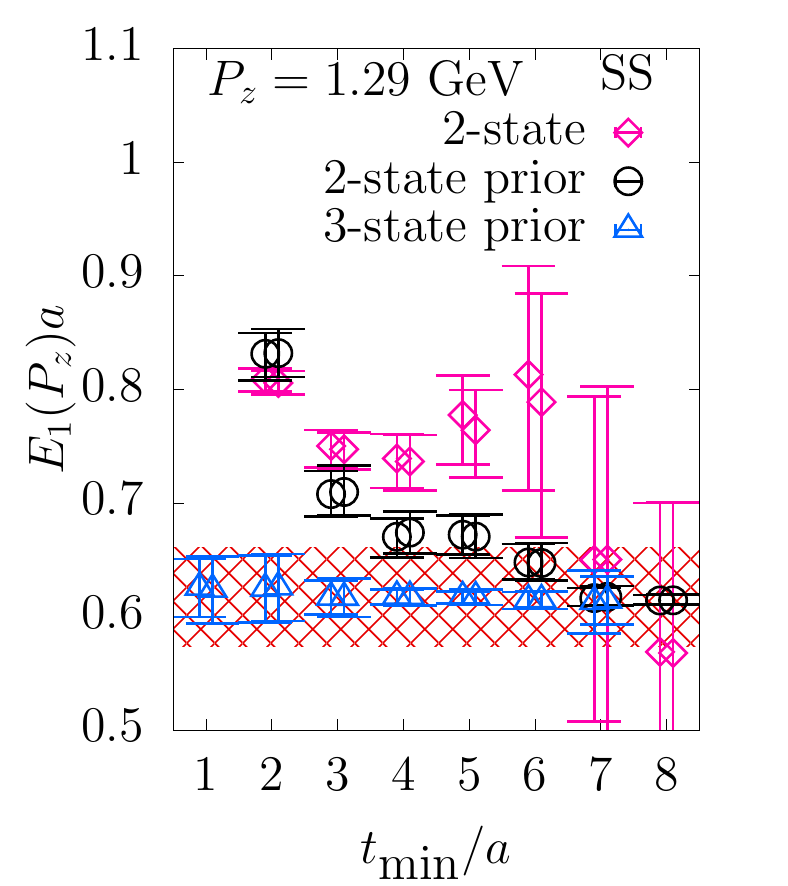}

\caption{
The systematical dependence of the ground state $E_\pi$ and the 
    first excited state $E_1$ on the fit range $[t_{\rm min},t_{\rm max}]$
    is shown.  In the left panels, we show such a dependence for the 
    pion SP correlator at two different $P_z$. For each $t_{\rm min}$, data from
    $t_{\max}=24a$ and $32a$ are shown. The black solid line is the value of $E_\pi$
    expected from the continuum dispersion relation. 
    The red patterned band is our best estimate of
    $E_1$ using the SP correlator.  In the right panels, the fit systematics 
    of $E_1$ for the SS correlator is shown. The red band is the prior used
    for $E_1$ from the SP correlator (same as the one in the left panels). The different symbols are
    the various fit strategies.}
\label{fig:energy}
\end{figure}

As we will discuss next, we used source-sink separations $t=8a,10a$
and $12a$ for the computation of three-point functions. Therefore,
we needed to analyze the excited state contribution to the SS and
SP correlators to perform the infinite source-sink extrapolations.
For this, we performed multi-exponential fits on the SS and SP pion
two point functions in the interval $t\in[t_{\rm min},t_{\rm max}]$
in order to extract the energy levels.  For fixed $t_{\rm min}$, we
varied $t_{\rm max}$ and checked the sensitivity of the result to
$t_{\rm max}$. Then, we repeated the procedure for different values
of $t_{\rm min}$.  We found that we were able to reliably extract
the ground state $E_{\pi}(P_z)$ as well as the first excited state
$E_1(P_z)$ using the four-parameter two-state fits to the SP correlator
instead of using the SS correlator.  This could be due to the fact
that the contributions from the high-lying energy levels are smaller
in the SP correlator compared to the SS correlator stemming from
the possible cancellations between the positive as well as the negative
amplitudes that are allowed in the SP correlator.  In the top-left
and bottom-left panels of \fgn{energy}, we have shown the systematics
of the two-state fits to the SP correlator.  In the top-left and
the bottom-left figures, we have shown the dependence of the best
fit values of $E_\pi$ (blue circles) and $E_1$ (black circles) as
a function of $t_{\rm min}$ used in the fits. For a given $t_{\rm
min}$, the data points from two values of $t_{\rm max}$ have been
clubbed together for $P_z=0$ and 1.29 GeV respectively, and it
demonstrates that there is no dependence on $t_{\rm max}$.  The
ground state is seen to compare well with the expectation from the
dispersion relation shown by the black solid lines.  The red band
shows the values of $E_1$ chosen as the best estimate of the first
excited state.

On the top-right and bottom-right 
panels of \fgn{energy}, we show similar plots for the
first excited state $E_1$ as estimated using the SS correlator.
The statistical errors of the excited state energy $E_1$ in the
simple two state fits (magenta diamond) quickly grow large with
increasing $t_{\rm min}$ and thus, these fits turned out to be of
limited use. Therefore, we performed constrained two- and three-state
exponential fits for the SS correlator with the ground state energy
fixed to $E_{\pi}=\sqrt{P_z^2+m_{\pi}^2}$ with $m_\pi=300$ MeV, 
and imposing a prior on
$E_1$ using its best estimate from the SP correlator -- that is,
we added the term $(E_1-E_{1,{\rm prior}})^2/\sigma_{\rm prior}^2$
to the $\chi^2$ with $E_{1,{\rm prior}}$ and $\sigma_{\rm prior}$ being
the mean and error of $E_1$, respectively, 
as determined from the SP correlator.  The $t_{\rm
min}$ dependence of the resulting $E_1$ from the constrained two-state
fit (black circles) and constrained three-state fit (blue triangles)
are shown in the top-right and bottom-right panels.  The two-state
fits of the SS two-point correlator largely overestimate the energy
of the first excited state for small $t_{\rm min}$ whether or not
priors are used, and there is a significant dependence on $t_{\rm
min}$. One should use $t_{\rm min} \ge 6a$ to obtain reliable results
for the first excited state from the SS correlator.  The three-state
fits with priors on $E_\pi$ and $E_1$ give energies of excited
states that are the same within errors for the SS two point correlators
and show almost no $t_{\rm min}$ dependence. In summary, we determined
the lowest three energy levels using the SP correlator and then
determined the corresponding amplitudes $|A_n| =
|\bra{0}\pi^+_S(0,\vec{P})\ket{E_n,P_z}|$ of these excited states
in the SS correlator through a constrained fit analysis.

In \fgn{disp}, we show the three energy levels obtained from the different
fits discussed above, as a function of $P_z$.  For $P_z=0$, we compare
our result with the energy levels that would correspond to the pion
resonances $\pi(1300)$ and $\pi(1800)$ from the PDG~\cite{Tanabashi:2018oca}.
In order to account for the 300 MeV pion mass, we shifted the PDG
values by 0.161 GeV as an approximation and these are shown as the
two arrows in \fgn{disp}.  Our estimate of the first excited state
energy agrees with this shifted mass of $\pi(1300)$. We also
show the expected $P_z$-dependence of $E_1(P_z)$ assuming a
particle-like dispersion and this describes the actual data very
well. The energy of the second excited state is much larger than
expected, meaning that the third state effectively parametrizes
several higher lying states.  As one can see from the figure the
energy gap between $E_\pi$ and $E_1$ shrinks with increasing $P_z$
as expected.
\begin{figure}
    \includegraphics[scale=0.7]{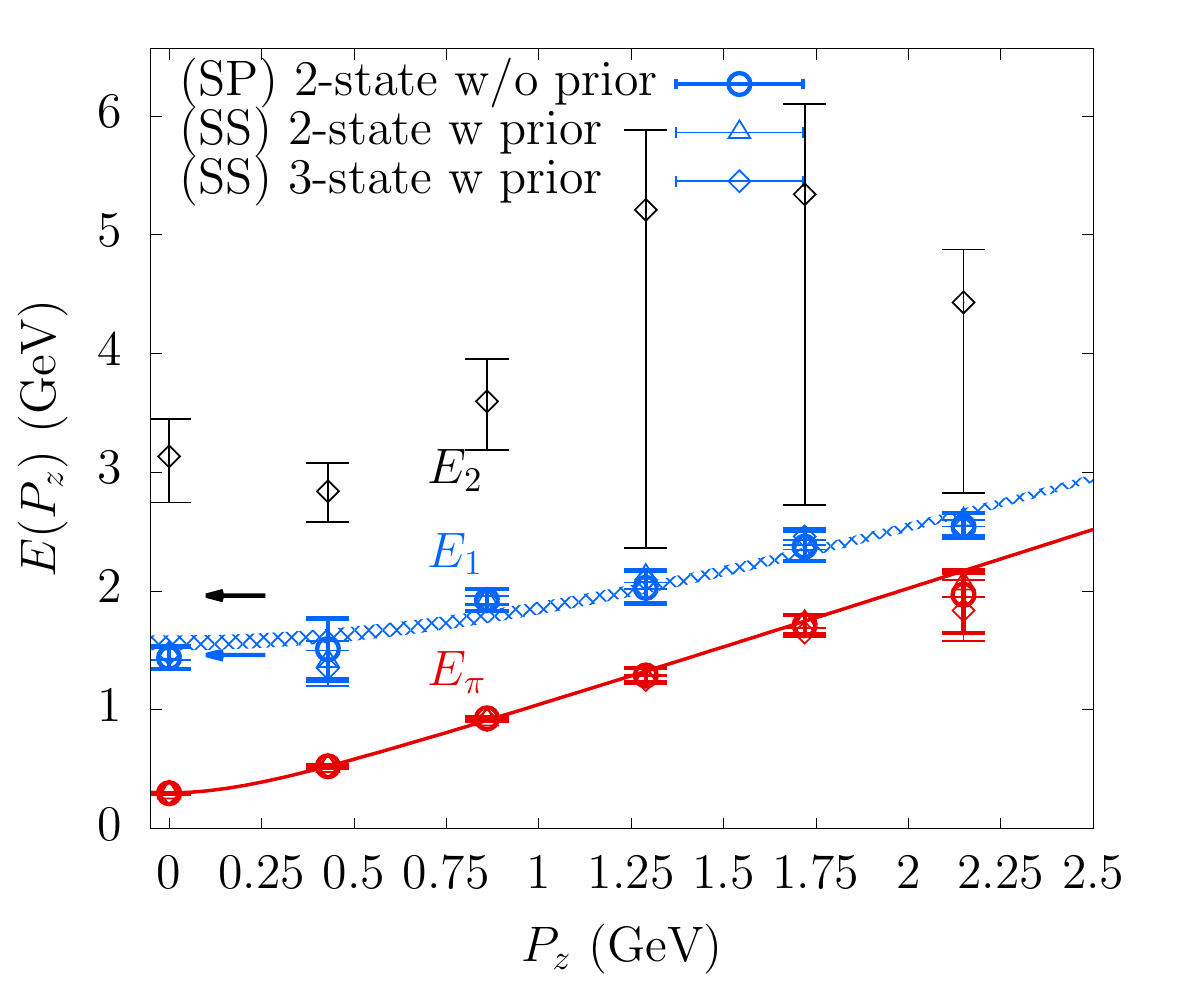}
    \caption{The energies of the ground state and the first two excited
    states as functions of $P_z$.
    The red, blue and black symbols correspond to $E_\pi,E_1$ and
    $E_2$ respectively.  For each color, the different symbols
    correspond to different fitting methods (2-state and 3-state
    fit with or without prior on the ground state) and the types of source-sink
    (SP or SS).  The lines show the expected dispersion relations
    for the pion and its first excited state.  The arrows are the
    PDG values of $\pi(1300)$ and $\pi(1800)$ which are shifted to account for 
    $m_\pi=300$ MeV.
    }
\label{fig:disp}
\end{figure}
The results on the excited state energies will be important for the
analysis of the pion three-point function discussed in the next
section.

\section{Extraction of the bare quasi-PDF matrix elements from the three-point functions}
\label{sec:3pt}

\begin{figure*}
\includegraphics[scale=0.585]{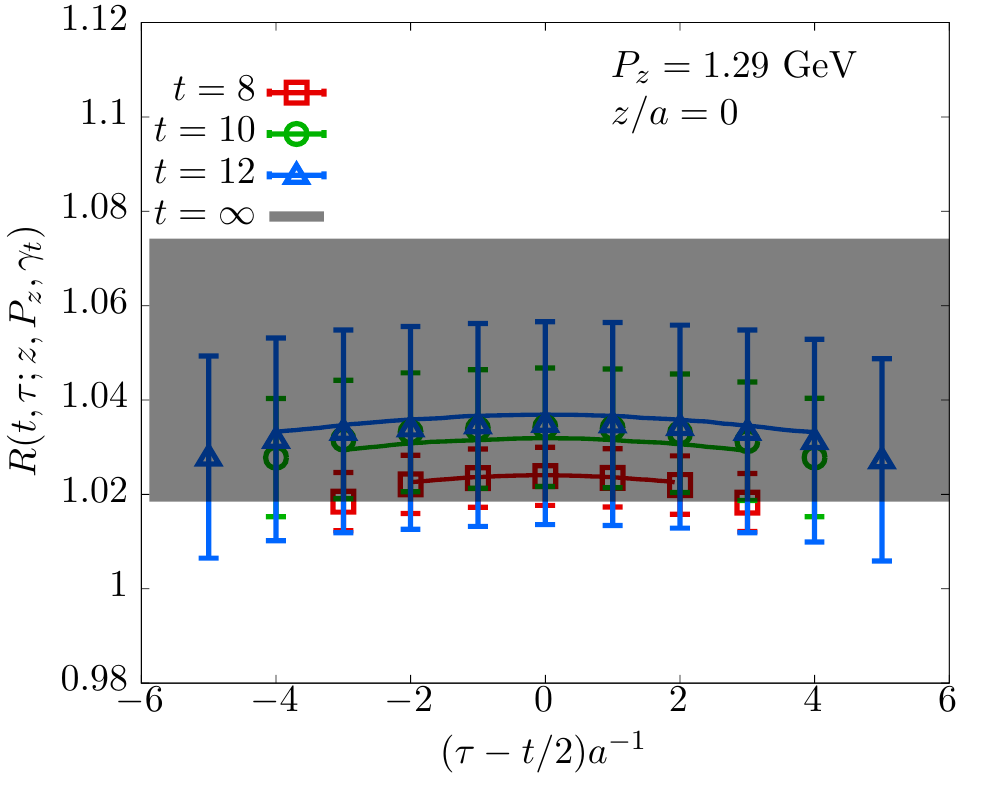}
\includegraphics[scale=0.585]{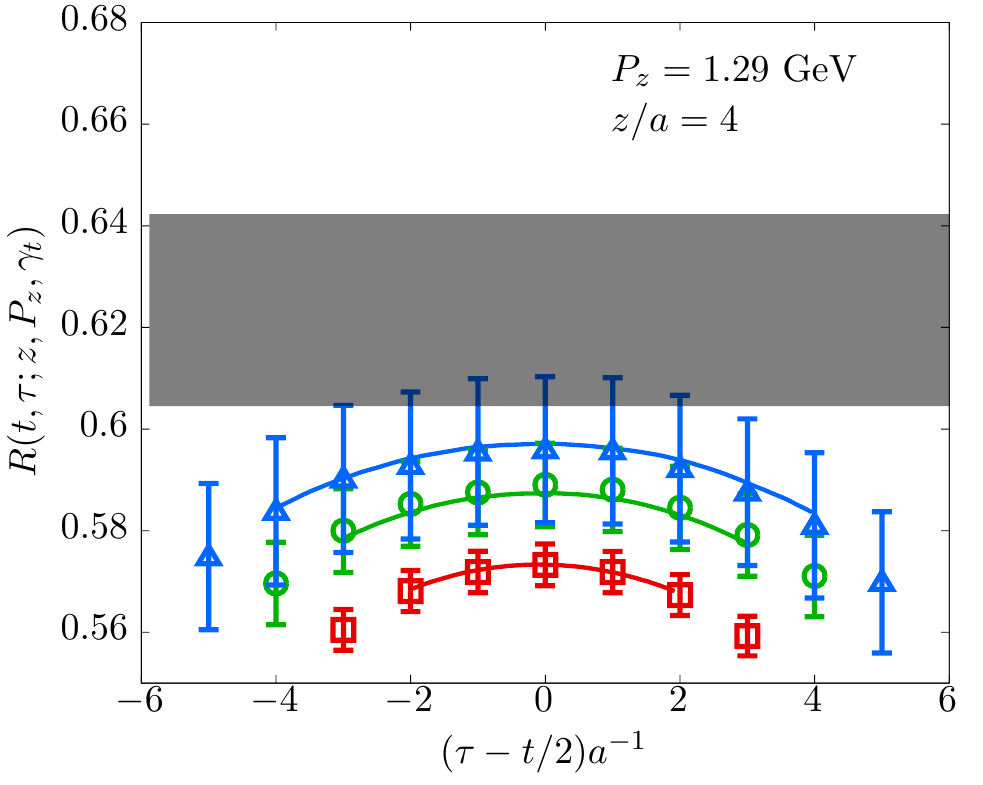}
\includegraphics[scale=0.585]{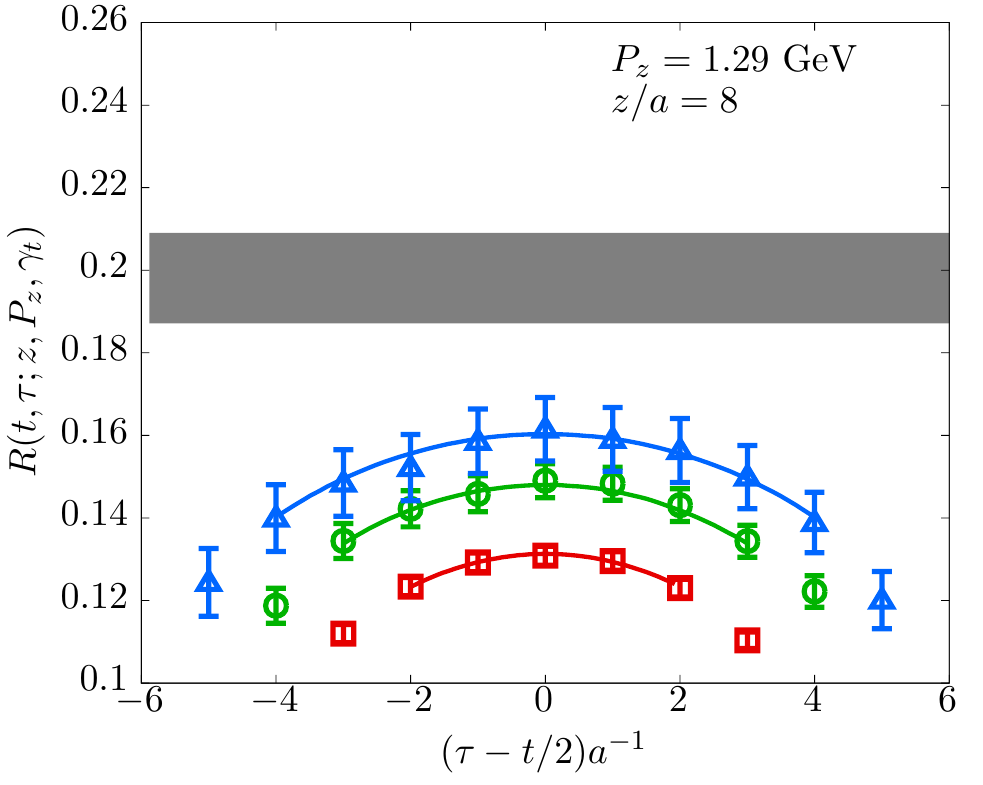}

\includegraphics[scale=0.585]{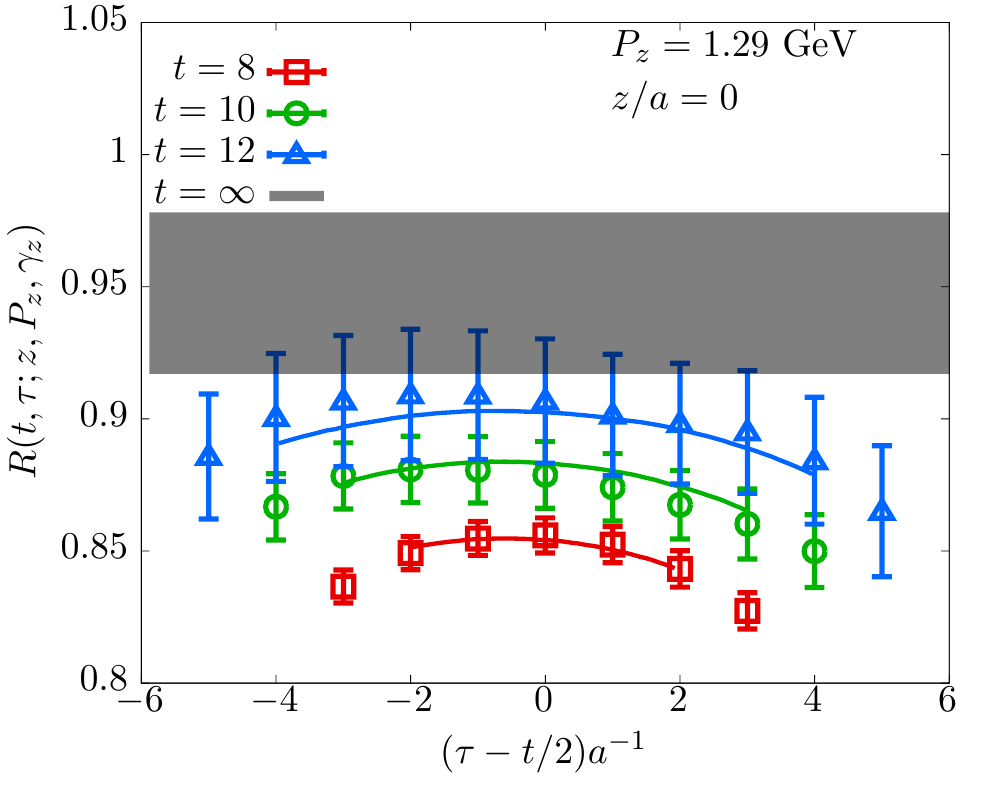}
\includegraphics[scale=0.585]{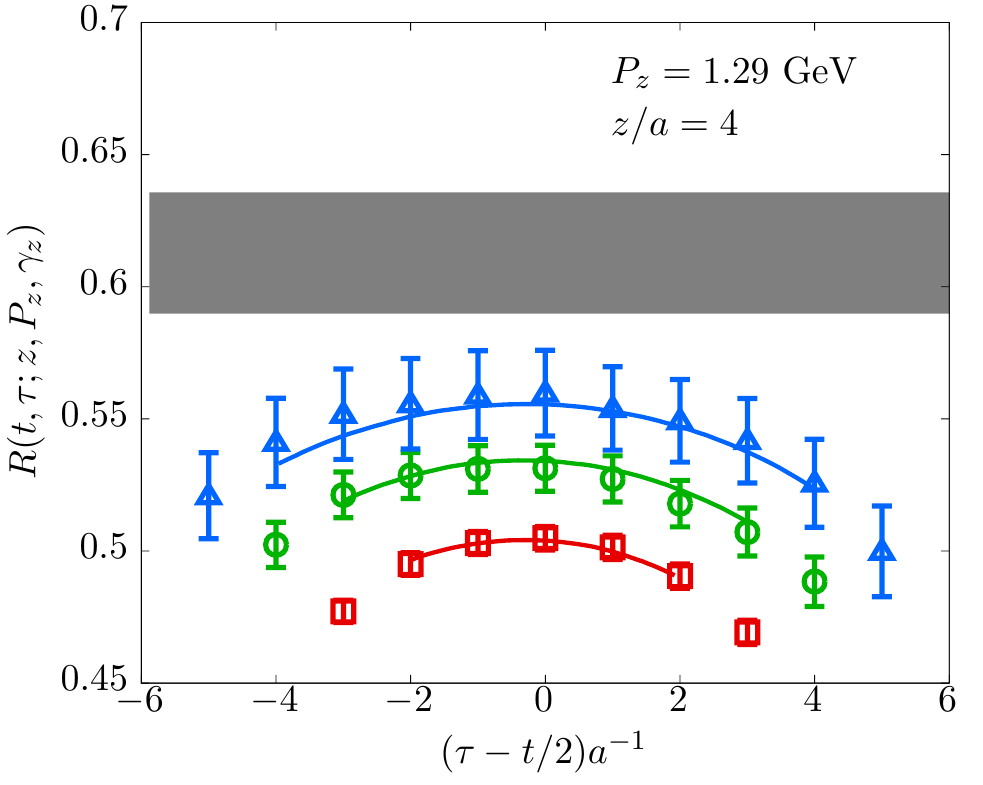}
\includegraphics[scale=0.585]{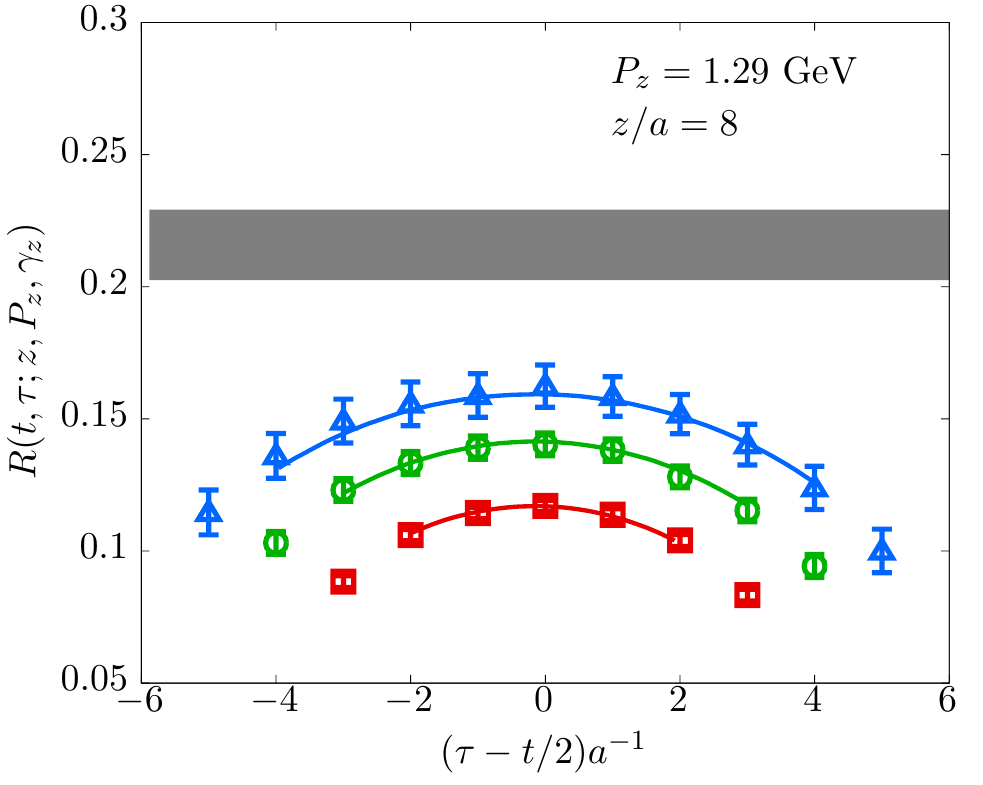}
\caption{
The ratio of the three point function to the two-point function,
$R(t, \tau; z, P_z, \Gamma)$ is shown as function of $\tau-t/2$ for
$\gamma_t$ (top row) and $\gamma_z$ (bottom row) for $z/a=0$, 4 and
8 (from left to right) and $P_z=1.29$ GeV.  The central values of
the two states fits of the lattice results for different source-sink
separations are shown as the curves. The horizontal band corresponds
to the extrapolated result for the infinite source-sink separation.
}
\label{fig:3pt_fit}
\end{figure*}

The next step is the calculation of the bare qPDF matrix element
\begin{equation}
    h^B_{\Gamma}(z,P_z)=\mel**{E_\pi, P_z}{\mathcal{O}_{\Gamma}(z;\tau)}{E_\pi,P_z}, 
\end{equation}
where the bilocal $u-d$ qPDF operator in a time-slice $\tau$
involving a quark and an antiquark separated along the $z$-direction by
${\cal L}=(0,0,0,z)$ is given by
\begin{eqnarray}
    &&\mathcal{O}_{\Gamma}(z;\tau)=\sum_{\vec{x}}\left(\overline{u}_x W_{x,x+{\cal L}}\Gamma u_{x+{\cal L}} - \overline{d}_x W_{x,x+{\cal L}}\Gamma d_{x+{\cal L}}\right),\quad\text{where}\cr
    &&\Gamma=\gamma_t,\gamma_z,1;\quad W_{x,x+{\cal L}}=\prod_{x'=x}^{x+{\cal L}} U_3(x'),
\label{Oz}
\end{eqnarray}
and it is made gauge-invariant by the Wilson line $W_{x,x+{\cal L}}$.
The Dirac $\gamma$ matrices in the qPDF operator are in the
Minkowskian convention.  The state $\ket{E_\pi,P_z}$ denotes the
onshell ground state pion with momentum $P_z$.  In addition to the
natural choices of $\Gamma=\gamma_t$ and $\gamma_z$ that approach
$\gamma^+$ in the light-cone limit, we also considered $\Gamma=1$.
This choice of $\Gamma$ is needed because under renormalization,
$\mathcal{O}_{\gamma_z}(z)$ mixes with
$\mathcal{O}_1(z)$~\cite{Constantinou:2017sej}.  We applied one-level
of HYP smearing to the links entering $W_{x,x+{\cal L}}$ in order
to reduce the noise. Since the qPDF calculation involves values
of $z\sim{\cal O}(a)$, we checked that there is no significant
difference between the renormalized matrix elements using the smeared
and unsmeared Wilson line.  To obtain the bare matrix element
$h^B_{\Gamma}(P,z)$, we computed the three-point function at different
source-sink separations $t$ and operator insertion point $\tau$
\begin{equation}
    C_{\rm 3pt}^{SS}(t,\tau;z,P_z)= \bigg{\langle}\big{[}\pi^+_S(t,\vec{P})\big{]}\mathcal{O}_{\Gamma}(z;\tau)\big{[}\pi^+_S(0,\vec{P})\big{]}^{\dagger}\bigg{\rangle},
\end{equation}
and constructed the ratio
of the three-point function to two-point function,
\begin{equation}
    R(t, \tau; z, P_z, \Gamma) = \frac{C^{SS}_{\rm 3pt}(t,\tau; z,P_z)}{ C^{SS}_{\rm 2pt}(t,P_z)}.
\end{equation}
The reader can refer to Appendix~\ref{app_contract} for a detailed
description of the construction of three-point functions.  The two-point 
function is always real when the source and sink are of the same
type. The three-point function for the $u-d$ qPDF operator
${\cal O}_{\Gamma}(z)$ in a pion external state is real at all $z$
for $\Gamma=\gamma_z,\gamma_t$ and purely imaginary for $\Gamma=1$
(refer Appendix~\ref{pionreal}).  Inserting a complete set of states
in the above equation,
\begin{equation}
    R(t, \tau; z, P_z, \Gamma)=\frac{\sum_{n,n'}A_nA_{n'}^* \mel**{E_n,P}{\mathcal{O}_{\Gamma}(z)}{E_{n'},P)} e^{-(E_{n'}-E_n)\tau-E_n t}}{\sum_m |A_{m}|^2 e^{-E_m t}},
\label{3pt_spectral}
\end{equation}
with $E_{n+1} \ge E_n$, and $E_0=E_\pi$.
It is easy to see that
in the infinite $t$ limit, $R(t, \tau; z, P_z, \Gamma)$ is equal
to $h^B_{\Gamma}(z,P)$. 
The above equation holds for the infinite time extent. For
a finite time extent, the effects of periodic boundary condition
should be taken into account. This turns out
to be important for $P_z=0$, while for non-zero $P_z$
the effect is negligible as discussed in Appendix~\ref{matrixelements}.
In practice, one truncates the sums in
\eqn{3pt_spectral} at some value $n$, and then obtain $h^B_{\Gamma}(z,P_z)$
by fitting the $t$ and the $\tau$ dependence of $R(t, \tau; z, P_z,
\Gamma)$ using $\bra{E_n,P}\mathcal{O}_{\Gamma}(z)\ket{E_{n'},P)}$
as fit parameters. In the fits, the values of 
$A_n$ and $E_n$ were held fixed at values determined from the two-state 
fit analysis on the SS correlators. In what follows, we will refer to this method
of fitting using $n$-state Ansatz to the data between $\tau/a>\tau_o$
and $\tau/a < t/a-\tau_o$ as \texttt{Fit}($n$,$\tau_o$).  In this
method, the excited states are suppressed by $\exp(-(E_n-E_\pi)
t/2)$.  For $z=0$, it is easy to see that $R(t, \tau; z, P_z,
\Gamma)$ is symmetric in $\tau$ around the mid-point $\tau-t/2$.
For $z \neq 0$ and $P_z\neq 0$, we only have the following relation
(see Appendix~\ref{matrixconj}),
\begin{equation}
    \bra{E_n,P_z}\mathcal{O}_{\Gamma}(z)\ket{E_{n'},P_z)}^*=\phi_\Gamma\bra{E_{n'},-P_z}\mathcal{O}_{\Gamma}(z)\ket{E_n,-P_z)},
\end{equation}
where $\phi_\Gamma=1$ for $\Gamma=\gamma_t,1$ and $\Phi_\Gamma=-1$
for $\Gamma=\gamma_z$.  Thus, generically the matrix elements
$\bra{E_n,P_z}\mathcal{O}_{\Gamma}(z)\ket{E_{n'},P_z)}$ and
$\bra{E_{n'},P_z}\mathcal{O}_{\Gamma}(z)\ket{E_n,P_z)}$ are independent
and the number of fit parameters is thereby increased. Based on the
above relation, we constructed appropriate averages using both
the positive and negative values of momenta to increase the statistics.
However, in practice the gain was marginal.

We demonstrate the extraction of the matrix element using the fit
method in \fgn{3pt_fit}, where the ratios $R(t, \tau; z, P_z, \Gamma)$
are shown for $\Gamma=\gamma_t$ and $\gamma_z$ qPDFs for the $P_z=1.29$
GeV pion.  The results on $R(t,\tau;z,P_z,\Gamma)$ and the extraction
of matrix elements for the other values of $P_z$ are given in
Appendix~\ref{matrixelements}.  In the figure, we show the data at
$t/a=8,10$ and $12$ along with the result of \texttt{Fit}(2,2).
Using the fit, the results for the $t \to \infty$ extrapolations
are shown with the horizontal bands.  We also performed the three states
fit of $R(t, \tau; z, P_z, \Gamma)$ and the picture looks similar. In
this case, the data points at all $\tau-t/2$ could be described by
the fit. The $t \to \infty$ extrapolations from the three state fit
gave results consistent with the two state ones, albeit with larger
errors. A closer look at \fgn{3pt_fit} (and also from
\fgn{3pt_fit_gammaz_all}) reveals that the excited state contribution
is larger for $\Gamma=\gamma_z$ than for $\Gamma=\gamma_t$.
Furthermore, the excited state contribution grows with increasing
$z$.  The non-symmetric nature of $R(t, \tau; z, P_z, \Gamma)$ for
$z \neq 0$ is also apparent in the figure. We expect that
$h_{\Gamma}(z=0,P)=1$ for $\Gamma=\gamma_t$ because of the charge
conservation once a proper  renormalization is implemented and the
continuum limit is taken.  Our extrapolation procedure gives a
result for the bare matrix element which is larger than one at
all the values of momenta as can be seen in \fgn{3pt_fit}, as well as from
\fgn{3pt_fit_gammat_all} 
in Appendix~\ref{matrixelements} where, in addition,
one can also see that $h^B_{\gamma_t}(z=0,P_z)$ is independent of $P_z$.
Thus, any deviation of  $h^B_{\gamma_t}(z=0,P_z)$ away from unity should be 
taken care of by the renormalization. We will see in \scn{pert} that this is indeed 
the case.

Alternatively  we can use the summation method \cite{Maiani:1987by}
to obtain $h^B_{\Gamma}$.  Here one sums over all $\tau/a$ minus a
certain number of end points $\tau_o$
\begin{equation}
  R_{\text{sum}}(t; z, \Gamma) =
  \sum_{\tau/a=\tau_o}^{t/a - \tau_o}R(t, \tau; z, P_z, \Gamma).
\end{equation}
We will refer to this method as \texttt{sum}($\tau_o$).
For large $t$, one would find a linear behavior in $t$ of $R_{\rm sum}$ as
\begin{equation}
R_{\text{sum}}(t; z, \Gamma) \simeq
   (t - 2\tau_o) h^B_{\Gamma}(z,P_z) +{\rm const}+{\cal O}(e^{-(E_1-E_\pi) t}).
\end{equation}
The advantage of this method is that the excited state
contributions are suppressed as $\exp[-(E_n-E_\pi) t]$ instead of
being suppressed as $\exp[-(E_n-E_\pi) t/2]$ in the fitting method.
We show a sample result using \texttt{sum}(1) and \texttt{sum}(2)
in \fgn{sum} for $\Gamma=\gamma_z$ and $z=0$.  We see that
$R_{\text{sum}}(t; z,\gamma_z)$ can be well fitted by a straight
line in $t$, and the slope gives the value of the matrix element.
As a cross-check, we also show the expected curve for $R_{\text{sum}}(t;
z, \Gamma)$ using our best fit from \texttt{Fit}(2,2) as the dashed
curves. It can be seen that the difference between a simple straight
line fit and the curve from \texttt{Fit}(2,2) is small.  One can
also note that \texttt{sum}(1) and \texttt{sum}(2) are almost
parallel, meaning that the extracted matrix element is independent
of $\tau_o$ confirming that the method works well.
\begin{figure}
\includegraphics[width=7cm]{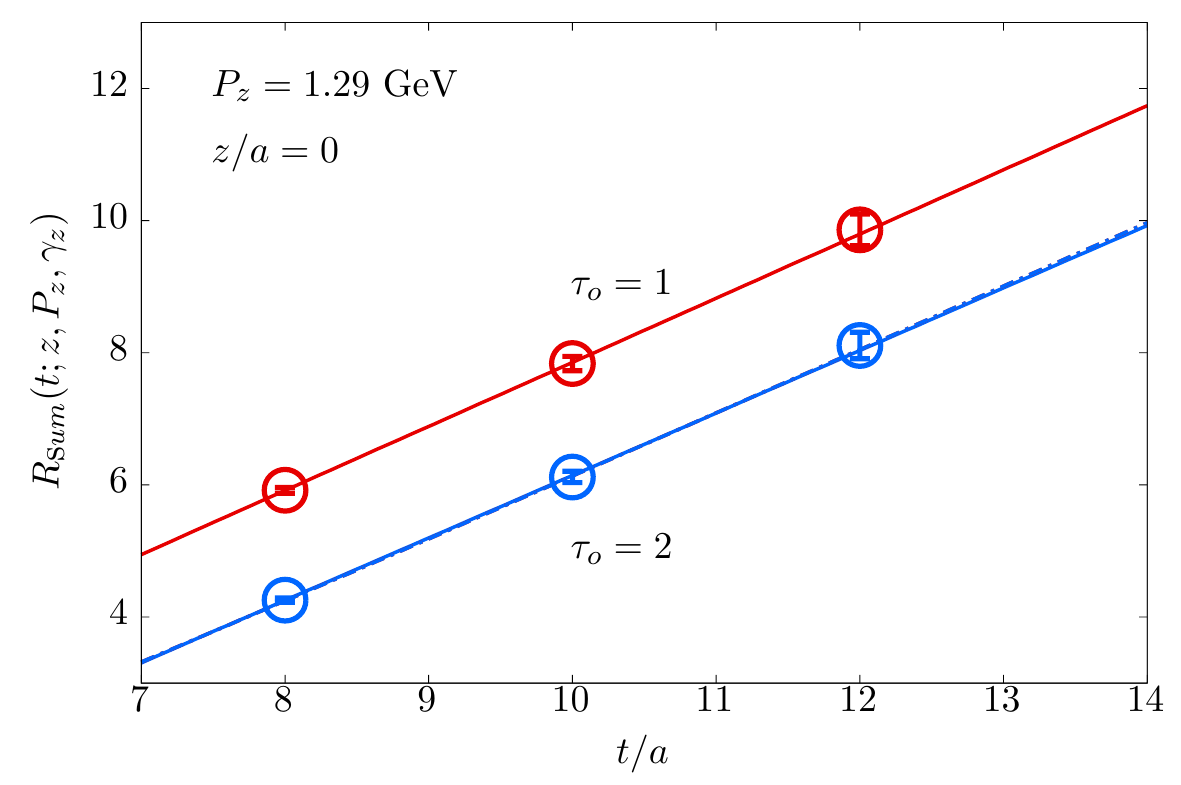}
\caption{The $t$ dependence of $R_{\text{sum}}(t; z=0, \gamma_z)$
    for $P_z=1.29$ GeV with $\tau_o=1$ (red) and $\tau_o=2$ (blue). 
    The solid lines are the straight line fit $h^B_{\rm \gamma_z}(z,P_z) t+{\rm const}$
    to the data. The dashed lines are the expected curve for $R_{\text{sum}}(t; z=0, \gamma_z)$
    using the \texttt{fit}(2,2) best fit parameters.
    }
\label{fig:sum}
\end{figure}

To better understand the systematic effects due to excited state
contaminations, one can look at the case $z=0$ in detail, where the
statistical errors are the smallest.  The bare matrix element
$h^B_{\gamma_z}(z=0,P_z)$ after renormalization is expected to be
proportional to the hadron velocity, $P_z/E_\pi(P_z)$. One can take
the ratio of matrix elements
$h^B_{\gamma_z}(z=0,P_z)/h^B_{\gamma_t}(z=0,P_z)$ to avoid the issues
of renormalization.  The results for this ratio of matrix elements
is shown in \fgn{gammaztogammat} along with the curve for
$P_z/\sqrt{P_z^2+m_\pi^2}$.  We see that our lattice results
follow the expectations reasonably with small $3-4\%$ deviations from
the expected result at small $P_z$. 
A reason for this could be the  large systematic
uncertainty in $\gamma_z$ matrix element due to the relatively larger
excited state extrapolations required.  We see that within errors,
the two-state fit, three-state fit and the summation methods are
consistent.
\begin{figure}
\includegraphics[width=8cm]{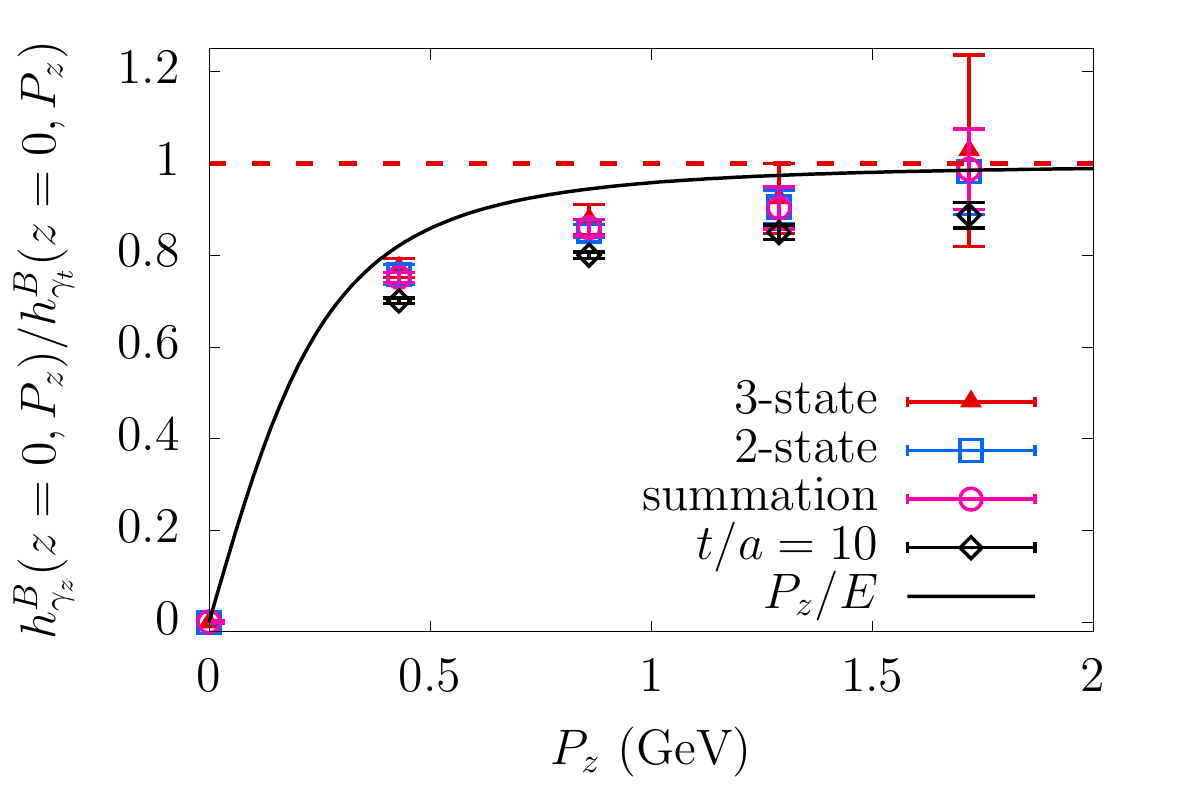}
    \caption{The ratio of the matrix elements for $\gamma_z$ to $\gamma_t$ as a function
    of $P_z$. The curve shows the expected result, $P_z/E_\pi(P_z)$.}
\label{fig:gammaztogammat}
\end{figure}

In \fgn{3pt_extrap}, we show the results for $h^B_{\Gamma}(z,P_z)$
as functions of $z$ for the two highest momenta $P_z=1.29$ and $1.72$
GeV determined using the HYP smeared Wilson line. Since the real
part is symmetric about $z=0$, we have only shown the data for $z\ge
0$. At each $z/a$, we have shown the resulting $t\to\infty$
extrapolated results using \texttt{Fit}(3,1), \texttt{Fit}(2,2),
\texttt{sum}(2) and \texttt{sum}(2) methods, and these points at a
given $z/a$ are slightly displaced for better visibility.  We see
that the results from all these methods agree with each other within
the errors. For $\gamma_z$, some tension between the summation
method, and the two and three state fits is observed at larger $|z|$.
At larger values of $z$ the matrix elements are suppressed partly
because of the larger value of $P_z$ and in part by the divergent
self-energy contribution in the spatial Wilson line. The latter
will be removed upon renormalization as we will see in the next
section.  Having demonstrated a robust determination of the matrix
element using multiple $t\to\infty$ strategies, we will use the
matrix elements obtained using \texttt{Fit}(2,2) in the rest of the
paper.
\begin{figure*}
\includegraphics[scale=0.55]{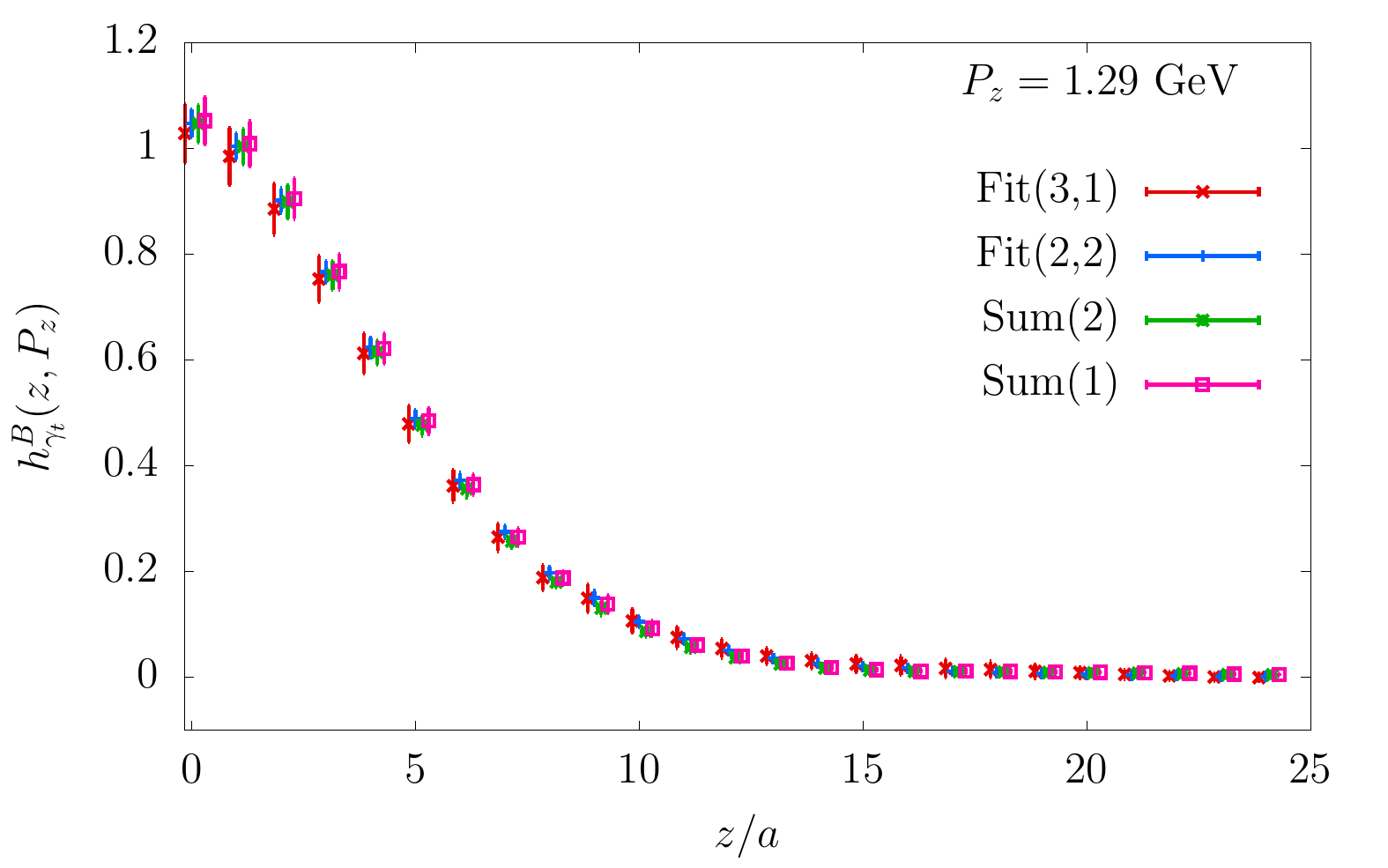}
\includegraphics[scale=0.55]{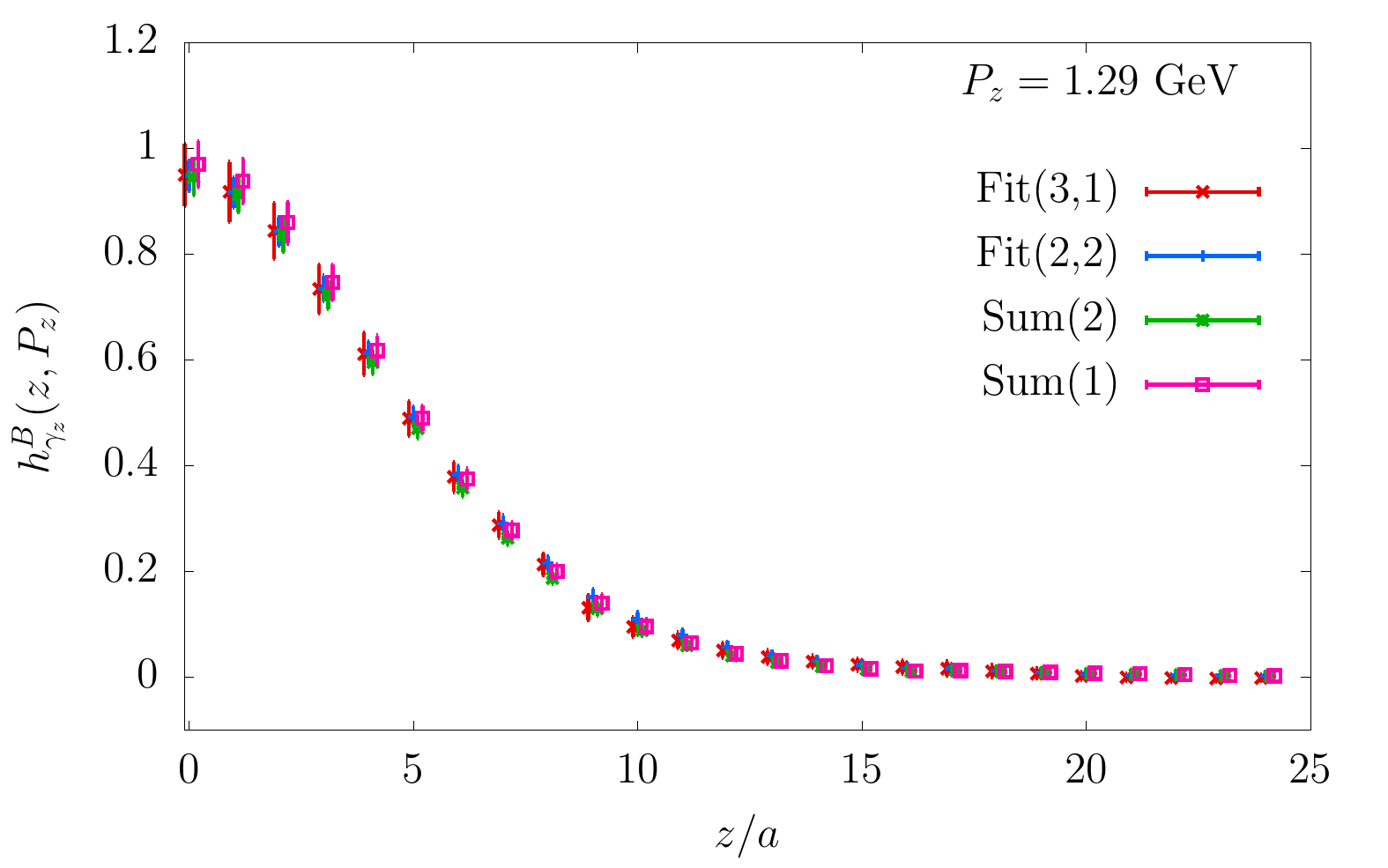}

\includegraphics[scale=0.55]{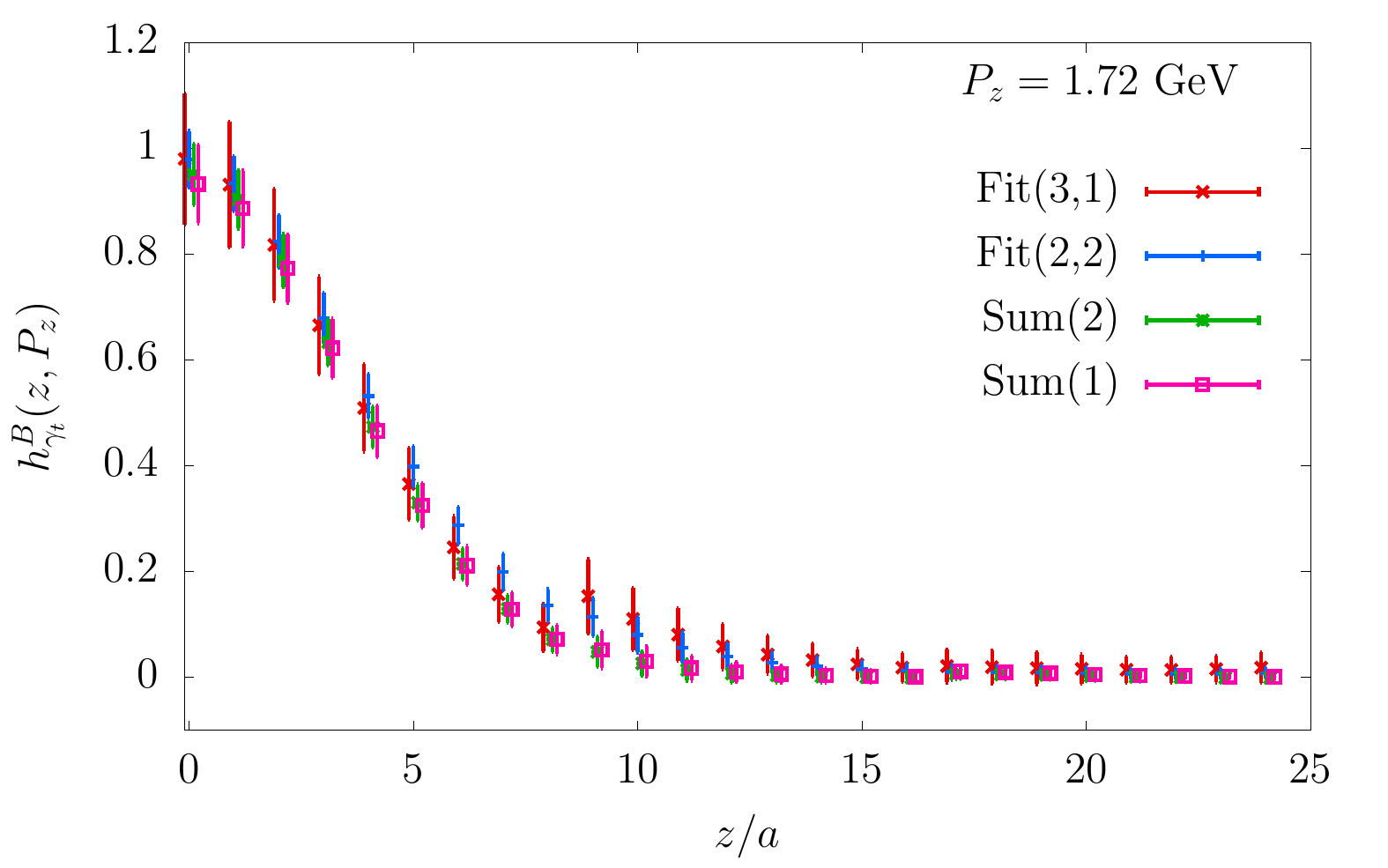}
\includegraphics[scale=0.55]{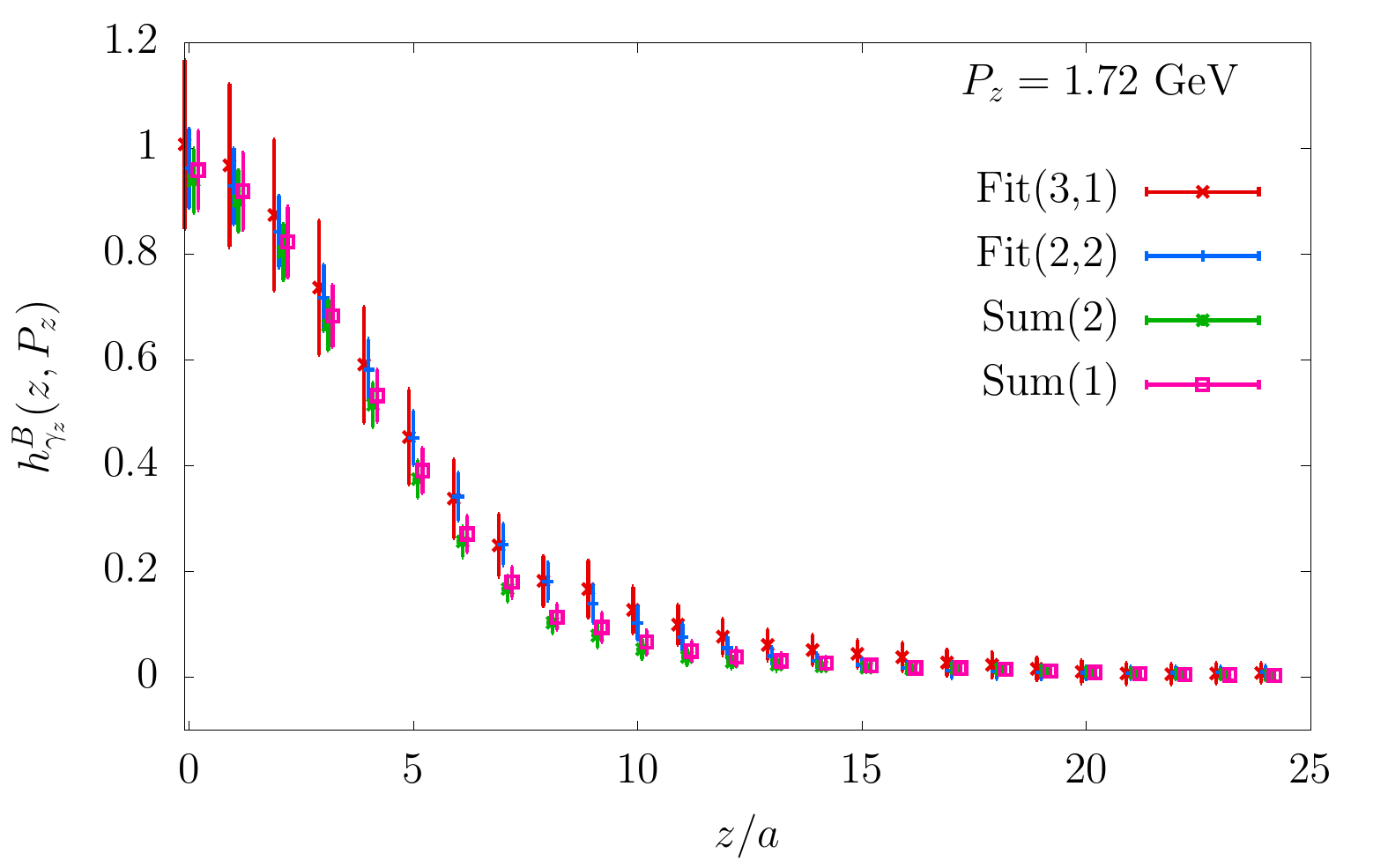}
    \caption{The bare matrix elements $h^B_{\gamma_t}(z,P_z)$ (left) and
    $h^B_{\gamma_z}(z,P_z)$
(right) as a function of quark-antiquark separation $z$.  The panels in the top row show results for $P_z=1.29$
GeV, while the panels in the bottom row show the results for
$P_z=1.72$ GeV. The different symbols are from various
    methods of $t\to\infty$ extrapolation.}
\label{fig:3pt_extrap}
\end{figure*}

So far we discussed results on the three-point function obtained
using 1-HYP smearing for the spatial link. We also performed
calculations using the unsmeared spatial Wilson line.  In this case, the bare matrix
element rapidly decreased with $z$ due
to the larger value of the Wilson line self-energy divergence.
However, we found that the results, after non-perturbative renormalization
(discussed next in \scn{pert}), were similar to those obtained with smeared
Wilson lines within errors.  The main difference between the
renormalized three-point function obtained with the smeared and unsmeared
Wilson line is that, for the latter the statistical errors at large
$z$ are significantly larger.

\section{RI-MOM non-perturbative renormalization and its comparison with 1-loop}
\label{sec:pert}

In the last section, we discussed the extraction of the bare qPDF
matrix element which has to be renormalized.  The renormalizability
of qPDF has been recently demonstrated to all orders of
perturbation theory~\cite{Ishikawa:2017faj,Ji:2015jwa}.  In addition
to the quark wavefunction renormalization $Z_q$ and the composite
operator renormalization required for $z=0$, the qPDF operator
at non-zero $z$ requires additional renormalization due to the the
UV divergence present in the Wilson line connecting the quark and
anti-quark~\cite{Polyakov:1980ca}.  When a lattice fermion that
breaks chiral symmetry at finite lattice spacings is used, as is
the case in this paper, it has been shown that only the renormalization
of $\gamma_t$ qPDF operator is purely multiplicative, while
the $\gamma_z$ qPDF operator mixes with the scalar
qPDF~\cite{Chen:2017mzz,Constantinou:2017sej}.  A renormalization
scheme that is implementable on the Euclidean lattice is the RI-MOM
scheme and it is now standard in the lattice QCD literature. The
corresponding RI-MOM counter-term for the qPDF operator in leading
order perturbation theory has been worked out using off-shell quark
external states~\cite{Stewart:2017tvs}, and it is one of the
ingredient used in the perturbative matching of the RI-MOM renormalized
qPDF to the $\msbar$ PDF. In this section, we discuss the
renormalization procedure, and then compare the running of the
renormalization constants as determined on the lattice with the
corresponding perturbative expectations. This allows us to quantitate
the validity of the leading order perturbation theory and matching.

For non-perturbative renormalization, we compute the expectation
value of qPDF operator between offshell quark external states
with momentum $p$. We refer to the momentum of quark in the direction
of Wilson-line as $p_z$ and the magnitude of the component perpendicular
to the Wilson-line as $p_\perp$.  For these computations, we use
Landau gauge fixing.  Let $\Lambda_{\Gamma}(z,p)$ be the quark-line
amputated bare qPDF,
\begin{equation}
    \Lambda_\Gamma(z,p)=\left\langle Q(p)\right\rangle^{-1}\left\langle \overline{u}(p)\mathcal{O}_\Gamma(z,\tau) u(p)\right\rangle \left\langle Q(p)\right\rangle^{-1},
\end{equation}
where $Q(p)$ is the quark propagator $\langle u(p) \overline{u}(p)\rangle$ and $u(p)=\sum_{x}u_x e^{-ip.x}$.
Let us define the bare qPDF after
projection as
\begin{equation}
q_{\Gamma}(z,p)\equiv{\rm Tr}\left[{\cal P} \Lambda_{\Gamma}(z,p)\right],
\end{equation}
consistent with the definition used in perturbative calculations.
Here, ${\cal P}$ is the operator used to project onto one of the
$\gamma$-matrices $\Gamma=\gamma_\alpha$, and ${\rm Tr}(\ldots)$
is a trace over both color and Dirac indices. Based on previous
works~\cite{Stewart:2017tvs,Chen:2017mzz}, we will use
$\slashed{p}$-projection for which ${\cal
P}=\slashed{p}/\left(12p_\alpha\right)$. Alternatively, one can use
${\mathcal P}=\Gamma$~\cite{Alexandrou:2017huk} or the minimal
projection~\cite{Stewart:2017tvs}.  In the case of $\Gamma=\gamma_t$,
since the renormalization is simply multiplicative, the renormalized
quark qPDF is given by
\begin{equation}
q^R_{\gamma_t}(z,p,p^R)=Z_{\gamma_t\gamma_t}(z,p^R) q_{\gamma_t}(z,p),
\label{zfacgt}
\end{equation}
where the $z$-dependent RI-MOM renormalization constant $Z$ is
determined using the renormalization condition set at momentum $p^R$
as
\begin{equation}
q^R_{\gamma_t}(z,p,p^R)\bigg{|}_{p=p^R} \equiv e^{i p^R_z z}.
\end{equation}
The right-hand-side of the above equation is the tree-level value
of $q_{\gamma_t}$.  The renormalization constant so obtained is in
general a complex number.  For $\Gamma=\gamma_z$, we have to take
care of mixing with the scalar $\Gamma=1$. Hence, the renormalized
qPDF is defined as
\begin{equation}
q^R_{\gamma_z}(z,p,p^R)=Z_{\gamma_z\gamma_z}(z,p^R) q_{\gamma_z}(z,p)+Z_{\gamma_z 1}(z,p^R) q_1(z,p).
\label{zfacgz}
\end{equation}
The diagonal part $Z_{\gamma_z\gamma_z}$ and the mixing term
$Z_{\gamma_z 1}$ are determined using the two RI-MOM conditions~\cite{Chen:2017mzz}
\begin{eqnarray}
&& q^R_{\gamma_z}(z,p,p^R)\bigg{|}_{p=p^R} \equiv e^{i p^R_z z},\cr
&& Z_{\gamma_z\gamma_z}(z,p^R){\rm Tr}\Lambda_{\gamma_z}(z,p)+\cr &&\qquad Z_{\gamma_z 1}(z,p^R) {\rm Tr}\Lambda_{1}(z,p)\bigg{|}_{p=p^R} \equiv 0.
\end{eqnarray}
Using the renormalization constants $Z$ determined above using quark
external states, the renormalized pion qPDF can also be
determined by
\begin{eqnarray}
    h^R_{\gamma_t}(z,P_z,p^R)&=&Z_q Z_{\gamma_t\gamma_t}(z,p^R)h^B_{\gamma_t}(z,P_z),\cr
    h^R_{\gamma_z}(z,P_z,p^R)&=& Z_q Z_{\gamma_z\gamma_z}(z,p^R)h^B_{\gamma_z}(z,P_z)+\cr
    &&Z_q Z_{\gamma_z 1} h^B_1(z,P_z),
\end{eqnarray}
where $Z_q$ is the quark renormalization, that can be determined
using the condition~\cite{Alexandrou:2010me}
\beq
Z_q(p^R)^{-1} \frac{1}{12}{\rm Tr}\left(\langle Q(p^R)\rangle^{-1}Q_{\rm tree}(p^R)\right)=1,
\eeq{zqdef}
where $Q(p)$ is the quark propagator determined using the Landau gauge
and $Q_{\rm tree}$ is the free quark propagator for which we use
the free massless Wilson-Dirac propagator.

\begin{figure*}
\centering
\includegraphics[scale=0.7]{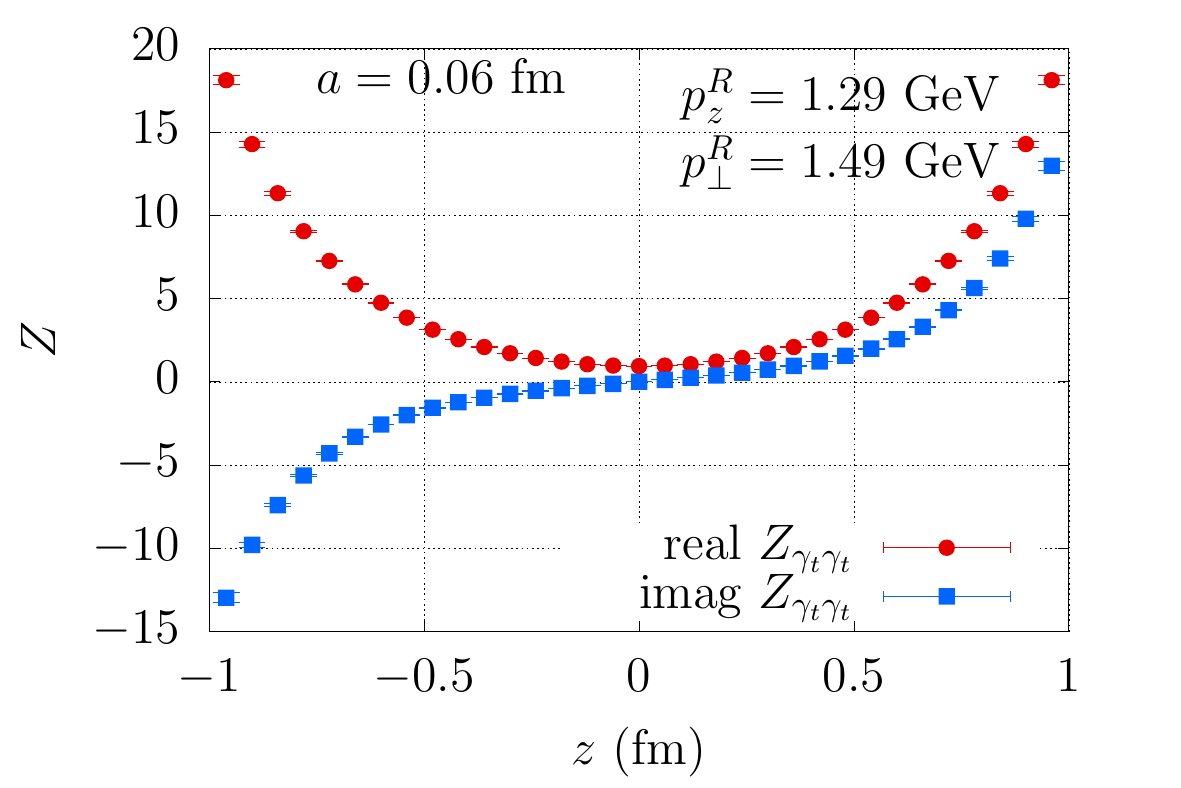}
\includegraphics[scale=0.7]{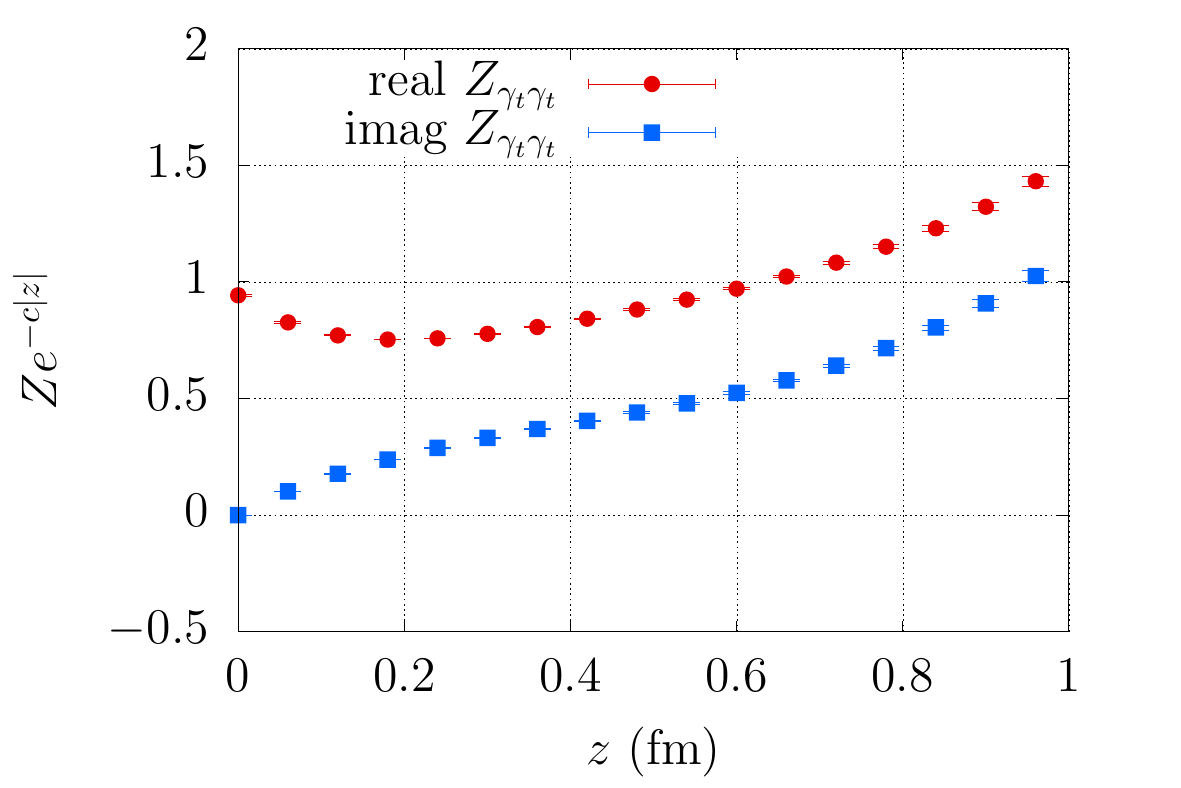}

\includegraphics[scale=0.7]{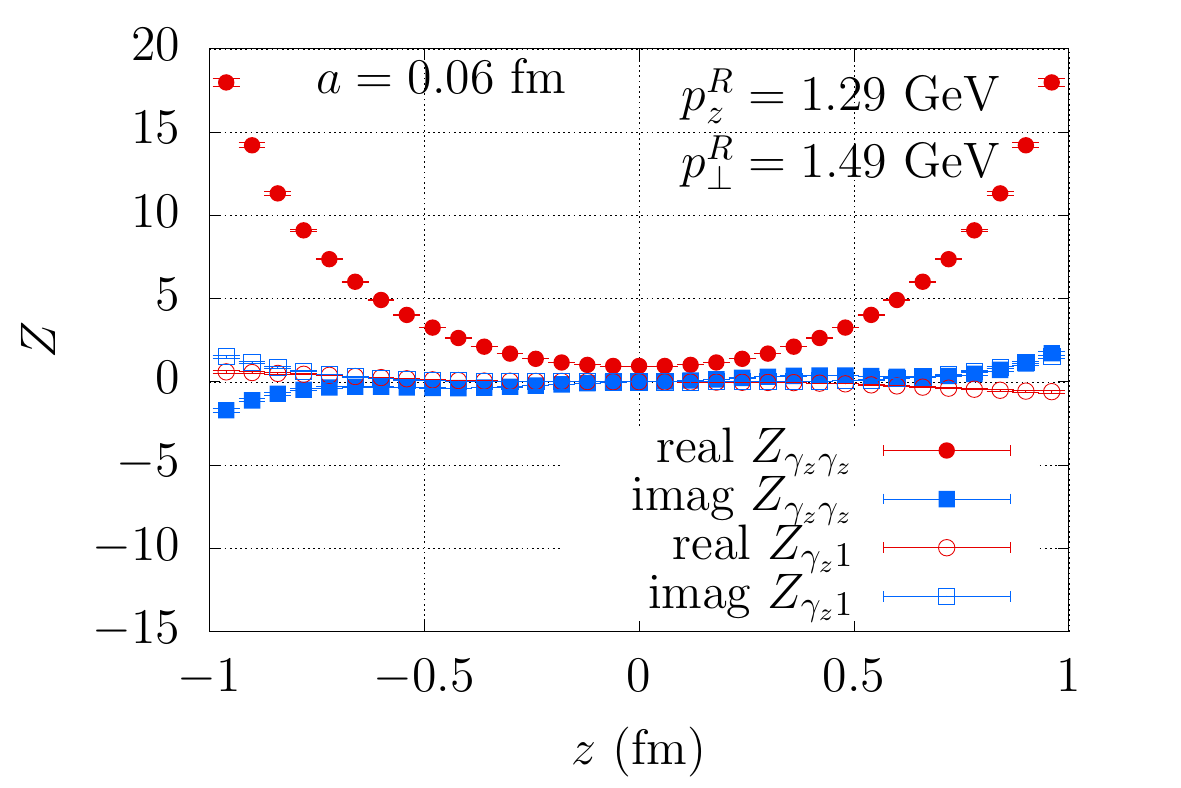}
\includegraphics[scale=0.7]{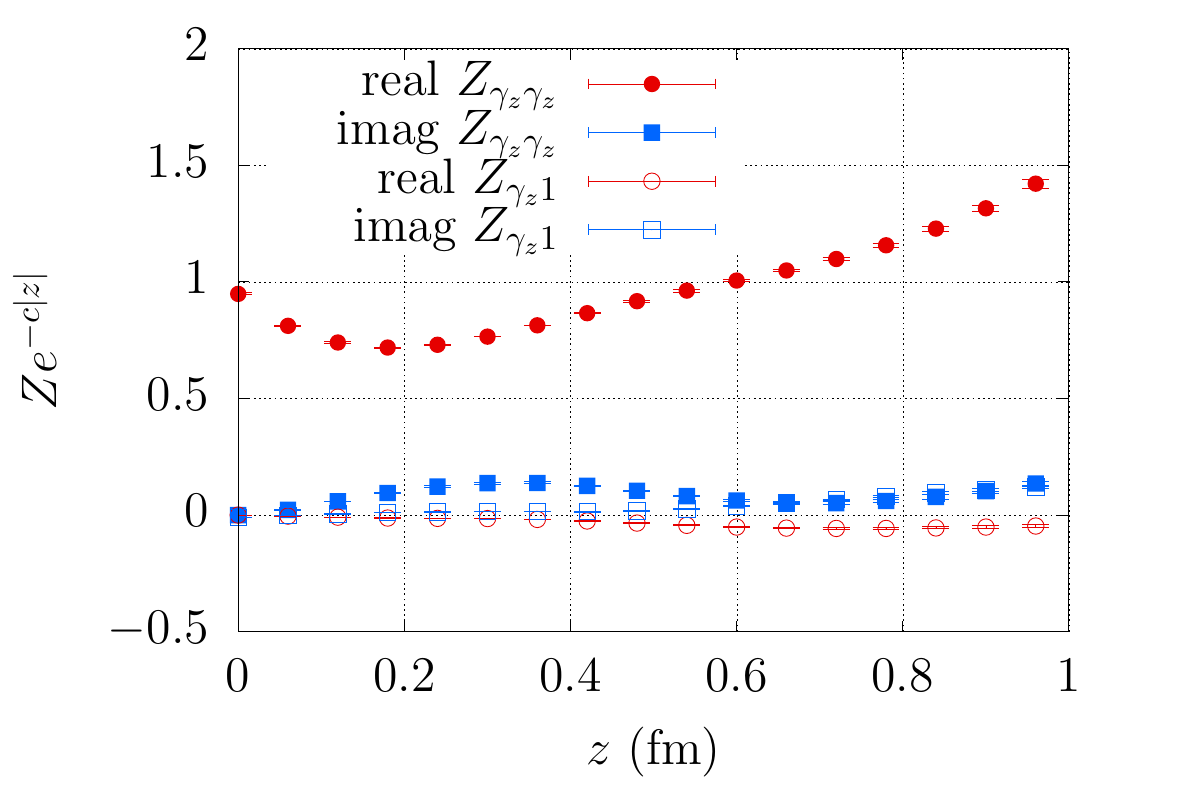}
\caption{
The RI-MOM renormalization constants using $\slashed{p}$-projection
at lattice spacing $a=0.06$ fm at renormalization scale $p_z=1.29$
GeV, $p_\perp=1.49$ GeV are shown. In the top left panel, the real
and imaginary parts of the renormalization constant for $\gamma_t$
qPDF operator is shown as a function of quark-antiquark
separation $z$ in physical units. On the top right panel, the
self-energy divergent part $e^{-c |z|}$ of the Wilson line is removed
from renormalization constant for $\gamma_t$ qPDF. Similarly,
in the bottom left panel, the diagonal part $Z_{\gamma_z\gamma_z}$
and the mixing term  $Z_{\gamma_z 1}$ are shown, and the corresponding
values after the removal of self energy divergence is shown in the
bottom right panel.  }
\label{fig:zfac}
\end{figure*}

In \fgn{zfac}, we show the renormalization factors using the above
RI-MOM renormalization conditions on the $0.06$ fm ensemble.  On
the top left panel of \fgn{zfac}, we show the real and imaginary
parts of $Z_{\gamma_t\gamma_t}$ determined at $p_z^R=1.29$ GeV and
$p_\perp^R=1.49$ GeV.  The rapid, almost exponential, increase in
$Z$ with $z$ is due to the self-energy divergence present in the
bare Wilson line that connects the quark and antiquark in the
qPDF operator. This divergent piece, $e^{c|z|}$, cannot be
captured perturbatively and it needs to be determined nonperturbatively
in a particular scheme. However, this might not be an issue for the
one-loop matching if $e^{c|z|}$ cancels exactly between the
renormalization factors and the bare qPDF operator.  Therefore,
we remove $e^{c |z|}$ from the renormalization constant that is
shown in the top-left panel, and display the result in the top-right
panel.  The value of $c$ for our $a=0.04$ fm ensemble was determined
in~\cite{Bazavov:2018wmo}, and for 1-HYP Wilson line $ca=0.1586$.  
This removal of Wilson line self-energy
reduces the almost exponential dependence of $Z(z)$ to a weak
dependence on $z$.  In fact, we see that both the real and imaginary
parts of $Z_{\gamma_t\gamma_t}$ remain ${\cal O}(1)$ even up to
$z=1$ fm, thereby providing a qualitative justification for the
usage of leading order perturbation theory to describe the lattice
data at short $z$ and at high quark momenta.  We show similar data
in the bottom left and right panels for the Z-factors for $\gamma_z$
qPDF. In this case, we have the diagonal factor $Z_{\gamma_z\gamma_z}$
as well as the off-diagonal factor $Z_{\gamma_z 1}$ to take care of
mixing with scalar on the lattice. We show $Z_{\gamma_z\gamma_z}$
and $Z_{\gamma_z 1}$ as the filled and unfilled symbols in the
bottom panels respectively. We observe that the imaginary part of $Z_{\gamma_z\gamma_z}$
is small compared to the real part. This is not the case for
$Z_{\gamma_t\gamma_t}$, which in turn will affect the asymmetry of
the $u-d$ qPDF $\tilde{q}_{u-d}(x)$ of pion about $x=0$. We
also note that the mixing of $\gamma_z$ with the scalar is a minor 5-10\%
effect, but we nevertheless take care of it in our calculation.

\begin{figure}
\centering
\includegraphics[scale=0.65]{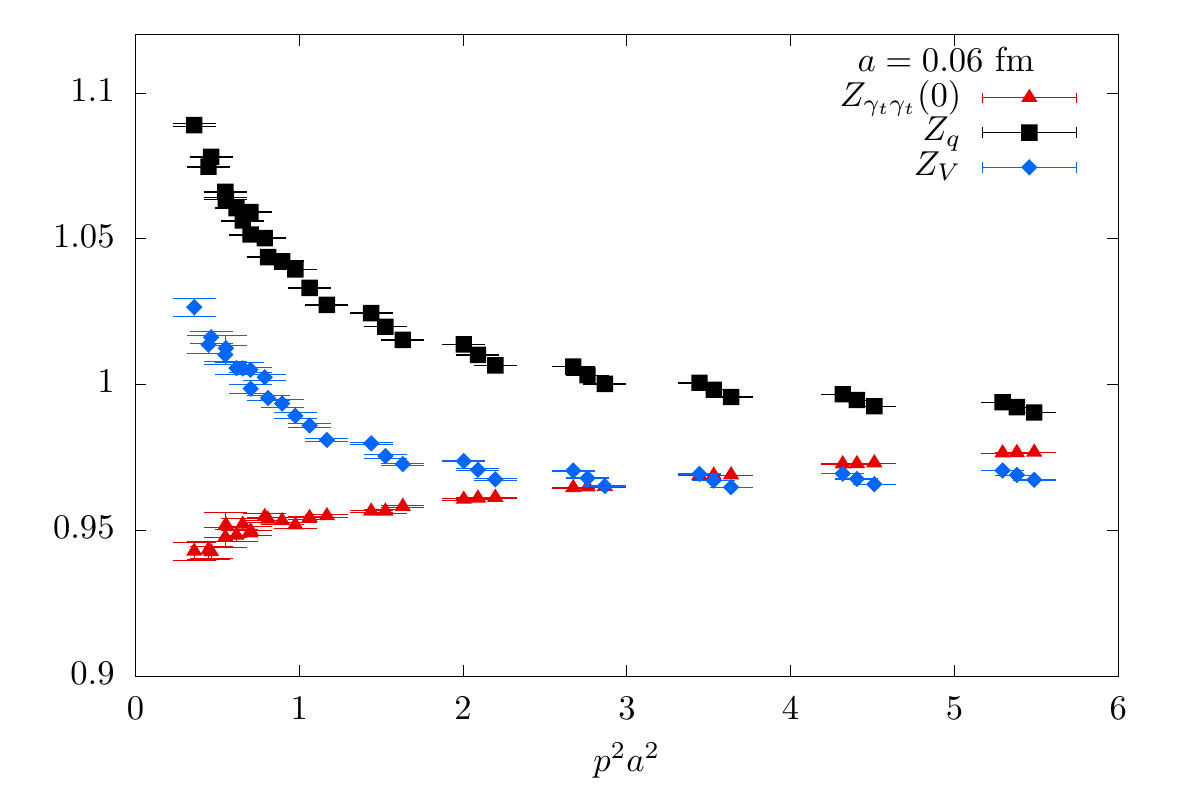}
\caption{The $(p a)^2$ dependence of the renormalization factors
$Z_q$ (black squares), $Z_{\gamma_t\gamma_t}(z=0)$ (red triangles)
and $Z_V=Z_q Z_{\gamma_t\gamma_t}(0)$  (blue diamonds) are shown
for the $a=0.06$ fm ensemble.
}
\label{fig:zq}
\end{figure}

As we discussed in the last section, the matrix element at $z=0$, 
$h_{\gamma_t}$ is the local current operator which will
be exactly conserved in the continuum limit. Hence, $Z_q
Z_{\gamma_t\gamma_t}(z=0)$ is the vector current renormalization factor
$Z_V$ and the dependence of $Z_V$ on $p$ will give us an idea
of the leading $(pa)^2$ perturbative lattice artifacts for values of 
$p \gg \Lambda_{\rm QCD}$ as well as the 
other higher order (or perhaps non-perturbative) contributions to this lattice 
correction to $Z_V$ at smaller renormalization
scales~\cite{Alexandrou:2010me}.  In \fgn{zq}, we show $Z_q$
determined using \eqn{zqdef}, the value of $Z_{\gamma_t\gamma_t}$
at $z=0$ as well as their product $Z_V$ as a function of $(p a)^2$.
One sees a reasonable plateau for $Z_V\approx 0.97$ 
only for $(pa)^2>2$.
For comparison, the value of $Z_V$ as obtained from the 
bare pion isospin charge $h^B_{\gamma_t}(z=0,P_z=0)$ is 0.961(3).
The values of $Z_V$ determined from $h^B_{\gamma_t}(z=0,P_z)$ at the other non-zero $P_z$
also give consistent values.
With the uncertainties of choosing the scaling region in $(pa)^2$
to take the $(pa)\to 0$ limit of $Z_V$, we expect the $Z_V$ to be in 
the range 0.97 to 0.99.
For relatively smaller values of renormalization momenta $(pa)\approx
1 - 1.5$, chosen such that the renormalization scales lie in the  
vicinity of the pion momenta used in this paper,
one sees noticeable, but small $~5\%$ dependence on $pa$.
We used the value of $Z_q$ estimated at the same value of $p$ as
used in $Z_{\gamma_t\gamma_t}$ for renormalizing our pion qPDF.

\begin{figure*}
\centering
\includegraphics[scale=0.55]{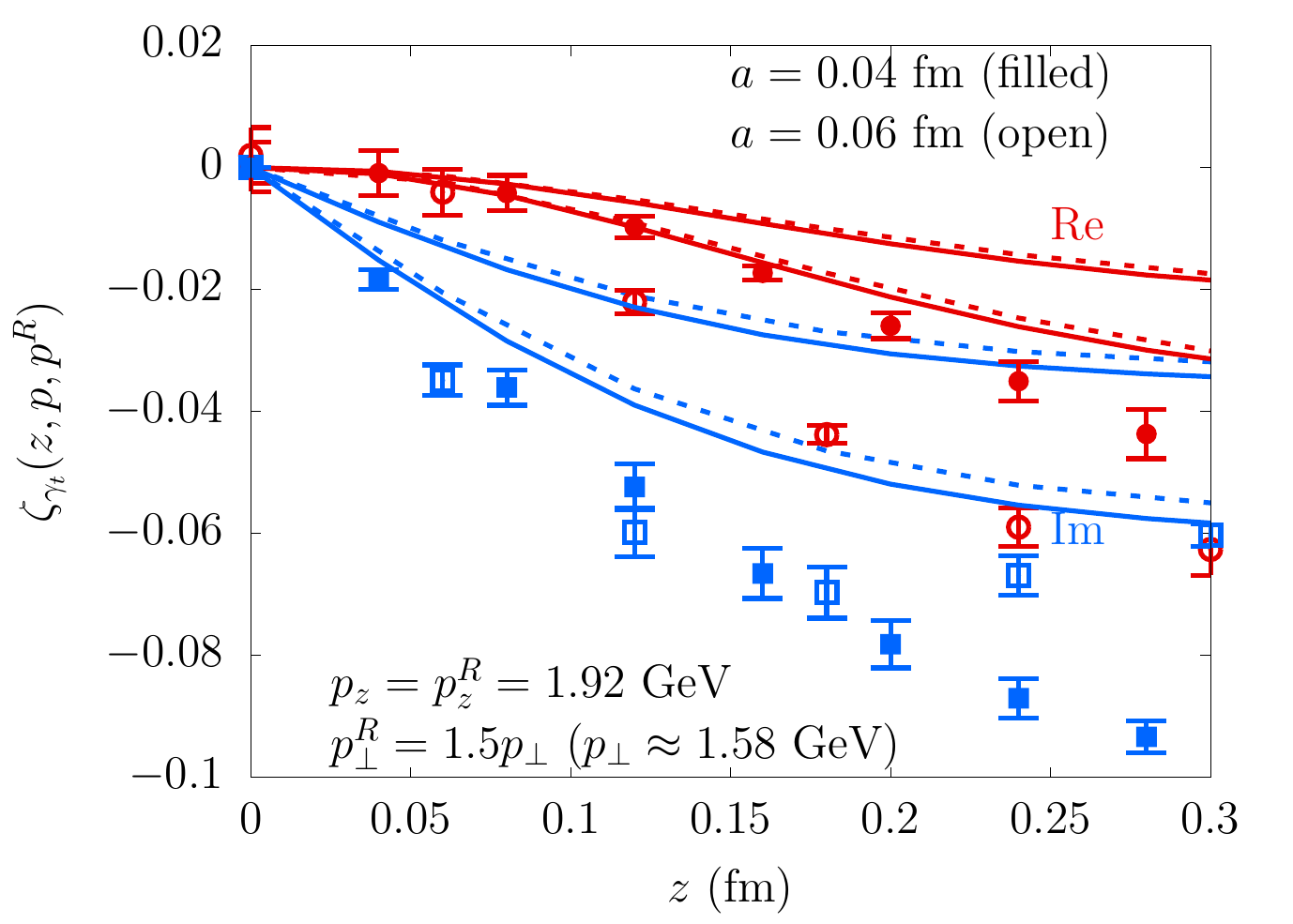}
\includegraphics[scale=0.55]{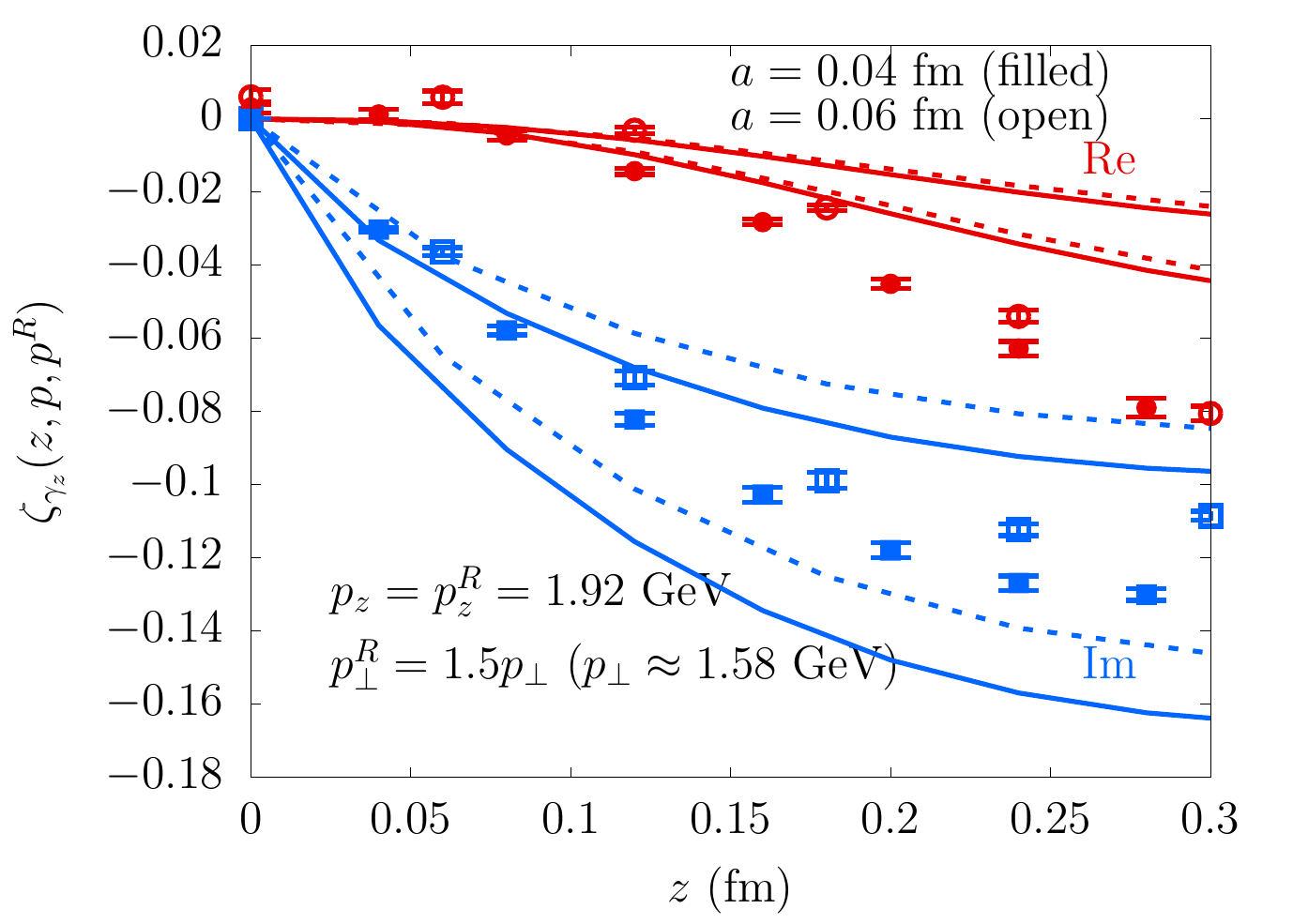}
\caption{
    The behavior of $\zeta(z)$ (symbols) with the quark-antiquark
    separation $z$, is compared with the expectation from 1-loop
    perturbation theory (bands) when $p_\perp^R$ is slightly
    away from $p_\perp$. The red symbol and bands are
    the real part of $\zeta$ while the blue ones correspond to
    the imaginary part. 
   The bands enclosed by solid
   curves corresponds to 
    the momenta $(p_z,p_\perp)$ for the 0.04 fm data,
    and similarly the band enclosed by the dashed curves corresponds
    to $(p_z,p_\perp)$ of the 0.06 fm data (see text). 
    On the left and right panels, the comparisons
    are made for $\zeta_{\gamma_t}$ and $\zeta_{\gamma_z}$ respectively.
    In each of the panels, the data from two different lattice
    spacings are also shown ($a=0.04$ fm as filled circles and
    $a=0.06$ fm as open circles).  For the data shown, $p_z=p_z^R=1.92$
    GeV. The transverse momentum of the quark $p_\perp=1.58$ GeV,
    and the transverse renormalization momentum is chosen to be 1.5
    times $p_\perp$.
} 
\label{fig:zetazdep}
\end{figure*}

\begin{figure*}
\centering
\includegraphics[scale=0.67]{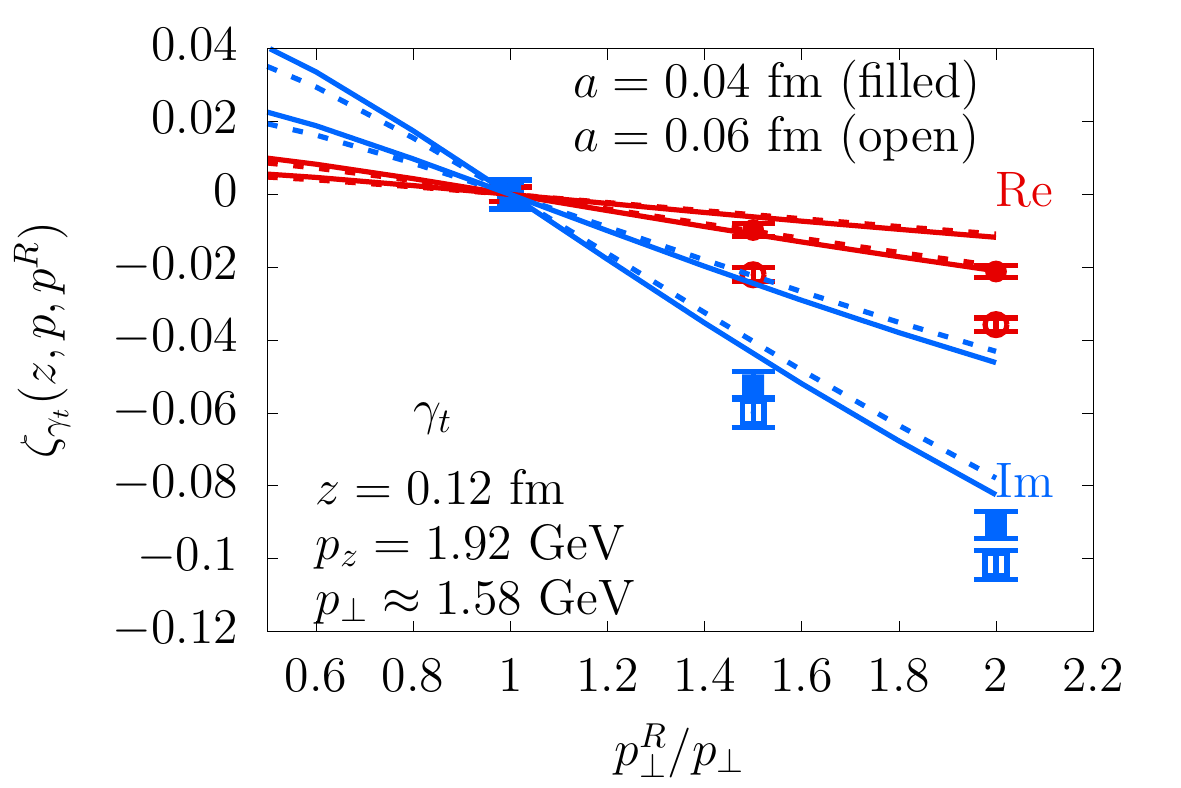}
\includegraphics[scale=0.67]{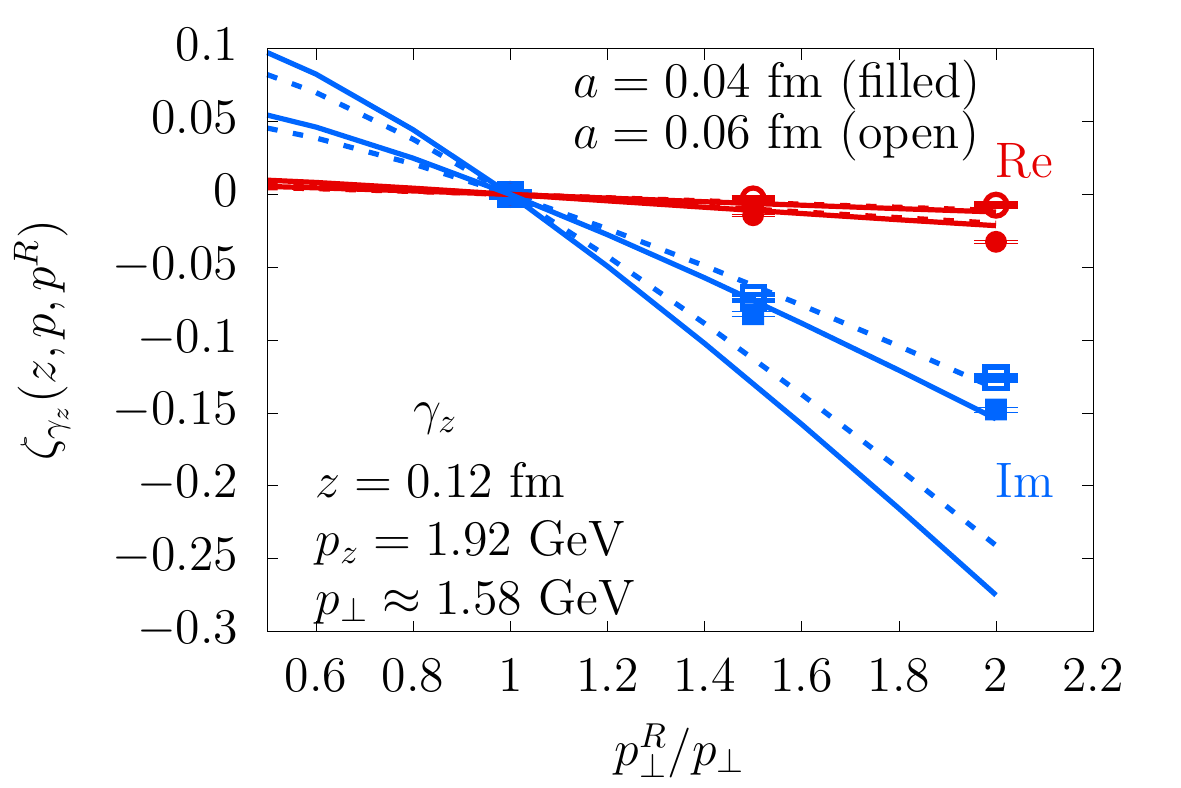}
\caption{
    The renormalization flow of $\zeta_{\gamma_t}$ (left panel) and
    $\zeta_{\gamma_z}$ (right panel) with the transverse renormalization
    scale, $p_\perp^R$ are shown at fixed $z=0.12$ fm. The data from two different lattice spacings,
    $a=0.04$ fm (filled circles) and $a=0.06$ fm (open circles),
    shown at fixed  $z=0.12$ fm and $p_z=p_z^R=1.92$ GeV. 
    The transverse momenta of the quark at $a=0.04$ and
    0.06 fm are $p_\perp=1.67$ and 1.48 GeV respectively, and they are chosen
    to be roughly equal for this comparison.  The real part of
    $\zeta$ is shown in red while the imaginary part is shown in
    blue. 
    The band enclosed by the solid red
    (blue) curves corresponds to the 1-loop result for real (imaginary)
    parts of $\zeta$ at $(p_z,p_\perp)$ for the 0.04 fm data,
    and similarly the band enclosed by the dashed curves corresponds
    to $(p_z,p_\perp)$ of the 0.06 fm data.
  }
\label{fig:flowz1p2fm}
\end{figure*}

\subsection{Comparison with leading order perturbation theory for $z < 0.3$ fm}

We will now investigate in a quantitative way the agreement/disagreement
of the lattice determination of the RI-MOM renormalized amputated
quark qPDF at $z < 0.3$ fm which one can expect to be in the
perturbative regime. For this, we construct a quantity
$\zeta_\Gamma(z,p,p^R)$ in the following way
\begin{equation}
    \zeta_\Gamma(z,p,p^R)=\frac{q^R_\Gamma(z,p,p^R)}{q^R_\Gamma(z,p,p)}-1,
\end{equation}
where $q^R_\Gamma(z,p,p)=e^{i p_z z}$ by renormalization condition.
In the case of $\Gamma=\gamma_t$, the above definition is simply 
\begin{equation}
    \zeta_{\gamma_t}(z,p,p^R)=\frac{Z_{\gamma_t\gamma_t}(z,p^R)-Z_{\gamma_t\gamma_t}(z,p)}{Z_{\gamma_t\gamma_t}(z,p)},
\end{equation}
which is similar to a discrete scale-dependent anomalous dimension
$\partial\log\left(Z_{\gamma_t\gamma_t}(z,p)\right)/\partial p$.
Through the dependence of $\zeta$ on $p^R$ slightly away from $p$,
we can understand how well the leading order perturbation theory
is able to describe the exact non-perturbative determination on the
lattice.  It is important to stress that apart from understanding
non-perturbative renormalization of qPDF in this way, we are
also essentially comparing one of the steps in the LaMET formalism that
is calculable on lattice. Hence, any agreement/disagreement
we observe quantifies the limitations of the leading order LaMET.
In perturbation theory, $\zeta$ is the ratio of the one-loop perturbative
correction to $q(z,p)$ to its tree-level value.  This expression
for $\zeta$ has been calculated, and it is given by~\footnote{The
formula differs from the one given in~\cite{Stewart:2017tvs} due
to the issue of order of $\epsilon=0$ limit in dimensional
regularization and the $z=0$ limit. We thank Yong Zhao for communicating
the corrected result to us.}
\begin{eqnarray}
    &&\zeta_\Gamma(z,p,p^R)=\frac{\alpha_s C_F}{2\pi}\int_{-\infty}^\infty dx (e^{i(1-x)p_z}-1) \bigg{[} H\left(x,p\right)\cr
    &&-\left|\frac{p_z}{p_R}\right|H\left(1+\frac{p_z}{p_z^R}(x-1),p^R\right)\bigg{]},
    \label{pertres}
\end{eqnarray}
where $H(x,p)$ is the 1-loop correction term to the bare qPDF,
and the two terms in the right hand side come from the bare and
RI-MOM renormalization counter terms respectively~\footnote{The
function $H(x,p)$ here is referred to as $h(x,p)$ in
Ref~\cite{Stewart:2017tvs}. We reserve $h$ to refer to qPDF matrix
element as is the convention.}. The functional forms of $H(x,p)$ for
$\gamma_z$ and $\gamma_t$ isovector qPDFs are given
in~\cite{Stewart:2017tvs,Liu:2018uuj} and therefore we do not provide
them here.  The asymptotic $3/(2|x|)$ behavior of the bare
and the RI-MOM counter term, that contributes to the UV divergence
when integrated over $x$, gets exactly canceled and we obtain a UV
finite and renormalized result for $\zeta$.  In the discussions
below, we will consider the cases with $p_z=p_z^R$ and $p_z\ne
p_z^R$ separately.  In the above leading order formula, the scale
at which $\alpha_s$ has to be evaluated is arbitrary. Therefore, we
vary $\alpha_s$ by changing the scale from $0.5 p_z^R$ to $2p_z^R$
through the 1-loop running, and quote this variation as an uncertainty
in the perturbative results below.  On the lattice side, we determine
$\zeta(z,p,p^R)$ using the non-perturbatively determined $Z$-factors.
In order to estimate the lattice spacing effects, we determined
$\zeta$ using two different lattice spacings; $a=0.04$ fm is shown as
filled symbols and $a=0.06$ fm is shown as open symbols in the various
plots that follow.

In the left and right panels of \fgn{zetazdep}, we show the 
typical dependence of $\zeta_{\gamma_t}(z)$
and $\zeta_{\gamma_z}(z)$ respectively,
as a function of $z$ when $p_\perp^R$ differs slightly from the
transverse quark momentum $p_\perp$, while the longitudinal components
$p_z$ and $p_z^R$ are the same.  Using \eqn{pertres}, we calculated
the prediction from leading order perturbation theory for $\zeta(z)$
at the same values of momenta.  The uncertainty bands for the
perturbative result are shown in \fgn{zetazdep} along with the
actual lattice data at the two different lattice spacings that are
shown using symbols.  For the data shown in \fgn{zetazdep}, the
longitudinal components $p_z$ for the two lattice spacings are
exactly 1.92 GeV, but the transverse components $p_\perp$ are only
approximately the same between the two due to the constraints of
allowed momenta on the two different lattice volumes {\sl i.e.},
$p_\perp=1.49$ GeV for $a=0.06$ fm and $p_\perp=1.67$ GeV for
$a=0.04$ fm. To take care of this slight offset in $p_\perp$ between
the two lattice spacings, we have distinguished the perturbative
results corresponding to $a=0.04$ fm as bands enclosed by solid
lines, and similarly for $a=0.06$ fm as bands enclosed by dashed
lines. It can be seen that the two perturbative results are not
very sensitive to this difference in $p_\perp$ assuring us that
whatever change we observe between the data at two different $a$
is mainly due to the change in $a$. We observe from the plots that the
leading order perturbation theory captures the qualitative
$z$-dependence of both the real and imaginary parts of $\zeta$ when
$p_\perp^R$ is changed from $p_\perp$. Surprisingly, the 1-loop
result seems to work better for $\zeta_{\gamma_z}$ than for
$\zeta_{\gamma_t}$. In the case of $\gamma_t$, one can certainly
see a large lattice spacing effect with the movement of data towards
the 1-loop result as the lattice spacing is reduced, while in the
case of $\gamma_z$, one can already see a consistency with the one
loop result at the lattice spacings that we use. Thus, it opens up
a question on whether the $\gamma_z$ qPDF fares worse compared to
the $\gamma_t$ qPDF simply due to the presence of mixing with
the scalar or whether $\gamma_z$ qPDF might eventually show better
perturbative convergence and lesser lattice spacing dependence in
spite of its other disadvantages.  

In \fgn{flowz1p2fm}, we concentrate on the renormalization flow of
$\zeta_\Gamma$ at fixed small value of $z=0.12$ fm.  The two panels
show the dependence of $\zeta_{\gamma_t}$ and $\zeta_{\gamma_z}$
on $p^R_\perp$ which is changed around $p_\perp$. As before, we
keep $p_z=p^R_z=1.92$ GeV. The 1-loop result is able to capture the
qualitative trend of the flow in both $\gamma_z$ and $\gamma_t$.
For both the cases, we can see that the reduction of lattice spacing
leads to a better agreement with the 1-loop result.  Having discussed the
cases where $p_z=p_z^R$, we now study the dependence of $\zeta$ on
$p_z^R\ne p_z$, while keeping $p_\perp=p^R_\perp$.  We show the
$z$-dependence of $\zeta_{\gamma_z}$ when $p_z^R=1.5 p_z$ in
\fgn{pzdep}. We find the perturbative result to have the same qualitative
behavior as the lattice data. Putting together the various
observations in this section, we found only an overall qualitative
agreement between the lattice results on $\zeta$ and the one-loop
perturbative results.  When the lattice spacing is reduced, we found
the agreement to get better.  It remains to be seen what the effect
of including higher order corrections in the perturbative result
for $\zeta$.

\begin{figure}
\centering
\includegraphics[scale=0.55]{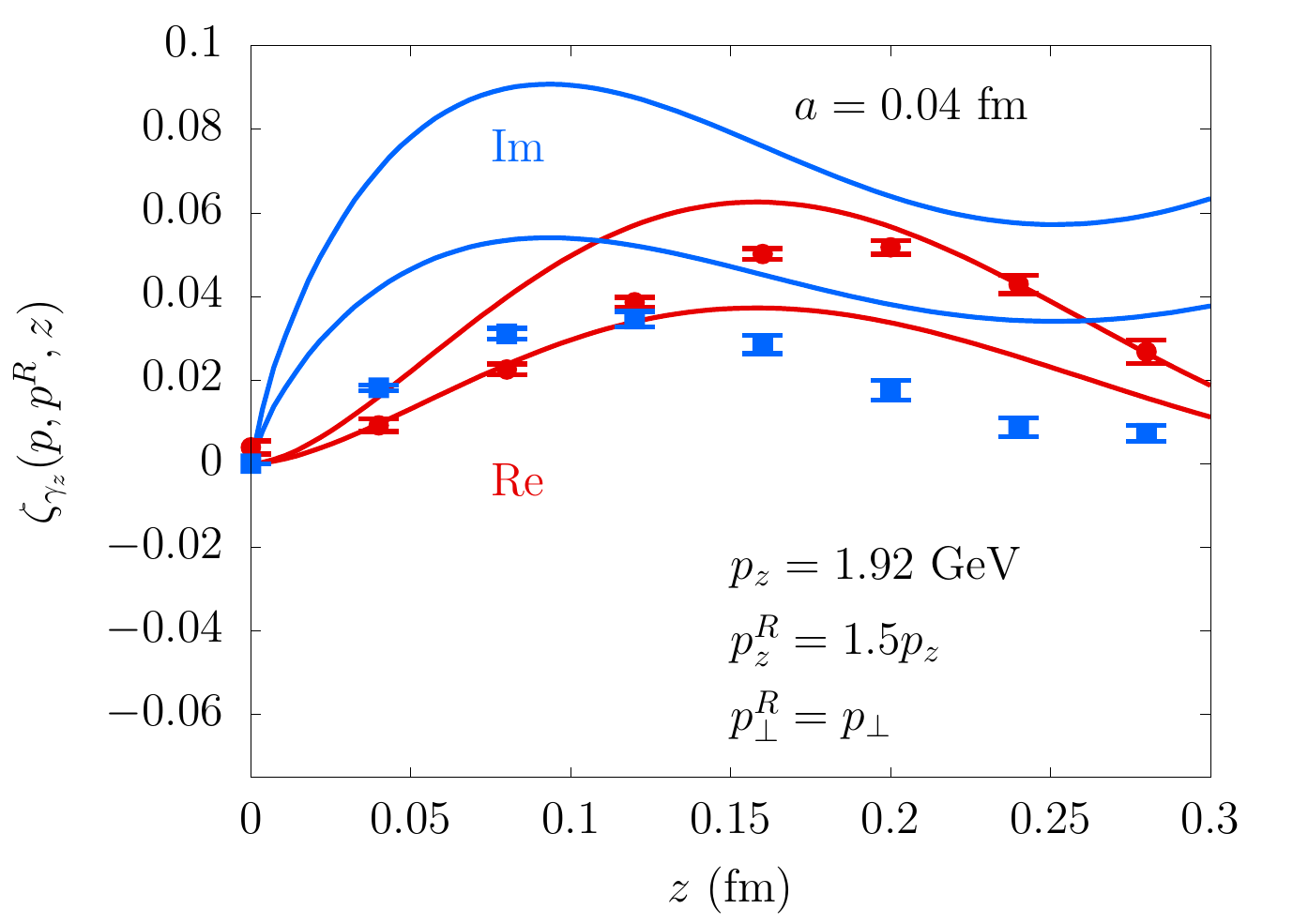}
\caption{The dependence of $\zeta$ on the longitudinal momentum
$p_z^R$.  $\zeta_{\gamma_z}$ is shown as a function of $z$ for a
specific choice of $p_z$ and $p_\perp=p_\perp^R$. The uncertainty
bands for the real and imaginary parts for the leading order
expectation are shown using bands enclosed by solid lines.  The
symbols are the lattice data determined at lattice spacing $a=0.04$
fm.
}
\label{fig:pzdep}
\end{figure}

\subsection{A way to classify quark-antiquark separations as perturbative or 
nonperturbative}

\begin{figure}
\centering
\includegraphics[scale=0.55]{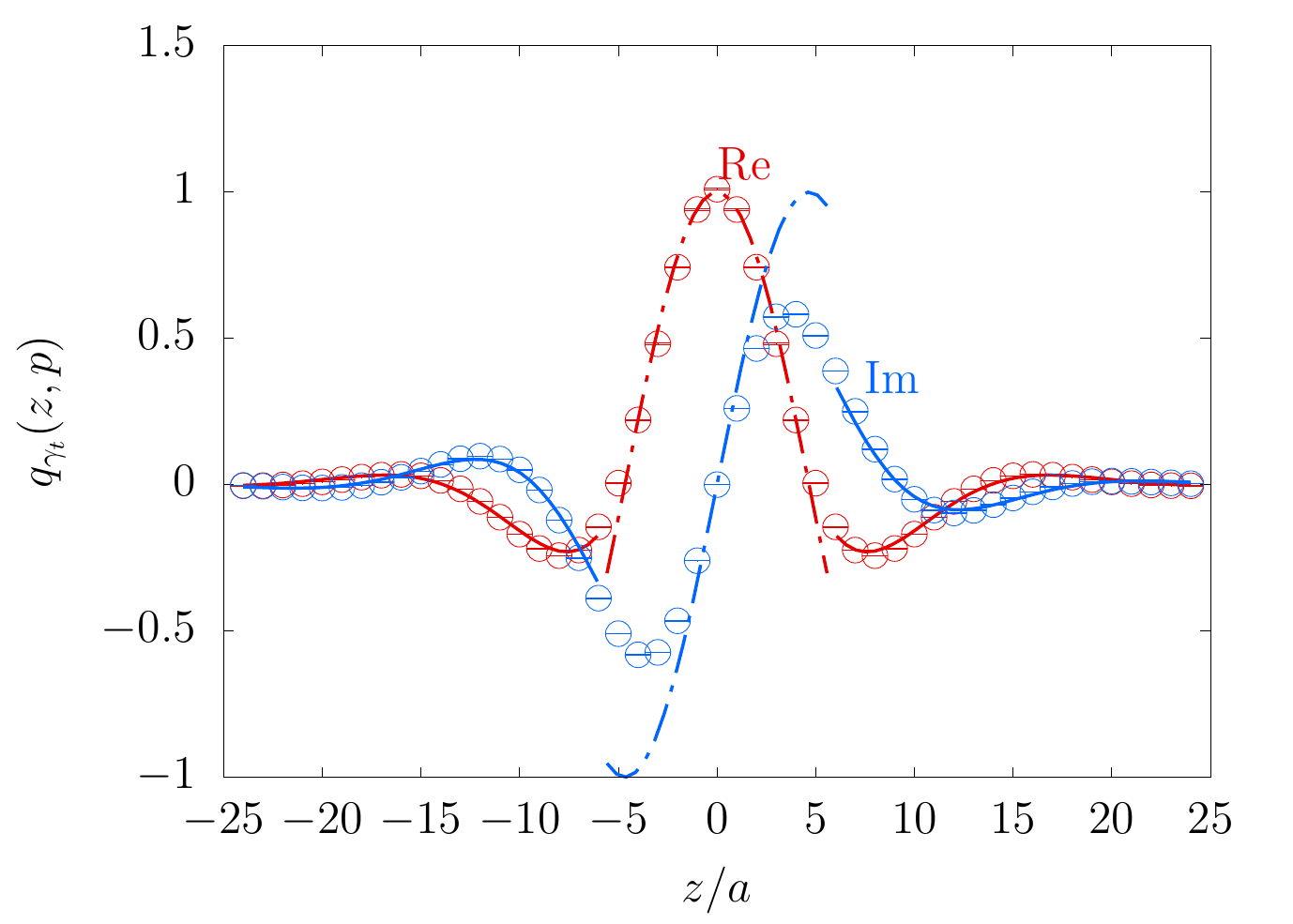}
\caption{The real and imaginary parts of the lattice data for
$q_{\gamma_t}(z,p)$ is compared with purely oscillatory
model (dashed curves) at short-distances, 
and damped oscillatory Ansatz (solid curves)
at larger $|z|$. The data corresponds to quark momentum $(p_z,p_\perp)=(1.29,1.49)$ GeV. }
\label{fig:expcos}
\end{figure}

\begin{figure}
\centering
\includegraphics[scale=0.5]{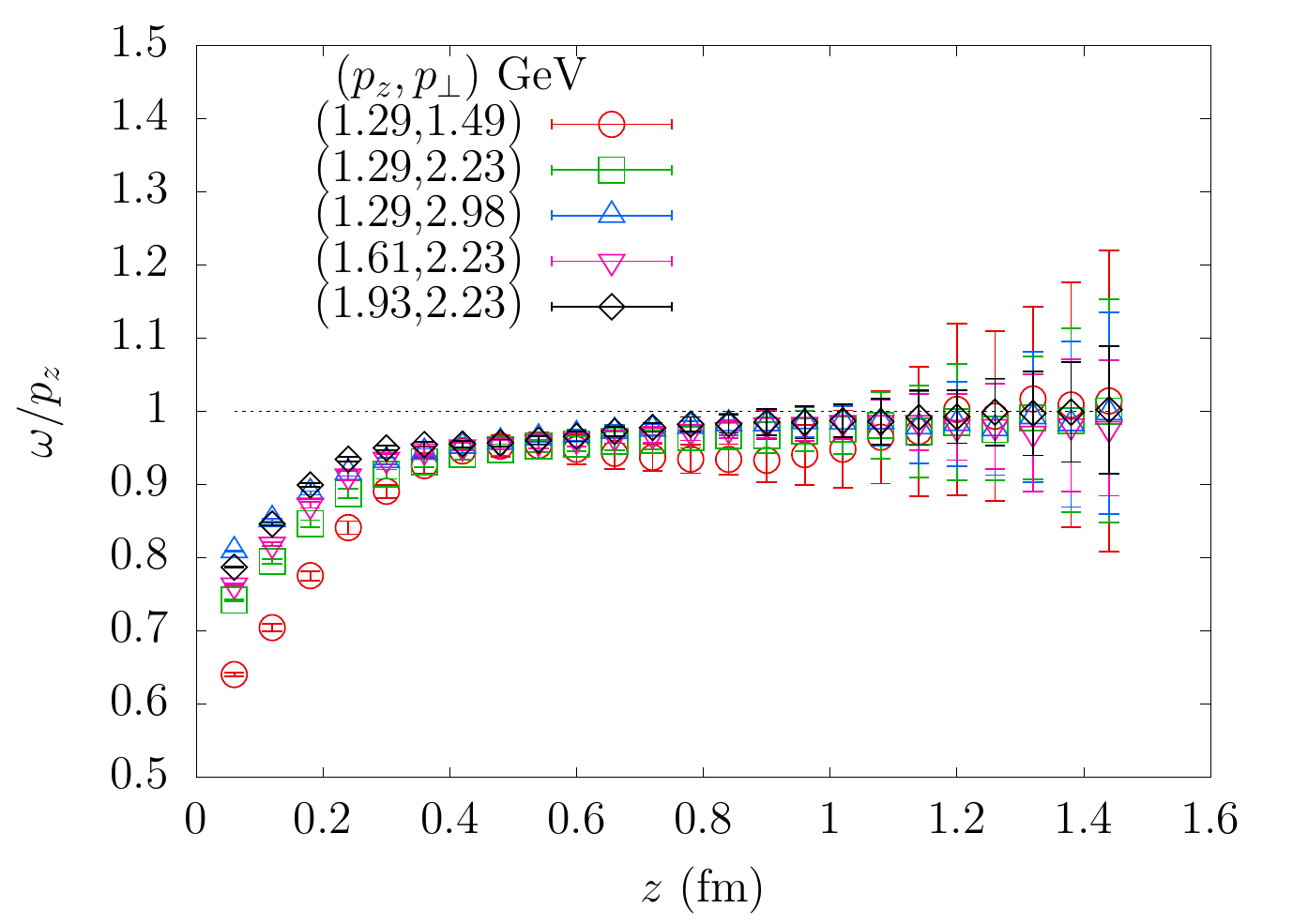}
\includegraphics[scale=0.5]{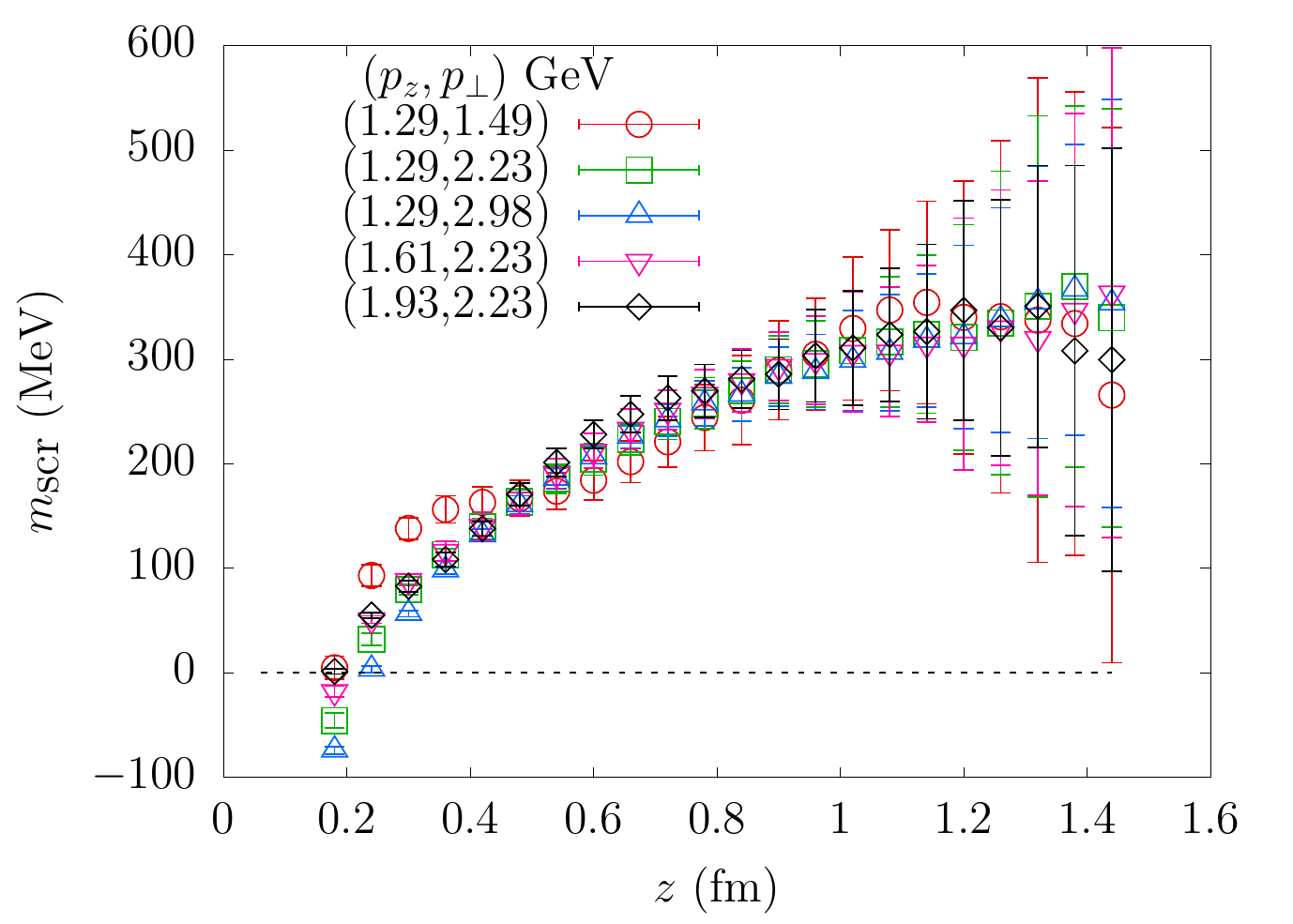}
\caption{
The effective frequency of oscillations $\omega$ (top) and
effective screening mass $m_{\rm scr}$ (bottom) as extracted
from $q_{\gamma_t}(z,p)$ are shown.  The various values of
$p=(p_z,p_\perp)$ for the data are tabulated in the plots. The
values of $\omega$ are normalized with respect to $p_z$.  1-HYP
smeared Wilson line was used in all the cases shown in the plots,
and the corresponding self-energy $ca=0.1586$ was subtracted to
obtain $m_{\rm scr}$.  }
\label{fig:effmass}
\end{figure}

Physically, one would expect that for a well-separated quark-antiquark
with $z > 1$ fm, one would start seeing traces of nonperturbative
physics in the qPDF. Quantifying the advent of non-perturbative
physics for large enough $z$ at finite quark/hadron momentum is
important with regard to the extraction of PDF since the real-space
qPDF at all $z$ enter the computation of its Fourier transform.
A simple first approximation to study this effect is the following.
In free theory, the qPDF with external quark states is a pure
wave $e^{i p_z z}$. We expect, to a first approximation, that the
effect of nonperturbative physics is to damp this pure wave via an
inverse screening length $m_{\rm scr}\sim
{\cal O}(\Lambda_{\rm QCD})$. Thus, we model the bare quark qPDF
as
\begin{equation}
q(z,p)=A e^{i \omega z} e^{-m_{\rm scr} |z|}  e^{-c |z|},
    \label{quarkans}
\end{equation}
where we have removed the UV divergent piece $e^{-c |z|}$ from the
qPDF and defined the left-over exponent $m_{\rm scr}$ as a
physical scale. We have also accounted for $\omega\ne p_z$ in the
interacting theory since the quark can lose momentum by emitting
gluons. There could be remnant non-trivial dependence of the amplitude
$A$ on $z$, which we assume to be sub-leading compared to the leading
damped oscillatory behavior and ignore them in the discussion here.
There is an ambiguity in $m_{\rm scr}$ depending on the scheme used
to determine the divergent piece $c$. Since the values of $c$
determined from the static quark potential method ensures that the
renormalization factors after the removal of $e^{-c|z|}$ are ${\cal
O}(1)$ at smaller $z$ in \fgn{zfac}, the choice of separation of
the exponential suppression factor into a divergent and physical
scales as defined in \eqn{quarkans} is well motivated in this scheme.
In \fgn{expcos}, we show the bare quark qPDF $q_{\gamma_t}(z,p)$
for $p_\perp=1.49$ GeV and $p_z=1.29$ GeV determined on the $a=0.06$
fm ensemble.  The short distance can simply be described by a pure
oscillatory $e^{i\omega z}$ behavior which is shown using the dashed
curves (with $\omega=0.85 p_z$ for the case shown). The solid curves
in the figure correspond to the ansatz in \eqn{quarkans} which
describes the data at larger $|z|$ well.  Without dwelling further
on finding the best parametrization of the lattice data that
asymptotically behaves like \eqn{quarkans}, we simply define an
effective $z$-dependent $\omega$ and $m_{\rm scr}$ through
\begin{equation}
m_{\rm scr}(z)-i\omega(z)\equiv -\frac{1}{a}\log\left(\frac{q(z+a,p)}{q(z,p)}\right)-c.
    \label{eff}
\end{equation}
In \fgn{effmass}, we show the behavior of $m_{\rm scr}$ and $\omega$
as a function of $z$ as extracted from $q_{\gamma_t}(z,p)$. We have
chosen different set of $p_z$ and $p_\perp$ to show the dependence
on $p_z$ at fixed $p_\perp$ and vice versa. From the top panel,
we see that $\omega/p_z$ is below 1 for $z < 0.4$ fm and seems to
approach a plateau closer to 1 for $z > 0.4$ fm. While the values
of $\omega$ at short distances depend on $p_z$ and $p_\perp$,
the approach to $\omega\approx p_z$ is universal. We observed this
behavior when we used $q_{\gamma_z}(z,p)$ as well.  A physical
reasoning for this observation could be that at shorter $z$, the
quark has the ability to radiate a gluon, and at distances $z >
0.4$ fm there is effectively a dressed quark carrying all the
momentum. In the bottom panel of \fgn{effmass}, we have shown the
effective screening mass $m_{\rm scr}$. In the plots, we have only
shown the data where 1-HYP smeared Wilson line was used.  For
this case, we subtracted $ca=0.1586$ in \eqn{eff} to get $m_{\rm
scr}$.  One can clearly see the emergence of non-zero $m_{\rm
scr}\approx 300$ MeV for $|z| > 0.5$ fm which is in the typical
${\Lambda}_{\rm QCD}$ scale. When we repeated this using quark
qPDF with unsmeared Wilson line, we found the results to be consistent
with the data shown in \fgn{effmass} after we subtracted out
$ca=0.3687$ corresponding to unsmeared Wilson line. This assures us
that the observed $m_{\rm scr}\approx 300$ MeV is a real physical
scale independent of the self-energy divergence of the Wilson line.
This signals the significant presence of a confinement scale beyond
$z\approx 0.5$ fm. Also, the near plateauing of both $\omega$ and
$m_{\rm scr}$ for these larger $z$ indicates that a simple physically
motivated ansatz in \eqn{eff} offers a surprisingly good
description of the actual non-perturbative data. One could have
expected this simply from observing the large $|z|$ part of
\fgn{expcos}.  It remains to be seen if this observation can be
used advantageously in improving the LaMET matching at
finite moderately large $p_z$.

\section{From renormalized quasi-PDF to PDF}
\label{sec:PDF}

\subsection{On obtaining the valence PDF using isovector $u-d$ qPDF of pion}

Having determined the renormalized qPDF we can now discuss the
matching between qPDF and PDF as well as the determination of pion
PDF from our lattice results. We computed the $u-d$ qPDF matrix
element of a pion which in practice we obtained from the real part
of the connected piece of the $u$ quark qPDF matrix element.
Now, we discuss how the $u-d$ qPDF and PDF are related to the
valence PDF of pion.

The $u$ and $d$ quark distributions, $f_u(x)$ and $f_d(x)$, as
determined using \eqn{soperform} has support from $x=-1$ to 1. One
can make connection with the conventional, separately defined quark
distributions $Q_{u,d}(x)$ and the anti-quark distributions
$Q_{\bar{u},\bar{d}}(x)$ that are non-zero only between $x=0$ and
1, through the relation
\begin{equation}
    f_{u,d}(x)=Q_{u,d}(x)\theta(x)-Q_{\bar{u},\bar{d}}(-x)\theta(-x).
\end{equation}
Therefore, $f_{u,d}(x)$ contains information on both the quark as
well as the antiquark distributions in the positive and negative
regions of $x$ respectively.  Let us first focus on $x\ge 0$.  In
the isospin symmetric case we are considering, $Q_u(x)=Q_{\bar
d}(x)$ and $Q_{\bar u}(x)=Q_d(x)$. Therefore, for the positively
charged pion $Q_u(x)-Q_d(x)=Q_u(x)-Q_{\bar u}(x)=f^{\pi,u}_v(x)$
is the valence u-quark distribution.  Again, due to the isospin symmetry,
the $u$ and $\bar d$ valence distributions are the same as
$f^{\pi,u}_v(x)=f^{\pi,{\bar d}}_v(x)=f^{\pi}_v(x)$.  However,
unlike the valence quark distribution, the isotriplet $u-d$ PDF
$f_{u-d}=f_u(x)-f_d(x)$ satisfies $f_{u-d}(|x|)=f_{u-d}(-|x|)$ and
it has support from $-1$ to 1.  That is,
\begin{eqnarray}
    f_{u-d}(x)&=&\begin{cases}Q_u(x)-Q_d(x), x>0\cr Q_{u}(-x)-Q_{d}(-x), x<0\end{cases};\cr f^\pi_v(x)&=&\begin{cases}Q_u(x)-Q_d(x), x>0\cr 0, x<0.\end{cases}
\end{eqnarray}
Therefore, one can obtain the $u-d$ quark distribution, and from it,
one can obtain $f_v^\pi(x)$ from $x\in[0,1]$, or equivalently, from
$[-1,0]$.

By applying the matching formula on $f_u(x)$ and $f_d(x)$ separately
and taking the difference to obtain the $u-d$ RI-MOM qPDF, we
now try to learn what is expected for this qPDF.  Writing down
only the $x/y$ dependence for the sake of brevity and keeping the
dependence on $yP_z/P_z^R$, $(P^R/P^R_z)^2$ and factorization
scale $\mu$ implicit, the one-loop contribution to the matching kernel
$C(x/y)$ from the RI-MOM to $\msbar$ scheme consists of two terms: ${\mathcal
F}_1(x/y)$ and ${\mathcal F}_2(1+\eta'(x-y))$ with
$\eta'=P_z/P_z^R$.  The expressions~\footnote{In~\cite{Stewart:2017tvs},
the terms ${\mathcal F}_1$ and ${\mathcal F}_2$ are referred to as
$f_1$ and $f_2$, respectively} for ${\mathcal F}_{1,2}$ depend on
the choice of $\Gamma$ ($\gamma_z$ or
$\gamma_t$)~\cite{Liu:2018uuj,Stewart:2017tvs}.  Furthermore,
${\mathcal F}_2$ depends on the projection method of the RI-MOM scheme.
Using the matching formula~\cite{Stewart:2017tvs,Liu:2018uuj} on
$f_u(x)$ and $f_d(x)$ to obtain the qPDFs $\tilde{q}_u(x)$ and
$\tilde{q}_d(x)$,

\begin{eqnarray}
    &&
\tilde q_{u,d}(x,P_z,p^R)=f_{u,d}(x,\mu)
    +\frac{\alpha_s C_F}{2 \pi} \int_{-1}^1 \frac{dy}{|y|} {\mathcal F}_1\left(\frac{x}{y}\right)_{+} f_{u,d}(y) \cr
    &&
    \quad-\frac{\alpha_s C_F}{2 \pi} \int_{-1}^1 dy |\eta'| {\mathcal F}_2\left(1+\eta' \left(x-y\right)\right)_{+} f_{u,d}(y)+\cdots.
\label{quddef}
\end{eqnarray}
The above equation includes both the sea and valence quarks, and 
there will be mixing with the gluon PDF which are included in the ``$\cdots$" part. 
In the above convolution, the vector current conservation is ensured 
by the plus function defined as 
\begin{equation}
    {\mathcal F}_{1,2}(\xi)_{+}={\mathcal F}_{1,2}(\xi)-\delta(1-\xi) \int_{reg} d\xi {\mathcal F}_{1,2}(\xi),
\end{equation}
such that any extra variable that ${\mathcal F}_{1,2}$ will depend
on are held fixed in the above integral.  Since the matching between
qPDF and PDF is linear, the $\tilde{q}_{u-d}=\tilde{q}_u(x)-\tilde{q}_d(x)$
is simply obtained as
\begin{eqnarray}
    &&    
\tilde q_{u-d}(x,P_z,P^R)=f_{u-d}(x,\mu)
    +\frac{\alpha_s C_F}{2 \pi} \int_{-1}^1 \frac{dy}{|y|} {\mathcal F}_1\left(\frac{x}{y}\right)_{+} f_{u-d}(y) \cr
    &&
    \quad-\frac{\alpha_s C_F}{2 \pi} \int_{-1}^1 dy |\eta'| {\mathcal F}_2\left(1+\eta' \left(x-y\right)\right)_{+} f_{u-d}(y),
\end{eqnarray}
with the terms in ``$\cdots$" in \eqn{quddef} exactly canceled between the $u$ and $d$ terms. This is the
matching relation we use to obtain the $u-d$ PDF from $u-d$ qPDF. Using the $u-d$ PDF, we 
obtained the valence PDF as discussed above. While $f_{u-d}(|x|)=f_{u-d}(-|x|)$, it is 
also true that $\tilde{q}_{u-d}(|x|)\ne \tilde{q}_{u-d}(-|x|)$ in the RI-MOM scheme.
One way to understand this is from the fact that the bare qPDF matrix element is
purely real while the RI-MOM renormalization factor is in general complex, thereby
making the
renormalized qPDF matrix element complex. One can see this by starting from the 
matching convolution above and find that,
\begin{eqnarray}
    &&\tilde{q}(|x|,P_z,P^R)-\tilde{q}(-|x|,P_z,P^R)=
    \frac{\alpha_s C_F |\eta'|}{2 \pi}\times\cr &&\int_{-1}^1 dy \left[-{\mathcal F}_2(1+\eta'(x-y)) + {\mathcal F}_2(1-\eta'(x-y))\right]f(y,\mu),\cr&&\quad
\end{eqnarray}
is non-zero due to an RI-MOM specific term ${\mathcal F}_2$, while
the terms containing ${\mathcal F}_1$ cancel due to their dependence only on 
$|P_z|$.  In
other schemes such as the $\msbar$, this symmetry about $x=0$ would
be preserved by matching because the corresponding factorization
formulas depend on renormalization/regularization scales through
combinations such as $\mu^2 z^2$~\cite{Izubuchi:2018srq}.  In the RI-MOM
scheme there are two renormalization scales, $P^R$ and $P_z^R$, and
since the $z$-direction is special the above statement does not
hold.  Thus, it is important to capture this asymmetry in the qPDF,
or equivalently to describe both the real and imaginary parts of
the RI-MOM renormalized pion qPDF from matching.  We use the
matching kernel corresponding to the $\slashed{p}$-projection in
the results to be discussed next.

\subsection{Numerical results on pion valence PDF from matching}

\begin{figure*}
    \centering
    \includegraphics[scale=0.65]{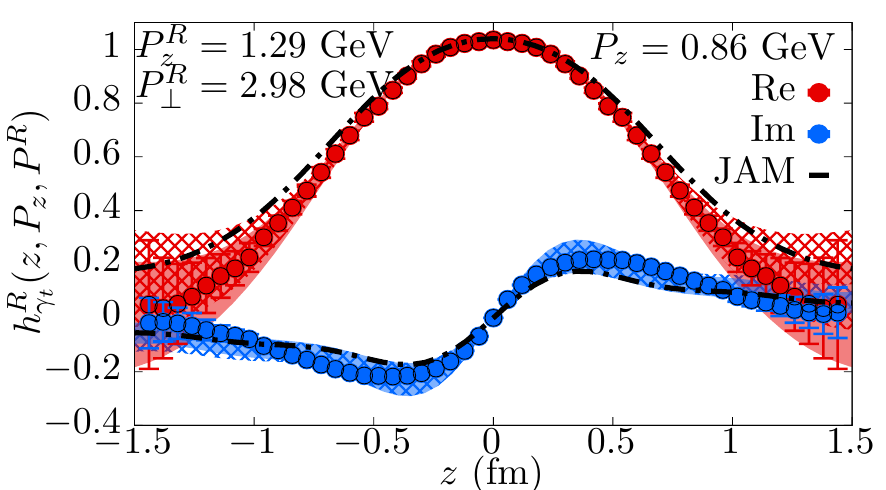}
    \includegraphics[scale=0.65]{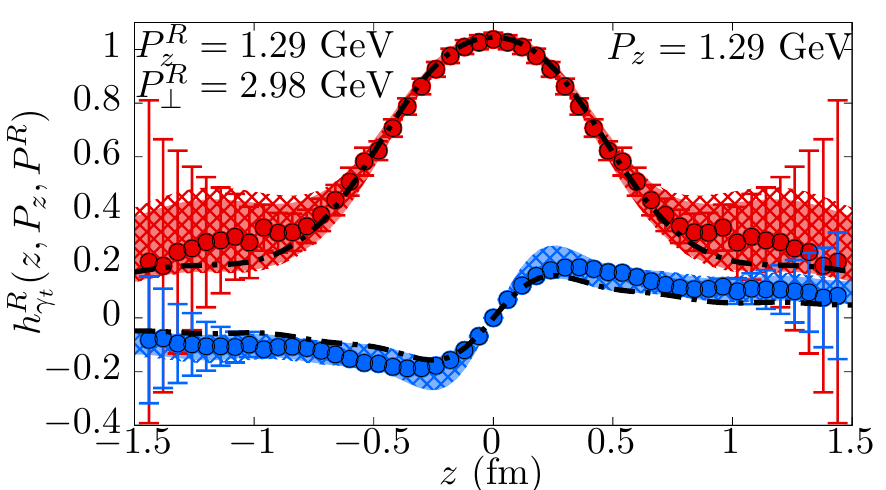}
    \includegraphics[scale=0.65]{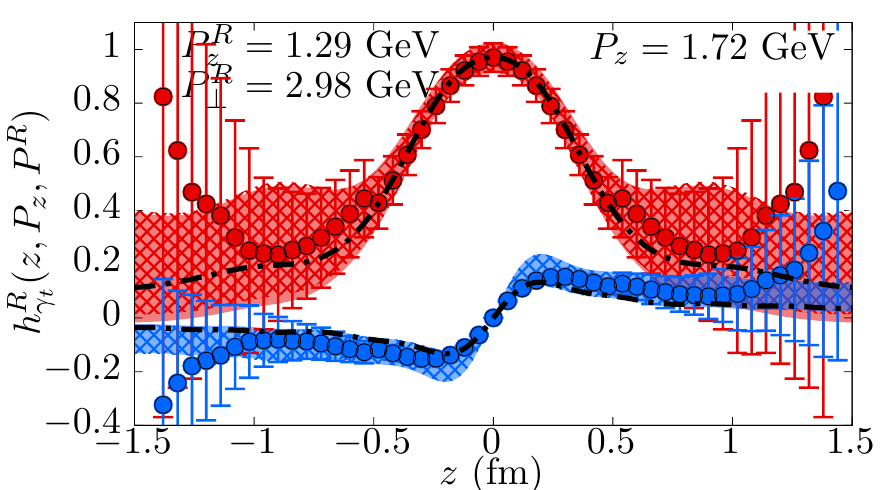}

    \includegraphics[scale=0.65]{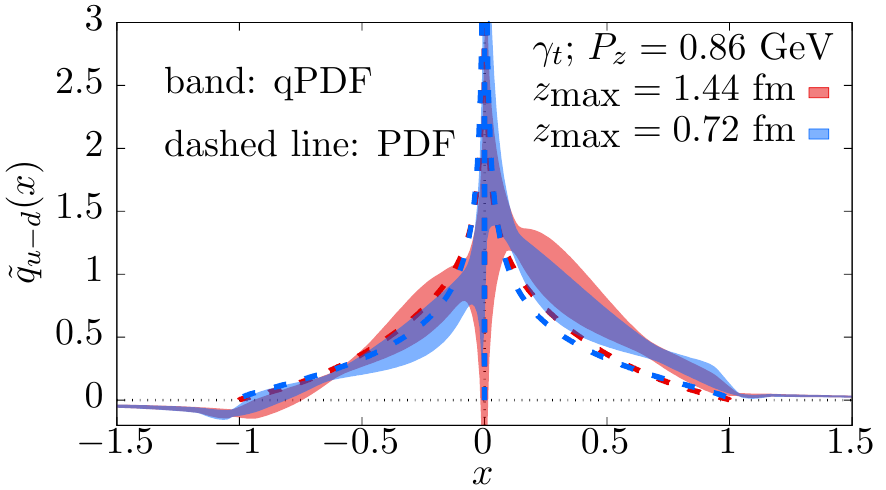}
    \includegraphics[scale=0.65]{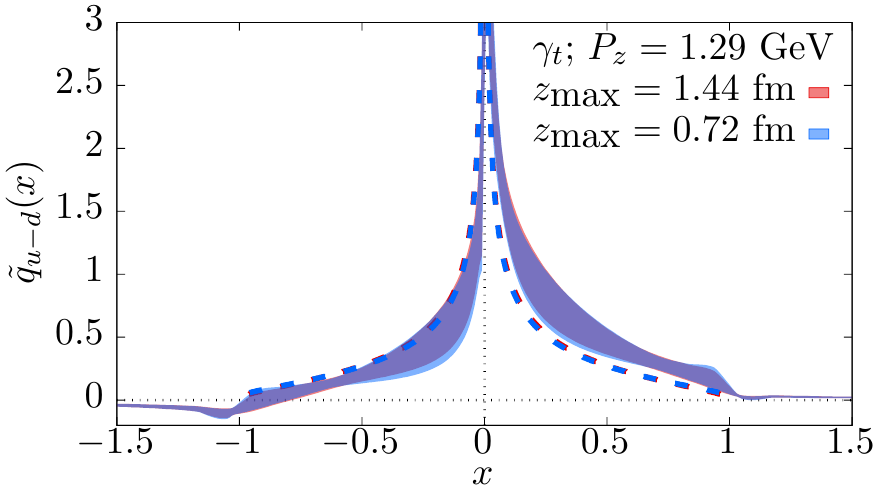}
    \includegraphics[scale=0.65]{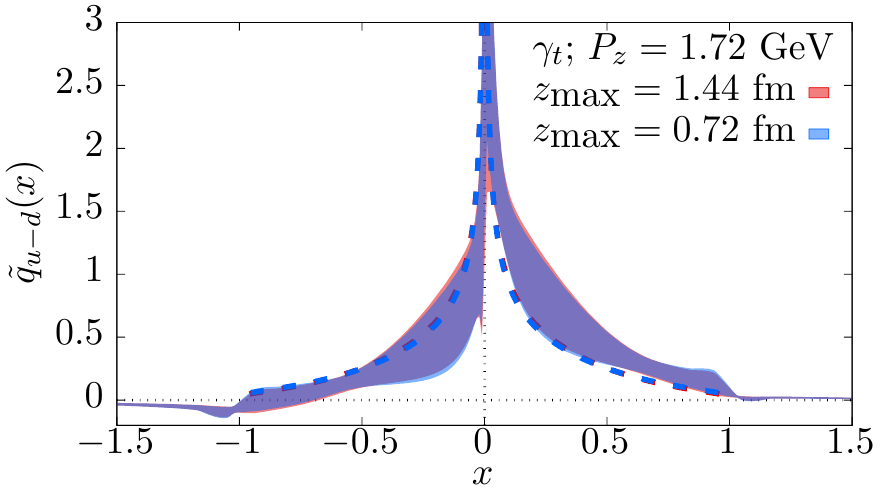}

    \caption{Top panels: The RI-MOM renormalized qPDF
    matrix element in real space $h^R_{\gamma_t}(z,P_z,P^R)$ at
    pion momenta $P_z=0.82, 1.29$ and 1.72 GeV are shown at fixed
    RI-MOM renormalization scale $(P^R_z,P^R_\perp)=(1.29,2.98)$
    GeV.  The red and blue points are the real and imaginary parts
    of the actual data respectively. The bands were obtained by
    fitting the two-parameter phenomenologically motivated real
    space qPDF matrix element to the data over a range
    $z\in[-z_{\rm max},z_{\rm max}]$ -- the solid band is for $z_{\rm
    max}=1.44$ fm and patterned one for $z_{\rm max}=0.72$ fm.
    Bottom panels: The two parameter $u-d$ PDF $f_{u-d}(x)$ (dashed
    lines) at $\mu=3.2$ GeV, and the matched qPDFs
    $\tilde{q}_{u-d}(x)$  (1-$\sigma$ error bands) that describe the real space
    qPDF on the top panels are shown.  To avoid clutter, only
    the central values of $f_{u-d}(x)$ are shown as dashed lines.  The results from
    different $z_{\rm max}$ are shown in red and blue.
    }
\label{fig:fitpdf}
\end{figure*}

The one-loop perturbative matching relates the Fourier transform
$\tilde{q}(x,P_z,P^R)$ of the renormalized RI-MOM real-space
qPDF matrix element $h^R(z,P_z,P^R)$,
and the $\msbar$ PDF $f(x,\mu)$ at factorization
scale $\mu$. The relation is through the convolution in \eqn{pdf2qpdf}.
There are two approaches to consider here:
\begin{enumerate}
    \item  One can parametrize the real space data $h^R(z,P_z,P^R)$
    over the range $z$ where one has the lattice data and then model
    the dependence of
	$h^R(z,P_z,P^R)$ over $z$ extending to infinity where data
	does not exist (c.f.,~\cite{Chen:2017lnm,Lin:2017ani}).
	Using such a parametrization, one can obtain its Fourier
	transform  $\tilde{q}(x,P_z,P^R)$. Since, the matching is
	only up to ${\cal O}({\alpha_s})$, one can invert the
	relation \eqn{pdf2qpdf} by replacing $f \leftrightarrow
	\tilde{q}$ and $\alpha_s\to-\alpha_s$.  Thereby, one can
	obtain $f(x,\mu)$.  In this method, one does not control
	what values of $z$ enter the Fourier transform and one could
	question the validity of perturbation theory for $z>1$ fm.
    \item One can start from a phenomenologically motivated
    $n$-parameter family of PDFs $f(x,\mu;a_1,\ldots a_n)$. Through
    \eqn{pdf2qpdf},
	one can obtain qPDF $\tilde{q}(x,\mu;a_1,\ldots a_n)$,
	and thereby, obtain a family of real space space qPDF matrix elements
	$h^R(z,P_z,P^R;a_1\ldots a_n)$. Using this, one can fit the
	parameters $(a_1,\ldots,a_n)$ so as to best describe the
	real space lattice data over a range $z$. This method was 
        used in the case of lattice cross-section approach in~\cite{Sufian:2019bol}.
        Since the model PDFs are not predictions from QCD, 
        the model dependence enters the analysis and one has to rely on the prior
        that experimentally determined PDFs are indeed very well described by such 
        a family of PDFs. However, the advantage of this method is that one 
        can precisely control the range of $z$ that enters the analysis, and 
        one also does not have to invert the matching convolution.
\end{enumerate}
From our observation on how the 1-loop perturbation theory fails
to capture the quark qPDF quantitatively even at short distances
and from the observation of significant nonperturbative screening
effects beyond $z=1$ fm, we think it is important to be in control
of what values of $z$ enter the convolution and hence, in this paper
we take the second approach. Also, due to the loss of signal to
noise ratio for $z>1$ fm, we found Fourier transforming the noisy
data to be challenging without introducing unwanted wiggles in
$\tilde{q}(x)$ at larger $x$.

To be on par with the experimental extraction of PDFs, one should
use sophisticated methods such as the usage of neural networks to
choose the set of model PDFs to start with (c.f.,~\cite{Karpie:2019eiq}).
We defer such an analysis to a future work and instead, we use a
simple two-parameter phenomenologically motivated functional form
for the valence PDF:
\begin{equation}
    f_v^\pi(x;a,b)=A x^a (1-x)^b,
\end{equation}
for $x\in[0,1]$ and zero elsewhere. As we will see below, such a
form is enough to describe our lattice data. One can fix the
coefficient $A$ through a stringent condition $\int_0^1 f^\pi_v(x)
dx =1$. Instead, we use a more conservative constraint on $A$ using
$\int_0^1 f^\pi_v(x) dx=h^R(z=0,P_z,P^R)$ to allow for sample by
sample fluctuations in $h^R(z=0,P_z,P^R)$ close to 1 and fold this
into the error estimate. It should be noted that the valence PDF
of pion determined from the experimental data by the JAM
collaboration~\cite{Barry:2018ort} can be well described by such a
two parameter ansatz, for example with $a=-0.407$ and $b=1.12$ at
$\mu=3.2$ GeV.

\begin{figure*}
    \centering
    \includegraphics[scale=0.85]{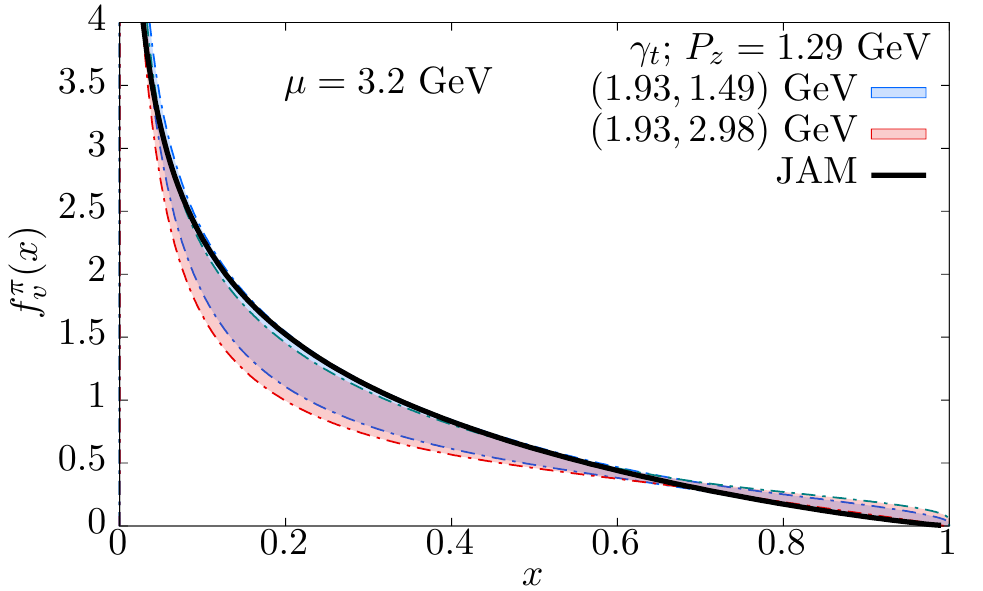}
    \includegraphics[scale=0.85]{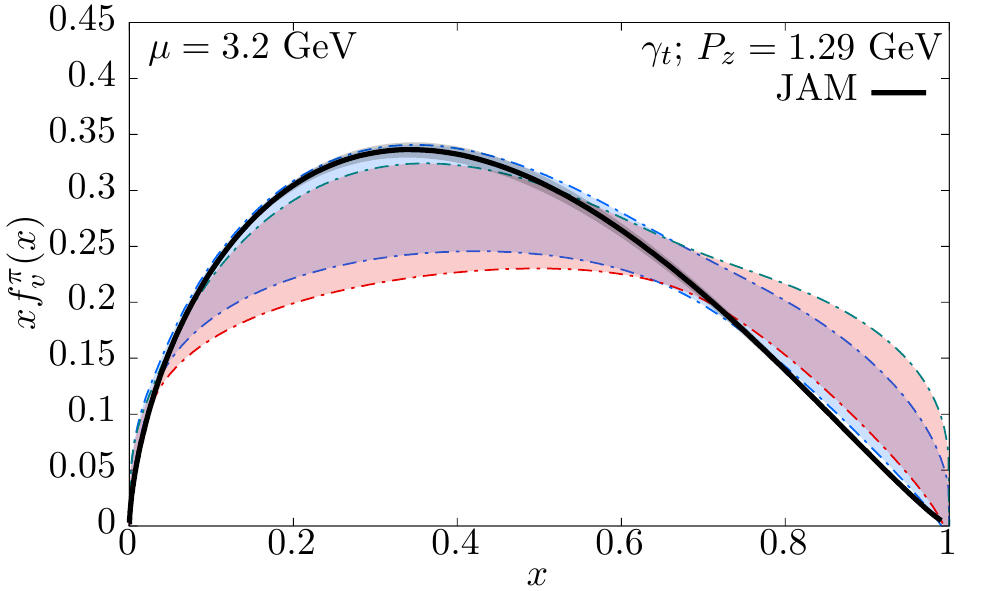}

    \includegraphics[scale=0.85]{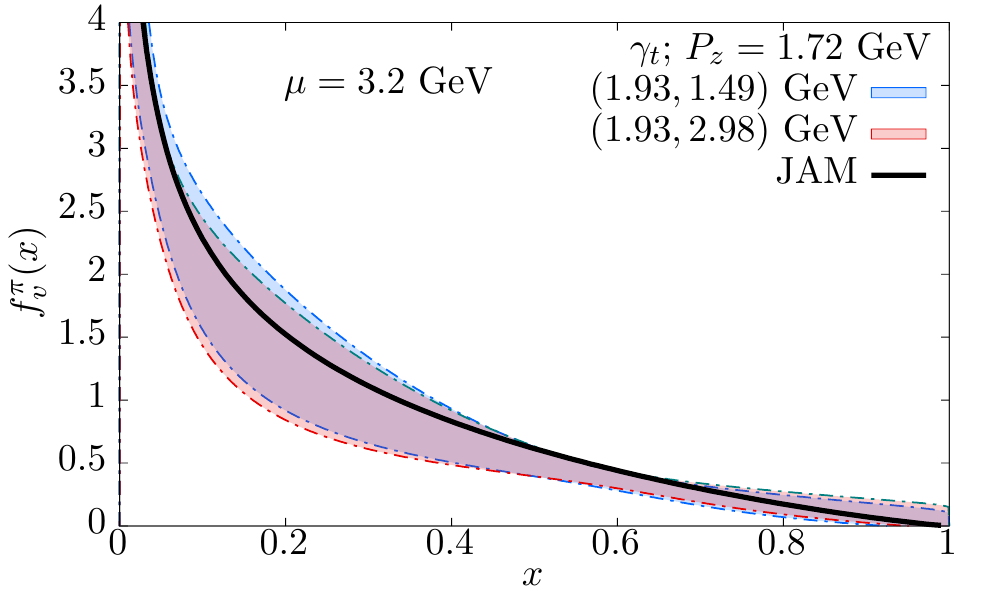}
    \includegraphics[scale=0.85]{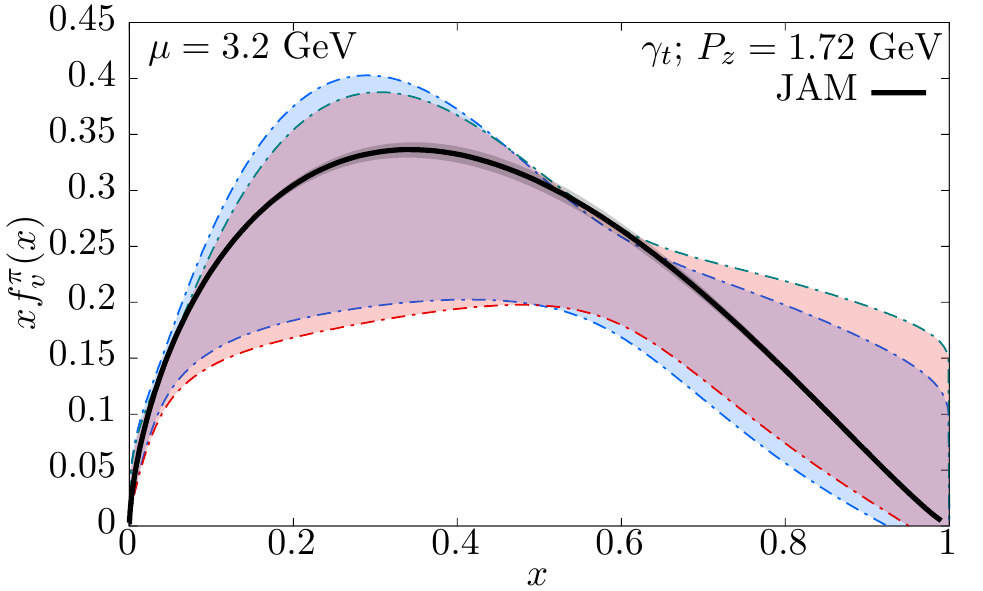}
    \caption{
	The top and bottom panels show our estimated pion valence
    PDF at $\mu=3.2$ GeV using $\gamma_t$ qPDF at $P_z=1.29$
    and 1.72 GeV respectively. The results using multiple RI-MOM
    scales $(P^R_z,P^R_\perp)$ are shown using different colored
    error bands.  On the left panels, the results for $f_v^\pi(x)$
    are shown, while on the right panels the results for $x f_v^\pi(x)$
    are shown.  For all the cases shown, the fit range was held
    fixed at $z_{\rm max}=0.98$ fm. The solid line (with a small
    error band around) is the JAM result~\cite{Barry:2018ort} for
    pion valence PDF at the same $\mu$.
    }
\label{fig:pdfprdep}
\end{figure*}

Using the above valence PDF, we construct the $u-d$ PDF
as
\begin{equation}
    f_{u-d}(x;a,b)=\frac{1}{2}\left(f^\pi_v(x;a,b)+f^\pi_v(-x;a,b)\right),
\end{equation}
with $x\in[-1,1]$ and zero elsewhere. Through the convolution of $f_{u-d}(x)$ with
the matching kernel, we obtain $\tilde{q}_{u-d}(x;a,b)$, which in turn we 
use to
construct the real space qPDFs $h(z;a,b)=\int_{-\infty}^\infty
\tilde{q}(x;a,b)e^{i x P_z z} dx$.  We will refer to these functions 
$h(z;a,b)$ as the two-parameter family of phenomenologically motivated
qPDF matrix elements. With the set of $h(z;a,b)$ from
a range of $a$ and $b$, we can fit the parameters $a$ and $b$ to
the data by minimizing either $\chi^2_r$ or $\chi^2_{ri}$ below:
\begin{eqnarray}
    \chi^2_r&=&\sum_{z=-z_{\rm max}}^{z_{\rm max}}\frac{\left({\rm Re}(h^R(z))-{\rm Re}(h(z;a,b))\right)^2}{\sigma_r(z)^2+\sigma^{\rm pert}_r(z)^2},\cr
    \chi^2_i&=&\sum_{z=-z_{\rm max}}^{z_{\rm max}}\frac{\left({\rm Im}(h^R(z))-{\rm Im}(h(z;a,b))\right)^2}{\sigma_i(z)^2+\sigma^{\rm pert}_i(z)^2},\cr
    \chi^2_{ri}&=&\chi^2_r+\chi^2_i.
\end{eqnarray}
In the above equations, $[-z_{\rm max},z_{\rm max}]$ specifies the
fit range. The statistical errors on the real and imaginary parts
of the lattice data $h^R(z)$ is $\sigma_r(z)$ and $\sigma_i(z)$
respectively. To account for any systematic errors coming from the higher
order corrections in $\alpha_s$ in the matching kernel, we determine
$h(z;a,b)$ from $f^\pi_v(x;a,b)$ by varying the value of $\alpha_s$
in the matching kernel from $\alpha_s(\mu/2)$ to $\alpha_s(2\mu)$ though
the 1-loop running. The corresponding changes in the real and imaginary
parts of $h(z;a,b)$ are denoted as $\sigma_r^{\rm pert}(z)$ and
$\sigma_i^{\rm pert}(z)$ respectively, and we include these uncertainties
in the matched result in the $\chi^2$. If the matching was exact,
then by fitting only the real part by minimizing $\chi^2_r$ would
automatically guarantee that the imaginary part also agree with the
data. Therefore at any finite order matching, the fits obtained by
minimizing $\chi^2_r$ and $\chi^2_{ri}$ will in general be different.
For the results shown below, we used $\chi_{ri}^2$ in order to
obtain the PDF that best describes both the real and imaginary parts
of the real space qPDF, but we also used $\chi^2_r$ and found
it to lead to consistent results, but with larger uncertainties.  We
did not include the correlations between the data at different $z$
for the primary reason that it is difficult to keep these correlations
intact in the process of excited state extrapolations. It also helps
us to easily incorporate the effect of $\sigma^{\rm pert}$ from
non-statistical origin in the analysis, and in treating the real
and imaginary parts of the renormalized matrix elements as two
distinct pieces of data as is the case in the context of matching.
We determined the errors on the fit parameters through the bootstrap
analysis.

In \fgn{fitpdf}, we show the fitting procedure for $\gamma_t$
qPDF.  In the top panels, we show the $P_z=0.86$, 1.29 and
1.72 GeV real-space RI-MOM pion qPDF matrix elements from left
to right. The symbols are the actual lattice data.  The solid and
patterned red (blue) bands are 1-$\sigma$ error-bands of the real
(imaginary) parts of the fitted real space qPDF
matrix element that best fits the data over the range $[-z_{\rm
max},z_{\rm max}]$ for $z_{\rm max}=1.44$ fm and 0.72 fm respectively.
The agreement with both the real and imaginary parts of the lattice
data is noteworthy. In fact, we find the qPDF matrix element
as inferred from the JAM PDF~\cite{Barry:2018ort} is able to explain
the lattice data well for the entire range of $z$ at the two largest
momentum.  In the bottom panels, we show the process leading from
model PDF to the real space qPDF matrix elements
shown in the top panels. In order to avoid cluttering the figure, 
we have shown only
the mean value of $f_{u-d}(x)$ (shown as dashed lines) while  we
have shown the error bands for the qPDF $\tilde{q}_{u-d}(x)$
as obtained through the 1-loop matching. The colors red and blue in the
bottom panels correspond to the fits with $z_{\rm max}=0.72$ and
1.44 in the top panels respectively.  As one can see, we started
from a symmetric $u-d$ PDF by construction and matching introduces
an $x\to-x$ asymmetry.  After Fourier transformation, this asymmetry
leads to the imaginary part in the real space data in the top panels
which captures the lattice data to a good accuracy.  For both the
real-space as well as in $x$ space, we find no significant difference
between using $z_{\rm max}=1.44$ fm and $0.72$ fm in the fits. We
could infer that within the precision of our numerical results, the
non-perturbative effects at $z\approx 1$ fm that we found using
quark qPDFs is not important.  Therefore, we show results for
an intermediate $z_{\rm max}=0.98$ fm in the results below. When
we repeated this analysis by minimizing $\chi^2_r$, we found the estimates
to be consistent with the above, but with larger uncertainties.

In \fgn{pdfprdep}, we show our results for $f_v^\pi(x,\mu)$ and $x
f_v^\pi(x,\mu)$ at the factorization scale $\mu=3.2$ GeV using the
procedure described above at our two largest pion momenta $P_z=1.29$
GeV and $1.72$ GeV starting from $\gamma_t$ qPDF.  For each
case, we overlay the results from two different RI-MOM scales $P^R$ in
order to show the scatter as a systematic error in our estimates.
We find the $P^R$ dependence to be minor compared to the error bands
(we repeated the analysis with multiple other values of $P^R$ that
are not shown and only minor scatter with respect to $P^R$ was
seen). We also show the result from the JAM collaboration~\cite{Barry:2018ort} 
for the
pion valence PDF at the same factorization scale as the black solid
line, which lies within the statistical and systematic uncertainties
of our estimates. In the left panels showing $f_v^\pi(x)$, this
overall agreement can be seen even up to smaller $x$, but one has
to be cautious of this agreement for $x\lessapprox \Lambda_{\rm
QCD}/P_z\approx 0.2$ for the two highest pion momenta we use.  By
construction, in our fitting procedure $f_v^\pi(x)$ has support
only from 0 to 1 without any necessity to recover this condition
in the infinite $P_z$ limit.  However, the values of exponent $b$
closer to zero are also allowed thereby leading to a wider error
band closer to $x$=1. This seems to be consistent with the observation
in Ref~\cite{Chen:2018fwa} that the PDF obtained from qPDF
through the inverse one-loop matching (approach-1) vanishes at about
$x\approx1.2$.  We see our $P_z=1.29$ GeV and 1.72 GeV estimates
to be consistent albeit with a significant increase in error at the
largest momentum.

\begin{figure}
\centering
\includegraphics[scale=1.0]{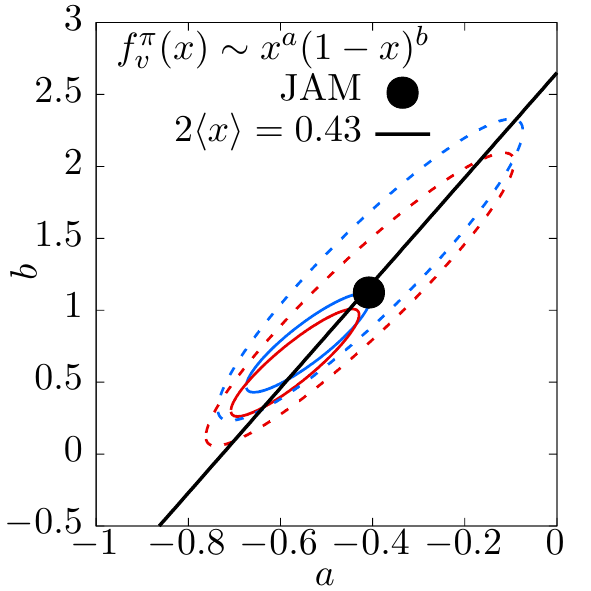}
    \caption{
The 1-$\sigma$ confidence region ellipse of the exponents $a$ and
$b$ in the model PDF at $\mu=3.2$ GeV that best describes
the real space RI-MOM  qPDF are shown. The solid lines and
dashed lines correspond to $P_z=1.29$ GeV and $P_z=1.72$ GeV. For
each of these pion momenta, the different colored lines correspond
    to different RI-MOM scale $P^R$. The black point is the JAM value~\cite{Barry:2018ort}
for valence pion PDF. The black straight line is the line of 
constant first moment of valence PDF, $\langle x\rangle =0.215$.
    }
\label{fig:covar}
\end{figure}

In \fgn{covar}, we summarize the information in \fgn{pdfprdep} by
showing the 1-$\sigma$ ellipses (whose $x$ and $y$ projections give
the marginal 68\% confidence intervals of the exponents $a$ and $b$
respectively).  In this figure, the dashed and continuous ellipses
are for $P_z=1.29$ and 1.72 GeV respectively.  The ellipses for
different $P^R$ are distinguished by the colors, with the color
code being the same as in \fgn{pdfprdep}.  The $P_z=1.29$ GeV data
offers a stronger constraint on allowed region of $(a,b)$ than the
noisier $P_z=1.72$ GeV.  In this plot, the JAM estimate is the black
point.  The JAM data is well within the $P_z=1.72$ GeV ellipses
while the $P_z=1.29$ GeV data seems to favor slightly smaller
exponent $b$. However, these differences are well within $2\sigma$.
Even though our lattice data has large errors on the exponents $a$
and $b$ individually, the data offers a tight constraint on the
combined allowed region.  In particular, the principal component
of this correlation between $a$ and $b$ points directly at the JAM
data implying that if one fixes the exponent $a$ to be from the
experiment, then the best value of $b$ would also be closer to that
from the experiment.  To understand this better, we have also shown
the line of constant value of first moment of the valence PDF, 
$\langle x\rangle=\int_0^1 x q_v^\pi(x)dx$, set
to 0.215 as inferred from the JAM data. It is clear that the 1$\sigma$
ellipses are oriented along this line, which means that qPDF
determines $\langle x\rangle$ robustly and this in turn provides a
strong constraint in the allowed PDFs.  Not surprisingly, we do
find consistent values of $\langle x\rangle=$ 0.21(2) and 0.22(3)
from the $P_z=1.29$ and 1.72 GeV estimates. It should be noted that
the moments of pion PDF have also been directly determined without
the usage of LaMET
formalism~\cite{Best:1997qp,Guagnelli:2004ga,Capitani:2005jp,Abdel-Rehim:2015owa,Oehm:2018jvm}
and similar values for the first moment for the pion were obtained,
but at slightly different values of $\mu^2$ than used here.

\begin{figure}
    \centering
    \includegraphics[scale=0.55]{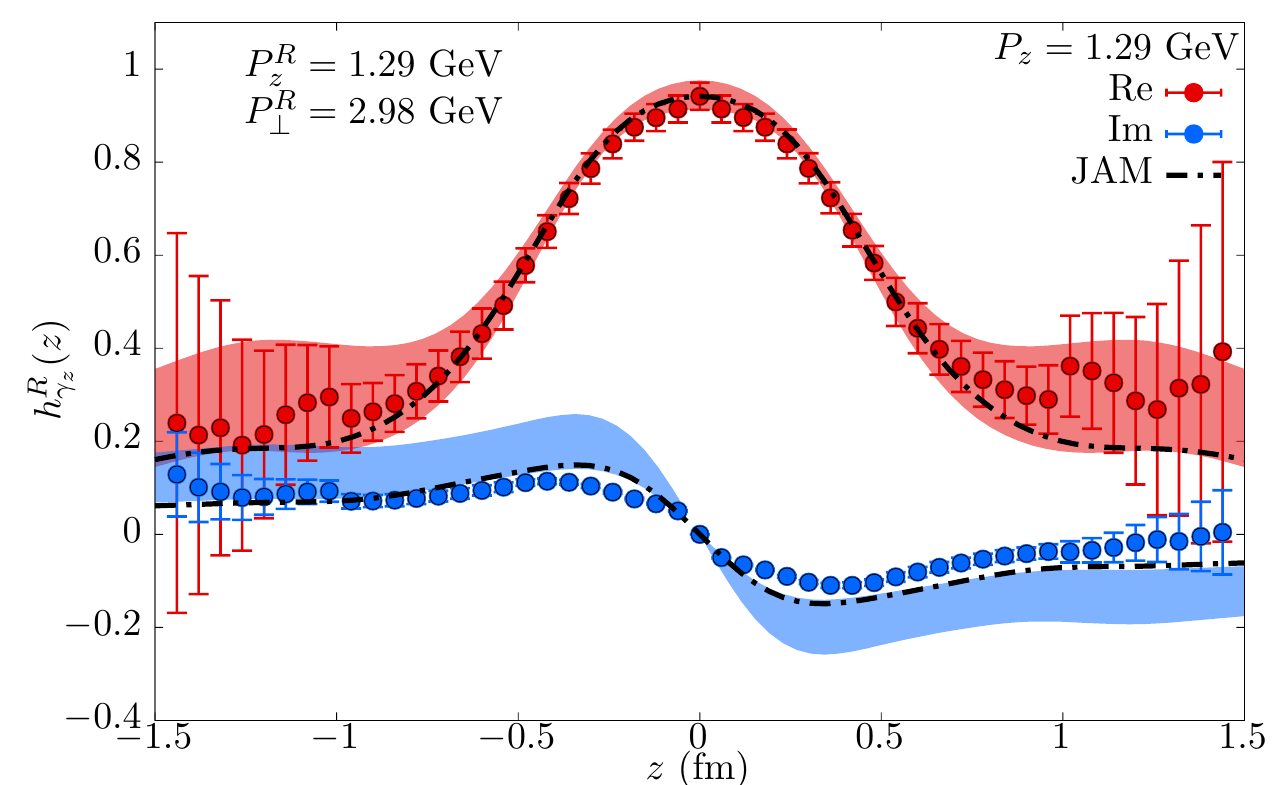}

    \includegraphics[scale=0.55]{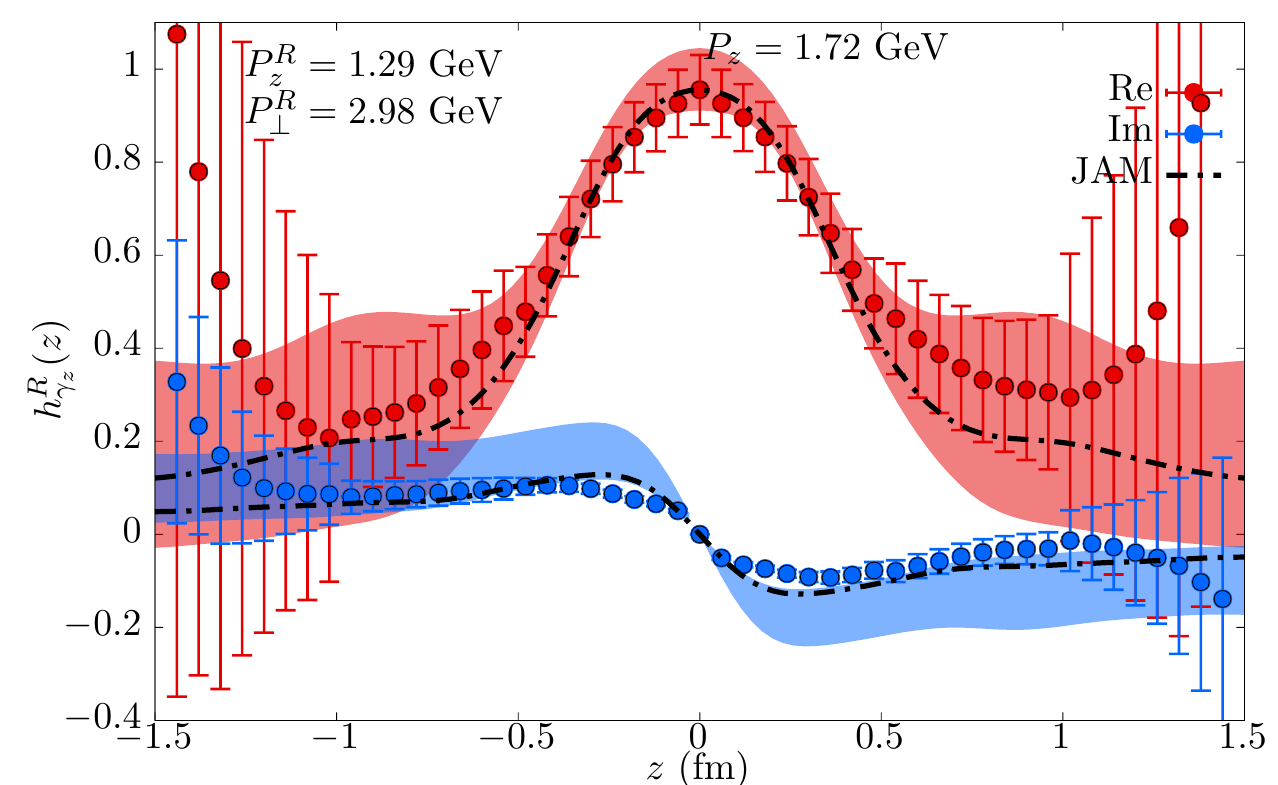}
    \caption{
        In the top and bottom panels, the real (red) and imaginary (blue) parts of the 
    renormalized real space $\gamma_z$ qPDF
    matrix element are shown for pion momenta $P_z=1.29$ GeV and 1.72 GeV
    respectively. The data points are the actual lattice data. 
    The bands are the expected matched $\gamma_z$ qPDF
    matrix element starting from our best estimate for valence pion PDF
    obtained using $\gamma_t$ qPDF analysis.
    }
\label{fig:gzpdf}
\end{figure}

The exponents $a$ and $b$ were also recently obtained using the
lattice cross-section approach~\cite{Sufian:2019bol} which used
current-current correlators, with the matching implemented at tree-level. 
Here, the exponents were estimated as
$a=-0.34(31)$ and $b=1.93(68)$ which are consistent with the region
allowed at the largest momentum in \fgn{covar}. It is worth noting
that there are indications from next-to-leading-logarithmic soft
gluon resummation calculation~\cite{Aicher:2010cb},
the Dyson-Schwinger equation~\cite{Nguyen:2011jy,Chen:2016sno,Bednar:2018mtf}
and light-front holographic QCD~\cite{deTeramond:2018ecg}
that the value
of exponent $b$ could be approximately 2 as expected from perturbative
counting rule (c.f.,~\cite{Brodsky:2005wx}), whereas a chiral quark model analysis~\cite{RuizArriola:2002wr,Broniowski:2017wbr}
suggests a value of $b$ closer to 1. It will be interesting
to see if a similar implementation of an improved matching kernel
could lead to a softer large $x$ behavior for the pion than what
is observed using the 1-loop qPDF matching here and perhaps
in~\cite{Chen:2018fwa}.  In fact, a general consideration of power
corrections to qPDF~\cite{Braun:2018brg} revealed the presence
of the form $\Lambda^2_{\rm QCD}/\left((1-x)x^2P^2_z\right)$ implying
higher values of $P_z$ might be required in order to correctly
describe physics close to $x=1$, and this might be the effect which we
are finding. Similar conclusions have also been obtained in 2d QCD~\cite{Ji:2018waw}.

Due to the larger errors in the $\gamma_z$ qPDF attributed mostly to
the steep excited state extrapolations,
we use the $\gamma_z$ qPDF to
provide a consistency check of our calculations instead.
For this, we use our above best estimates of the PDF obtained using
the $\gamma_t$ qPDF to get the corresponding prediction for
the real space $\gamma_z$ qPDF matrix element through a
convolution with the appropriate matching kernel.  In the top and
bottom panels of \fgn{gzpdf}, we show such a comparison between the
actual real space data of $\gamma_z$ qPDF (data points) along
with the prediction from our estimated PDF (bands) for pion momenta
$P_z=1.29$ and 1.72 GeV.  We find good descriptions of the real
part of the RI-MOM $\gamma_z$ qPDF at both the pion momenta
with a slight tension between the imaginary parts.  From our
discussion on the excited state contamination, it is important to first
gain better control of the larger excited state contamination in
$\gamma_z$ qPDF before one can investigate the effect of
one-loop matching on this rather small discrepancy.

\section{Conclusions}
\label{sec:concl}

We studied pion PDF in the framework of LaMET, which relates the
qPDF to PDF through the matching convolution in Eq. (\ref{pdf2qpdf}).
For this, we used a small lattice spacing $a=0.06$ fm.  We carefully
examined the effects of excited states using the two- and three-state exponential
fits of the relevant 2-point and 3-point functions as well as the
summation method. For our final analysis we used two momenta
$P_z=1.29$ GeV and $P_z=1.72$ GeV. We found the qPDF defined using
$\Gamma=\gamma_t$ was better determined compared to the $\gamma_z$
qPDF in the lattice calculation for the following reasons: smaller
statistical error, relatively smaller excited state extrapolation
leading to more robust result for the matrix element as well as due
to the absence of mixing. Therefore, we focused on the analysis of
the $\gamma_t$ matrix element.

The pion qPDF was non-perturbatively renormalized using the RI-MOM
scheme by calculating the matrix elements of qPDF operator with
off-shell quark states in the Landau gauge for different separations
$z$.  For these calculations we also used finer lattices with the lattice
spacing $a=0.04$fm.  We performed the comparison of this matrix
element in the Landau gauge with 1-loop perturbative calculations in
the RI-MOM scheme and found a qualitative agreement for $z<0.3$ fm. For
the smaller lattice spacings, $a=0.04$ fm we even found quantitative
agreement with the 1-loop result for sufficiently small $z$. We
also explored the role of non-perturbative effects in the calculation
of the off-shell matrix element.  The real part of the RI-MOM
renormalization coefficient is close to one, while the imaginary
part is close to zero once the divergent self energy part of the
Wilson line is removed.  We pointed out that the RI-MOM renormalization
procedure leads to an asymmetry in the iso-vector pion qPDF $\tilde
q(x,P_z,p_z^R,\mu_R)$ around $x=0$, while other renormalization
procedures lead to qPDF that is symmetric around $x=0$.

From the renormalized qPDF, we determined the valence quark pion
PDF using the 1-loop perturbative matching of the $\gamma_t$ qPDF, which
we implemented through a fit to the phenomenologically motivated
$x^a(1-x)^b$ functional form for the valence PDF.  We found our
results for the pion valence PDF using the two largest pion momenta
were consistent with each other, though the statistical errors are
rather large.  An overall agreement with the results obtained
recently by the JAM collaboration~\cite{Barry:2018ort} was seen.
We found our result for the PDF to capture the first moment $\langle
x\rangle$ more robustly than the small-$x$ and large-$x$ exponents,
$a$ and $b$ themselves.  We used the $\gamma_z$ qPDF matrix
elements to provide an internal consistency check by comparing to
the expectation from our estimates of the PDF and a satisfactory
agreement was seen.  From our analysis it is clear that the dominant
source of errors in the PDF determination is the statistical error of
the lattice calculations. It will be necessary to significantly
increase the statistics in the future lattice calculations. Future
high statistics lattice calculations will be important for an
accurate determination of the pion PDF as well as testing of the
LaMET approach around small $x$.

\section*{Acknowledgments}
We thank P.C. Barry for sharing the JAM valence PDF data, and
J. H. Weber for providing us the Wilson line data.  We thank R. S.
Sufian and Y. Zhao for fruitful discussions.  
This material is based upon work supported by: (i) 
The U.S. Department of Energy, Office of Science, Office of Nuclear Physics and High Energy Physics through the Contract No. DE-SC0012704; 
(ii) The U.S. Department of Energy, Office
of Science, Office of Nuclear Physics and Office of Advanced
Scientific Computing Research within the framework of Scientific
Discovery through Advance Computing (ScIDAC) award Computing the
Properties of Matter with Leadership Computing Resources; (iii) The
Brookhaven National Laboratory's Laboratory Directed Research and
Development (LDRD) project No. 16-37.
S.S. is supported by the National Science Foundation under CAREER 
Award PHY-1847893.
This research used awards of computer time provided by: (i) The
INCITE program at Oak Ridge Leadership Computing Facility, a DOE
Office of Science User Facility operated under Contract No. DE-AC05-
00OR22725; (ii) The USQCD consortium at its Brookhaven National
Laboratory and Jefferson Laboratory computing facilities.
The gauge configurations used in this study were generated using
awards of computing time provided by: (i) The INCITE program at
Argonne Leadership Computing Facility, a U.S. Department of Energy
Office of Science User Facility operated under Contract No.
DE-AC02-06CH11357.

\appendix

\section{Coulomb-gauge Gaussian and momentum (boosted) smearing}
\label{app_momsmear}
In order to create hadron interpolating operators that have a good
overlap with the corresponding ground states, quark field smearing
is typically required.  The amount of applied smearing is tuned to
produce spatial quark distributions of roughly of the same spatial
size as the hadron.  Gauge-covariant Wuppertal (Gaussian)
smearing~\cite{Gusken:1989qx} is commonly used for this purpose.
However, calculation of quasi- and pseudo-PDFs and high-momentum
hadron structure and spectrum in general requires lattices with
small lattice spacing.  Keeping the physical size of smeared quark
distributions the same becomes a numerical challenge on finer
lattices because it requires larger numbers of smearing iterations.
For this reason, we use Gaussian shape smearing in a fixed (Coulomb)
gauge that can be performed efficiently through a convolution with
a Gaussian profile kernel,
\begin{equation}
\label{coulgauss_smearing}
\mcS^\text{CG}_{x,y} \propto e^{-\frac{(\vec x-\vec y)^2}{2w_\text{CG}^2}}
  \propto \sum_{\vec p} e^{i\vec p (\vec x - \vec y)} \, e^{-\frac12 w_\text{CG}^2 \vec p^2}\,.\\
\end{equation}
In the free-field case, this kernel corresponds to the Wuppertal
smearing operator $(1+\frac{w^2}{4N}\Delta_\text{sp})^N$, where
$\Delta_\text{sp}$ is the spatial Laplacian and $w^2=2w_\text{CG}^2$.
The value for the width $w_\text{CG}$ is chosen to match the
mean-squared radius $\langle r^2\rangle = 3w_\text{CG}^2$ to that
of the optimal Wuppertal-smeared quark sources.  First, we fix the
Coulomb gauge
\begin{equation}
\psi^\text{C}_x = \Omega^\text{C}_x \psi_x\,,\quad
U^\text{C}_{x,\mu} = U^{\Omega^\text{C}}_{x,\mu} 
  = \Omega^\text{C}_x U_{x,\mu}\Omega^{\text{C}\dag}_{x+\hat\mu}\,,
\end{equation}
where $\Omega^C_x$ is the gauge transformation to the Coulomb gauge, 
which minimizes the functional
\begin{equation}
\Omega^\text{C} = \Omega\,:\quad
  \min_{\Omega_x} \, F^\text{C}[U^{\Omega}]
  = \min_{\Omega_x} \, \sum_x \sum_{\mu\ne t} \big[-\Re\Tr U^\Omega_{x,\mu}\big]\,,
\end{equation}
(for the Coulomb gauge, $\mu\ne t$ and the functional is minimized
independently on each time slice).  The numerical implementation
is identical to the algorithm used for fixing the Landau gauge in
NPR calculations.  Application of the smearing kernel requires two
3D Fourier transformations
\begin{equation}
\label{cgauss_conv}
[\mcS^\text{CG}\psi]_x
  = \Omega^{\text{C}\dag}_x \, \sum_{\vec p} e^{i\vec p \vec x}
      \, e^{-\frac12 w_\text{CG}^2 \vec p^2} \,
      \Big[\frac1V \, \sum_{\vec y} e^{-i\vec p \vec y} \, \Omega^\text{C}_y \, \psi_y\Big]\,,
\end{equation}
which is accelerated with offloading matrix-matrix products to GPU.

Incorporating momentum (boosted) into Coulomb-gauge Gaussian smearing
amounts to translation of the kernel in the momentum space,
\begin{equation}
\label{cgkernel_boosted}
(\mcS^{(\vec k)}\psi)_x 
  = e^{i\vec k\vec x} \, \big[\mcS^{(\vec 0)}\big]_{x,y} \, e^{-i\vec k\vec y} \, \psi_y
  = \big[e^{i\vec k\vec x} \, \mcS \, e^{-i\vec k\vec y} \big]_{x,y} \, \psi_y
\end{equation}
In a periodic finite volume, care must be taken to avoid spatial
discontinuities in the boosted smearing
kernel~(\eqn{cgkernel_boosted}).  Such discontinuities may arise
because the optimal boosted smearing momentum $\vec k$ typically
does not conform to finite-volume momentum quantization $\vec k =
2\pi\frac{\vec n}{\vec L}$ and the phase factors $e^{i\vec k\vec
x},\,e^{i\vec k\vec y}$ do not satisfy periodic boundary conditions.
The solution is to define the smearing kernel in the momentum space
as $\mcS^{(\vec k)}_{\vec x, \vec y} = \sum_{\vec p} e^{i\vec p
(\vec x - \vec y)} \, e^{-\frac12 w_{CG}^2 (\vec p-\vec k)^2}$,
where the momentum difference $(\vec p-\vec k)$ is understood as
the shortest distance between $\vec p$ and $\vec k$ in the Brillouin
zone.  Such choice leads to a smooth distribution in the momentum
space and respectively smooth and continuous smearing kernel in the
coordinate space.

Finally, it is important to note that the smearing
kernel in~\eqn{cgkernel_boosted} is Hermitian (as an operator
acting in the [coordinate $\otimes$ color] space),
\begin{equation}
\label{boostsm_herm}
\mcS^{(\vec k)\dag}_{x,y} = \big[ \mcS^{(\vec k)}_{y,x}\big]^\dag 
  = \mcS^{(\vec k)}_{x,y}\,.
\end{equation}
which is similar to the (boosted) Wuppertal smearing operator and
important for computing symmetric hadron correlation functions.

\section{Meson correlation functions with boosting}
\label{app_contract}

We use the interpolating operator for the $\pi^+=\bar{d}u$ meson 
\begin{equation}
\big[\pi^{+,(2\vec k)}\big]_{\xa} = \bar{\tilde{d}}_{\xa}\Gamma_M \tilde u_{\xa}
  = \bar{d}_{\xc} \, \mcS^{(-\vec k)}_{\xc,\xa} \, \Gamma_M \, 
      \mcS^{(\vec k)}_{\xa,\xb}\, u_{\xb} \,,
\end{equation}
which is constructed from smeared quark fields 
\begin{equation}
\bar{\tilde{d}}_{\xa} = \bar{d}_{\xc} \mcS^{(-\vec k)}_{\xc,\xa}\,,\\
\quad\tilde{u}_x      = \mcS^{(\vec k)}_{\xa,\xb} u_{\xb}\,,
\end{equation}
where the spinor matrix $\Gamma_M=\gamma_5$.
The Hermitian-conjugated (creation) meson operator is
\begin{equation}
\big[\pi^{+,(2\vec k)}\big]^\dag_{\xa} = \tilde{u}_{\xa}^\dag\, \Gamma_M^\dag \, \gamma_4 \tilde{d}_{\xa} 
  = \bar{u}_{\xb} \, \mcS^{(\vec k)}_{\xb,\xa} \, \overline{\Gamma}_M \,
      \mcS^{(-\vec k)}_{\xa,\xc} d_{\xc}\,,
\end{equation}
where $\overline{\Gamma}_M = \gamma_4\Gamma_M\gamma_4=(-\gamma_5)$.
The meson two-point correlation function with boost-smeared source
and sink and momentum projection at the sink is\footnote{
  Unless explicit summation is performed, implicit summation over
  repeated coordinate indices $\xb,\xc,\yb,\yc$ is assumed, as well
  as all over omitted spin and color indices.  }
\begin{equation}
\begin{aligned}
&C_\text{2pt}(\ya_4,\vec p;\,x) 
    = \sum_{\vec y}e^{-i\vec p(\vec y-\vec x)} \big\langle \big[\pi^{+,(2\vec k)}\big]_y \, 
        \big[\pi^{+,(2\vec k)}\big]^\dag_x \big\rangle \\
    \quad  &= \sum_{\vec y}e^{-i\vec p(\vec y-\vec x)} \, \Big(-\Tr\big[
        \mcS^{(-\vec k)}_{\xa,\xc} \, Q^d_{\xc,\yc} \, \mcS^{(-\vec k)}_{\yc,\ya} 
        \, \Gamma_M \,
        S^{(\vec k)}_{\ya,\yb} \, Q^u_{\yb,\xb} \, S^{(\vec k)}_{\xb, \xa}
        \, \overline{\Gamma}_M
      \big]\Big) \\
  \quad  &= \sum_{\vec y}e^{-i\vec p(\vec y-\vec x)} \, \Tr\big[
        \tilde{Q}^{d,(-\vec k)}_{\xa,\ya}\, \Gamma_M \,
        \tilde{Q}^{u,(\vec k)}_{\ya,\xa} \, \big(-\overline{\Gamma}_M\big)
\big]\,, 
\end{aligned}
\end{equation}
where $Q^q_{x,y}=\langle q_x \bar q_y\rangle$ and $\tilde Q^{q,(\pm\vec
k)} = \mcS^{(\pm\vec k)} Q^q \mcS^{(\pm\vec k)}$ are unsmeared and
smeared quark propagators, respectively.  Note that the meson
two-point function is constructed from the $u$-quark propagator
$\ya\leftarrow\xa$ and the $d$-quark propagator $\xa\leftarrow\ya$
smeared with momenta $(\vec k)$ and $(-\vec k)$, respectively.
Therefore, separate propagators for $u$ and $d$ quarks are required
to construct meson correlation functions ``boosted'' with the total
momentum
$(2\vec k)$, 
\begin{equation}
\begin{aligned}
\tilde{Q}^{u,(\vec k)}_{\ya,\xa} 
  &= S^{(\vec k)}_{\ya,\yb} \, Q^u_{\yb,\xb} \, S^{(\vec k)}_{\xb,\xa}
  \propto e^{i\vec k(\vec\ya -\vec\xa)}\,,\\
\tilde{Q}^{d,(-\vec k)}_{\xa,\ya}
  &= \mcS^{(-\vec k)}_{\xa,\xb} \, Q^d_{\xb,\yb} \, \mcS^{(-\vec k)}_{\yb,\ya}
  \propto e^{i\vec k(\vec\ya -\vec\xa)}\,,
\end{aligned}
\end{equation}
where ``$\propto$'' sign stands for additional coordinate dependence
due to the boosting.  The $d$-quark $\xa\leftarrow\ya$ propagator,
as usual, is computed using $\gamma_5$-Hermiticity of the Dirac
operator,
\begin{equation}
\label{bkwprop_calc}
\tilde{Q}^{d,(-\vec k)}_{\xa,\ya} 
  = \gamma_5 \big[\tilde{Q}^{d,(-\vec k)}_{\ya,\xa}\big]^\dag \gamma_5
  = \gamma_5 \big[ \mcS^{(-\vec k)}_{\ya,\yb} \, Q^d_{\yb,\xb} \, 
      \mcS^{(-\vec k)}_{\xb,\xa}\big]^\dag \gamma_5
\end{equation}
where the sign of the boosting momentum is preserved due to
the Hermiticity of the (boosted) smearing operator
$\mcS$ in~\eqn{boostsm_herm}.

Repeating similar steps to for the meson three-point function with
the insertion of the operator $\big[\bar u\,W\,\Gamma\,u\big]_z$
with arbitrary $\Gamma$-matrix and Wilson line
$W_{z,z+\hat\mcL}=(\prod_\mcL U)_{z,z+\hat\mcL}$ along path $\mcL$,
we get
\begin{equation}
\label{c3pt}
\begin{aligned}
& C^{W\Gamma}_\text{3pt}(\ya_4,\vec p^\prime; z_4,\vec q; x)\\
    &\quad= \sum_{\vec y,\vec z}e^{-i\vec p^\prime (\vec y-\vec x) + i\vec q\vec z} \big\langle 
        \big[\pi^{+,(2\vec k)}\big]_y \, 
        \big[\bar u_z \,W_{z,z+\hat\mcL} \, \Gamma \, u_{z+\hat\mcL} \big]
        \big[\pi^{+,(2\vec k)}\big]^\dag_x \big\rangle \\
&\quad= \sum_{\vec z}e^{i\vec q \vec z} \, \Tr\big[
        B^{\bar{d}\Gamma_M u(\ya_4,\vec p^\prime)}_{\xa,z}\, 
        W_{z,z+\hat\mcL} \, \Gamma \,
        F^{u}_{z+\hat\mcL,\xa} 
    \big]\,,
\end{aligned}
\end{equation}
where the forward propagator $F^u=Q^u \, S^{(\vec k)}$ and the meson
sink-sequential (backward) propagator $B^{\bar{d}\Gamma_M u(\ya_4,\vec
p^\prime)}$ is defined as
\begin{equation}
\label{bkwprop}
\begin{aligned}
B^{\bar{d}\Gamma_M u(\ya_4,\vec p^\prime)}_{\xa,z}  
    &= \sum_{\vec y} e^{-i\vec p^\prime (\vec y-\vec x)} \,
    \big(-\overline{\Gamma}_M\big)
\tilde Q^{d,(-\vec k)}_{\xa,\ya} \, \Gamma_M \, \mcS^{(-\vec k)}_{\ya,\yb} Q^d_{\yb,z} \,.
\end{aligned}
\end{equation}
which is also computed using the $\gamma_5$-conjugation.

\section{Explicit calculation to show that 
bare pion $u-d$ three point function is purely real or imaginary}
\label{pionreal}
In the previous appendix, we constructed the connected piece of the
three point function of $\overline{u}W \Gamma u$ operator in $\pi^+$.
If one repeats the computation using $\overline{d}W \Gamma d$
operator, one will find the disconnected piece to be the same as
the one in the full $\overline{u}W \Gamma u$ three point function
and hence such quark line disconnected terms will cancel in the
$\overline{u}W \Gamma u - \overline{d}W \Gamma d$ isospin nonsinglet
operator that we are interested in. Below, we further explain as
to why only the real part of the connected $\overline{u}W \Gamma
u$ three point function for $\Gamma=\gamma_t,\gamma_z$ and imaginary
part for $\Gamma=1$ contributes to the total isospin nonsinglet
three point function. For the sake of simplicity let us take the
case of point source and point sink, and take $\Gamma=\gamma_t$.
The full expression for the $u-d$ qPDF three point function is
\begin{eqnarray}
    &&C^{u-d}_{\rm 3pt}(t,\tau,{\mathcal L})=\sum_{\vec{y},\vec{z}}e^{-i\vec{p}.\vec{y}}{\rm Tr}\left[ (-\gamma_5)Q_{x,y}\gamma_5 Q_{y,z}W_{z,z+{\mathcal L}}\gamma_t Q_{z+{\mathcal L},x}\right]+\cr
    &&\quad\sum_{\vec{y},\vec{z}}e^{-i\vec{p}.\vec{y}}{\rm Tr}\left[ (-\gamma_5)Q_{x,y}\gamma_5 Q_{y,z+{\mathcal L}}W^\dagger_{z,z+{\mathcal L}}\gamma_t Q_{z,x}\right]^*,\cr
    &&\qquad\qquad\quad\equiv \sum_{\vec{y},\vec{z}}\left(e^{-i\vec{p}.\vec{y}} T_1 + e^{-i\vec{p}.\vec{y}} T^*_2\right),
\label{eq1}
\end{eqnarray}
where we do not make distinctions between $u$ and $d$ quark propagators due to 
isospin symmetry.
Let us call the trace in first term on the right hand side as $T_1$
and the second trace before being conjugated as $T_2$.  One can go
from $T_2$ to $T_1$ by parity transformation $x=(\vec{x},x_4)\to
x_p=(-\vec{x},x_4)$ , followed by a spatial translation $x\to
x+{\mathcal L}$ by making use of the transformation of the Dirac
propagator to be $Q_{x,y}\to \gamma_t Q_{x_p,y_p}\gamma_t$ and
$W_{x,x+{\mathcal L}}\to W^\dagger_{x_p-{\mathcal L},x_p}$ under
parity. In this case, the $\gamma_t$ from parity transformation for
$Q$ commutes with $\Gamma=\gamma_t$. In other cases, one should
take care of the $\pm$ factor.  Thus $C_{\rm 3pt}$ becomes
\begin{equation}
    C^{u-d}_{\rm 3pt}(t,\tau,{\mathcal L})=\sum_{\vec{y},\vec{z}}\left(T_1 e^{-i \vec{p}.\vec{y}}+T^*_1 e^{i \vec{p}.\vec{y}}\right),
\end{equation}
and therefore proportional to the connected piece of $\overline{u}\Gamma
W u$, which is the first term in the above equation. We normalize
the three point function such that the $u-d$ isospin charge of the
pion is 1.  By going through the similar calculation, one can show
that the three point function is real also for $\Gamma=\gamma_z$ while it is
is purely imaginary for $\Gamma=1$ $u-d$ pion qPDF.

\section{Relation between $P_z$ and $-P_z$ matrix elements}
\label{matrixconj}
In this appendix, we derive the relation between conjugates of the
matrix elements $\mel**{E_{n'},
P_z}{\mathcal{O}_{\Gamma}(z;\tau)}{E_n,P_z}$ that enter the excited
state contributions to the qPDF three-point function.  For
this, let us consider the conjugate of the simplest component of
the qPDF matrix element:
\begin{eqnarray}
&&\sum_z \mel**{E_{n'}, P_z}{\bar{u}_z\Gamma W_{z,z+{\mathcal L}} u_{z+{\mathcal L}}}{E_n,P_z}^*=\cr
&&\quad\sum_z \mel**{E_{n}, P_z}{\bar{u}_{z+{\mathcal L}}\Gamma W^\dagger_{z,z+{\mathcal L}} u_{z}}{E_{n'},P_z},
\end{eqnarray}
for $\Gamma=\gamma_t,\gamma_z,1$.
Using the parity operator $\Pi$, the right-hand-side of the above equation becomes
\begin{eqnarray}
&&\sum_z \mel**{E_{n}, P_z}{\bar{u}_{z+{\mathcal L}}\Gamma W^\dagger_{z,z+{\mathcal L}} u_{z}}{E_{n'},P_z}\cr
&&\quad=\sum_z \mel**{E_{n}, -P_z}{(\Pi \bar{u}_{z+{\mathcal L}}\Pi) (\Pi W^\dagger_{z,z+{\mathcal L}}\Pi) (\Pi u_{z}\Pi) }{E_{n'},-P_z}\cr
&&\quad=\sum_z \mel**{E_{n}, -P_z}{\bar{u}_{z-{\mathcal L}}\gamma_t \Gamma  W_{z-{\mathcal L},z} \gamma_t u_{z} }{E_{n'},-P_z}\cr
&&\quad=\sum_z \mel**{E_{n}, -P_z}{\bar{u}_{z}\gamma_t \Gamma \gamma_t  W_{z,z+{\mathcal L}}  u_{z+{\mathcal L}} }{E_{n'},-P_z}.
\end{eqnarray}
Defining, $\gamma_t \Gamma \gamma_t=\Phi_\Gamma \Gamma$ with
$\Phi_\Gamma=\pm 1$, we have the relation
\begin{eqnarray}
&&\sum_z \mel**{E_{n'}, P_z}{\bar{u}_z\Gamma W_{z,z+{\mathcal L}} u_{z+{\mathcal L}}}{E_n,P_z}^*=\cr
&&\quad\Phi_\Gamma \sum_z \mel**{E_{n}, -P_z}{\bar{u}_{z} \Gamma   W_{z,z+{\mathcal L}}  u_{z+{\mathcal L}} }{E_{n'},-P_z},
\end{eqnarray}
with $\Phi_\Gamma=1$ for $\Gamma=\gamma_t,1$ and $\Phi_\Gamma=-1$
for $\Gamma=\gamma_z$ with all $\Gamma$ being in the Minkowskian
convention.  Thus, we can average over $P_z$ and $-P_z$ data after
taking care of appropriate factor of $\Phi_\Gamma$.

\section{Pion two point functions and energy levels}
\label{app_2pt}

\begin{figure}
\includegraphics[scale=0.4]{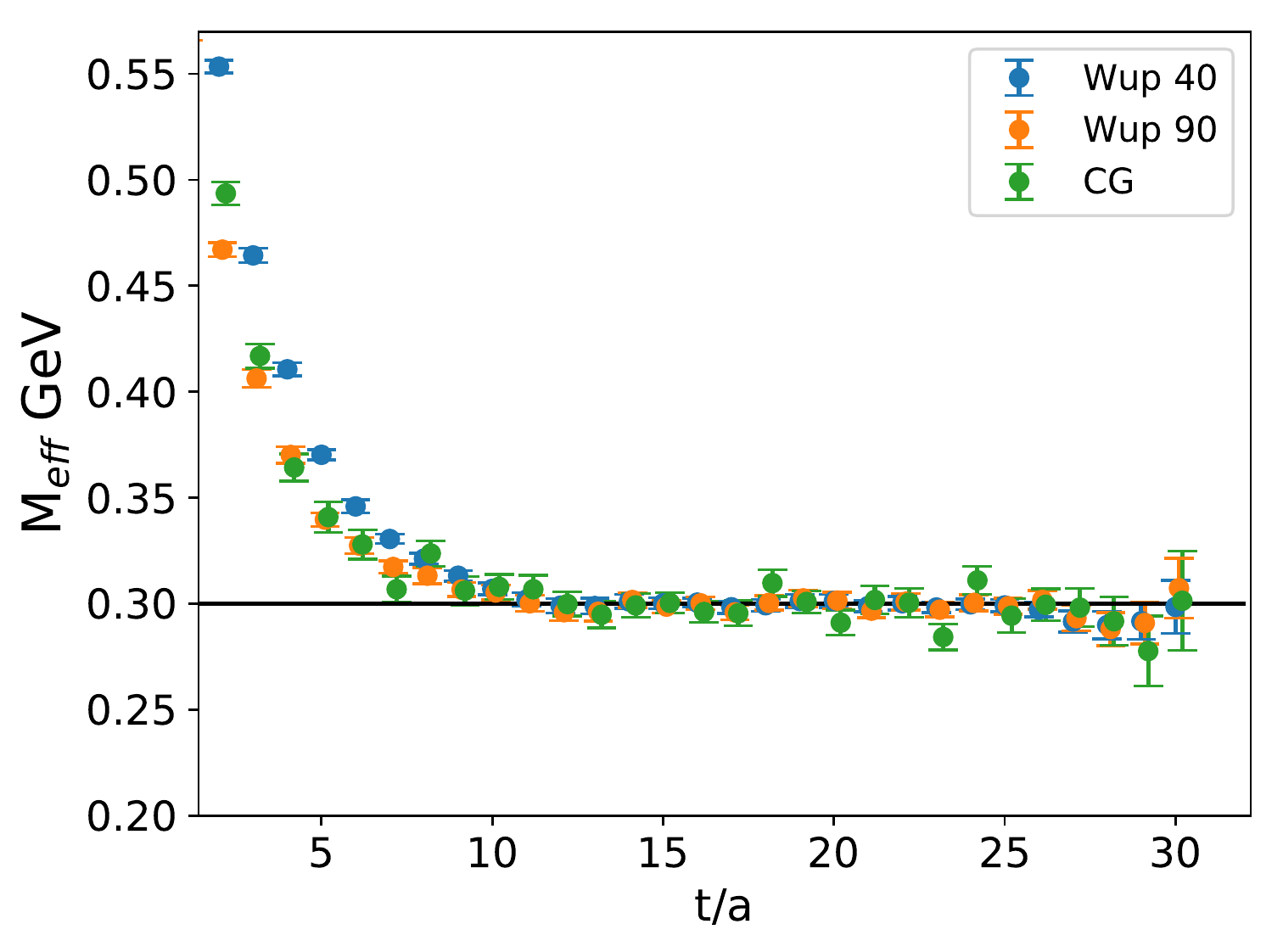}
\caption{Effective masses for $P_z=0$ using
Gaussian sources with 40 steps of Wuppertal smearings,
90 steps of Wuppertal smearings and Coulomb gauge.}
\label{fig:comp_wupp}
\end{figure}
\begin{figure}
\includegraphics[scale=0.4]{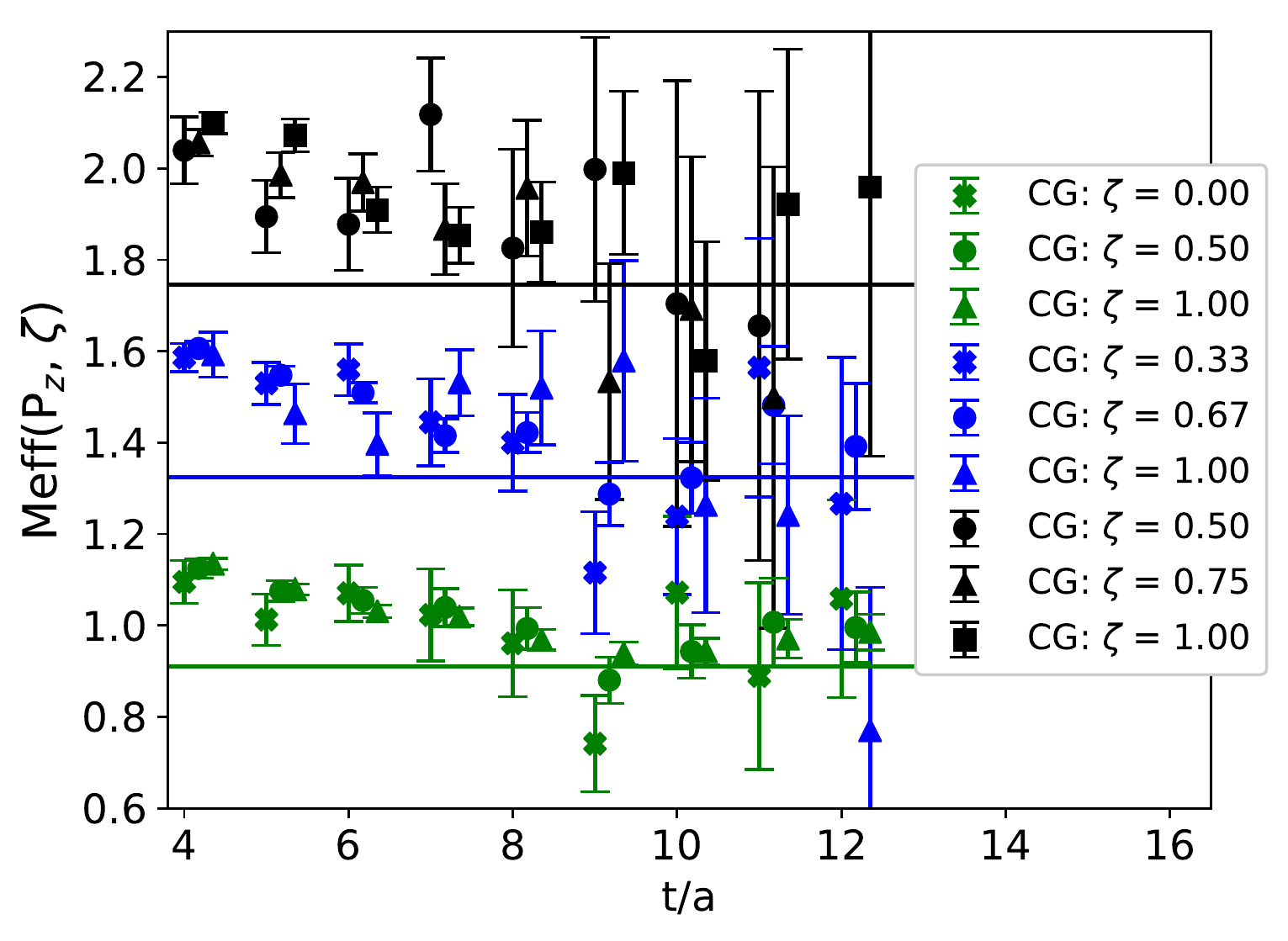}
\caption{Effective masses for different values of $\zeta$ with 50
configurations, Green, blue, and black points correspond to momentum 0.86,
1.29, and 1.72 GeV respectively.} 
\label{fig:comp_zeta}
\end{figure}
In this appendix we discuss some details of the calculations of the
pion two point function. We tested several different sources for
the pion.  In these tests we used 50 gauge configurations.  We used
Gaussian sources with several steps of Wuppertal smearings as well
as in Coulomb gauge (see main text). In Fig. \ref{fig:comp_wupp}
we show the effective mass for 40 and 90 steps of Wuppertal smearings
as well as the Coulomb gauge Gaussian sources of size 0.3 fm. We see
that 90 steps of Wuppertal smearings and Coulomb gauge Gaussian
sources give similar effective masses, while the excited state
contamination is larger for 40 steps of Wuppertal smearings.  We
also studied the two point functions for different boosted Gaussian
sources, with momentum boost $k_z$. The corresponding effective
masses are shown in Fig. \ref{fig:comp_zeta} for $P_z=0.86,~1.29$
and $1.72$ GeV.  for different values of $\zeta=k_z/P_z$. We clearly
see that non-zero value of $\zeta$ improves the signal for all
$P_z$. We also see that $\zeta=0.5$ is too small, while $\zeta=1.0$
is too large for $P_z=1.29$, but works well for $P_z=0.86$ GeV.

\section{Implementation of matching convolution}
\label{match}
Here, we describe the implementation of the plus function in the matching
formula such as to ensure current conservation. The matching kernel
is of the form
\begin{equation}
C\left(\frac{x}{y},y P_z\right)=\delta\left(\frac{x}{y}-1\right)+\frac{\alpha_s C_F}{2\pi}C^{(1)}_+\left(\frac{x}{y},y P_z\right),
\end{equation}
where the dependence on $P^R$ and $\mu$ are implicit. The first
perturbative correction is a plus function that ensures the vector
current conservation.  The property we know of the plus-function
is that
    $\int_{-\infty}^\infty dx C^{(1)}_+\left(\frac{x}{y},yP_z\right) =0$,
since the second dependence of the function is independent of $x$.
In order to
implement the plus-function correctly, we can use the following procedure:
\begin{equation}
    C^{(1)}_+(\xi,y P_z) = C^{(1)}(\xi,y P_z) - N(y P_z) \delta(\xi-1).
\end{equation}
The $x$-independent but momentum dependent coefficient $N(yP_z)$ is
\begin{equation}
    N(y P_z)\equiv \int_{reg} C^{(1)}(\xi,y P_z) d\xi,
    \label{normdef}
\end{equation}
where $ \int_{reg} d\xi$ involves an integration over the intervals
$[-\Lambda,-\epsilon]\cup[\epsilon,1-\epsilon]\cup[1+\epsilon,\Lambda]$
for some upper cut-off $\Lambda$ and a small exclusion parameter
$\epsilon$.  The above definition gives the usual result that
\begin{equation}
    \int_{-\infty}^{\infty} d\xi C^{(1)}_+(\xi,y P_z) f(\xi) =\int_{-\infty}^{\infty} C^{(1)}(\xi ,y P_z) (f(\xi)-f(1)),
\end{equation}
with $y$ held fixed as $\Lambda\to\infty$ and $\epsilon\to 0$.  The
following is then true for any function $f$:

\begin{eqnarray}
    \int\int dx dy C^{(1)}_+(x/y,y P_z) f(y)&=& \int dy \left(\int dx C^{(1)}_+(x/y,y P_z)\right) f(y)\cr
    &=&0,
\end{eqnarray}
leading to the vector current conservation or equivalently to the
total area preservation between the qPDF and PDF.  With this
prescription, the matching formula becomes
\begin{eqnarray}
    &&\int_{-\infty}^{\infty} \frac{dy}{|y|} C^{(1)}_+(x/y, yP_z) q(y)=\cr
    &&\int_{reg}  \frac{dy}{|y|} C^{(1)}(x/y, yP_z)  q(y) - N(x P_z) q(x) .
\end{eqnarray}

It is convenient to write the above formula in an explicitly vector current
conservation preserving form as
\begin{eqnarray}
    && \int_{-\infty}^{\infty} \frac{dy}{|y|} C^{(1)}_+(x/y, yP_z) q(y)=\cr
    &&\quad\int_{reg} \frac{dy}{|y|} C^{(1)}\left(\frac{x}{y}, yP_z\right)  q(y) -\int_{reg}\frac{dy'}{|x|} C^{(1)}\left(\frac{y'}{x}, xP_z\right)  q(x).\cr &&\qquad
\end{eqnarray}
However, care has to be taken in the numerical regularization of
the above equation to be consistent with the one in \eqn{normdef}.
That is, in the above equation, $\int_{reg} dy$ in the first integral
in the right-hand side involves the range $y\in
[-x/\epsilon,-x/\Lambda]\cup[x/\Lambda,x/(1+\epsilon)]\cup[x/(1-\epsilon),x/\epsilon]$
when $x>0$ and the range reversed when $x<0$. A consistent prescription
for $\int_{reg} dy'$ in the second integral in the right-hand side
involves $y' \in [-\Lambda x, -\epsilon x]\cup [\epsilon
x,x(1-\epsilon]\cup[x(1+\epsilon),x \Lambda]$.

\section{Results on two-state extrapolations to obtain the matrix elements at all $P_z$}
\label{matrixelements}

In \fgn{3pt_fit} in the main text, we showed some sample results
for the $t-\tau/2$ behavior and the $t\to\infty$ extrapolations of
the three-point function to two-point function ratio $R(t, \tau;
z, P_z, \Gamma)$ for $\Gamma=\gamma_t$ and $\gamma_z$ at a specific
intermediate value of $P_z=1.29$ GeV.  In \fgn{3pt_fit_gammaz_all}
and \fgn{3pt_fit_gammat_all} of this appendix, we show similar
results at all $P_z$ for $\Gamma=\gamma_z$ and $\gamma_t$ respectively
using \texttt{Fit}(2,2).

For the case of $P_z=0$, special care needs to be taken. 
For a finite temporal extent $L_t$ of the lattice, 
ignoring the effect of periodicity due to the presence of the terms
$e^{-E_\pi(L_t-t)}$ in the denominator of \eqn{3pt_spectral} is justified
when $e^{-E_\pi t} \ll e^{-E_\pi(L_t-t)}$. But, one should include the
effect of boundary condition if the two terms become comparable.
For the largest source-sink separation $t/a=12$ we use, the
contribution from the wrapping-around term, $e^{-E_\pi(L_t-t)}$,
relative to $e^{-E_\pi t}$ for $P_z=0$ is 2.7\%, whereas for higher
$P_z$ it is negligible e.g., for the smallest non-zero momentum
$P_z=0.43$ GeV, this effect is 0.2\%.  Hence, we included the term
$e^{-E_\pi(L_t-t)}$ in the denominator of \eqn{3pt_spectral} for the
extrapolation of $R(t,\tau,z;P_z)$ for $P_z=0$, and we also checked that the
effect of the periodicity of lattice was indeed negligible for any
of the non-zero $P_z$ we used. 

For the case of $P_z=0$ displayed in the top-most panels of
\fgn{3pt_fit_gammat_all}, we have shown the data in two
ways to make the fits and the extrapolated value easier to understand.
The unfilled symbols are the data for $R(t,\tau;z,P_z,\gamma_t)$
defined as the ratio of $C_{3{\rm pt}}(t,\tau;z,P_z)$ to $C_{2{\rm
pt}}(t;P_z)$, and the solid curves are the fits including the
$e^{-E_\pi(L_t-t)}$ term in the denominator of \eqn{3pt_spectral}.
While the fits describe the data well, the trend in the data with
increasing $t$ can be seen to be away from the extrapolated value.
To make the reason clearer, we have shown the modified ratio of
$C_{3{\rm pt}}(t,\tau;z,P_z)$ to the two-point function without the
wrap-around term, $C_{2{\rm pt}}(t;P_z)-A_0e^{-E_\pi(L_t-t)}$, as the
filled symbols. The dashed curves are now the fits using just
\eqn{3pt_spectral}. The values of the amplitude $A_0$ and the energy
$E_\pi$ were obtained by the two-state fit as described in the main
text. Now, the trend with increasing $t$ is clearer.

\onecolumngrid

\begin{figure}[H]

\centering

\includegraphics[scale=0.48]{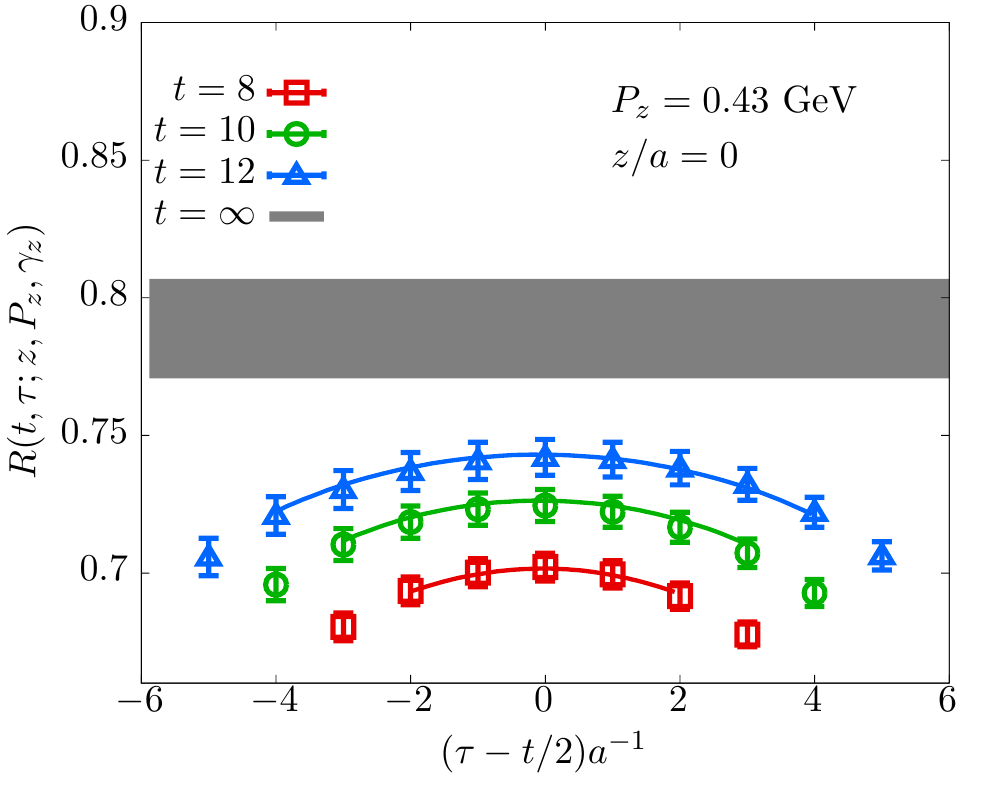}
\includegraphics[scale=0.48]{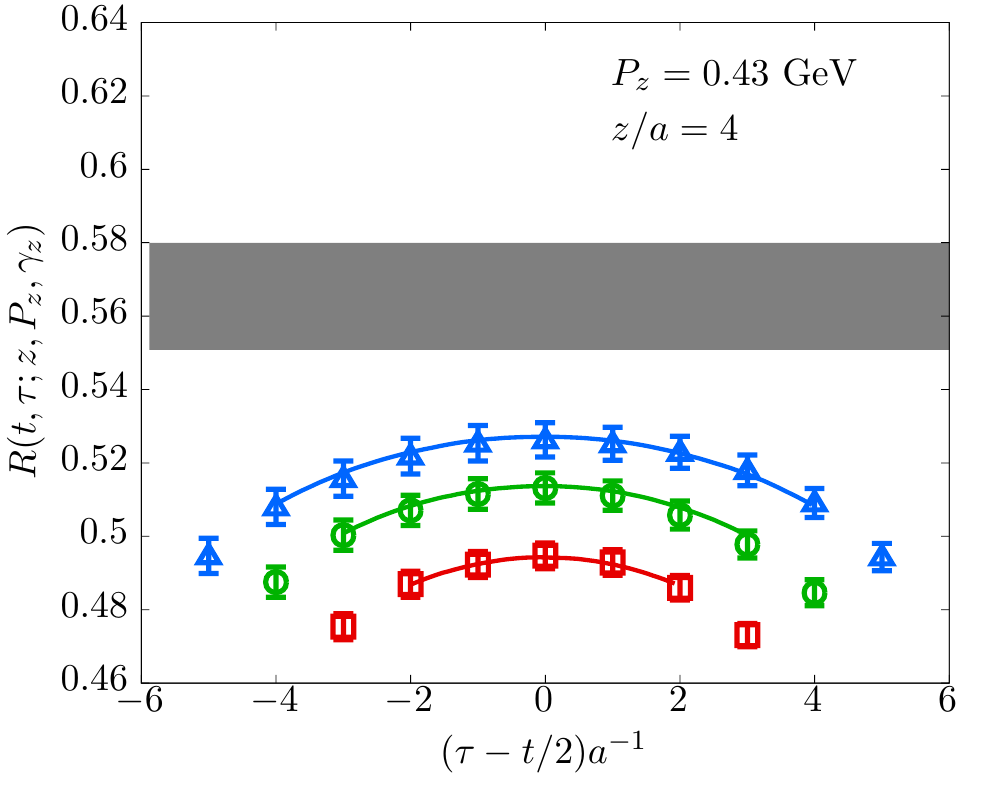}
\includegraphics[scale=0.48]{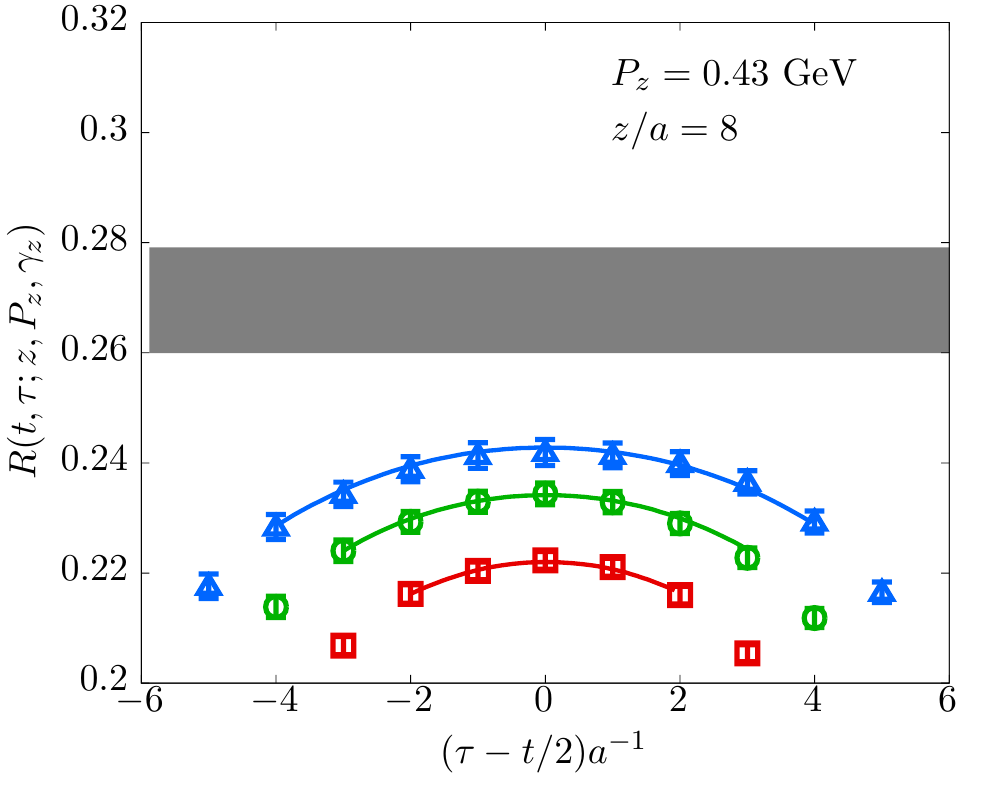}

\includegraphics[scale=0.48]{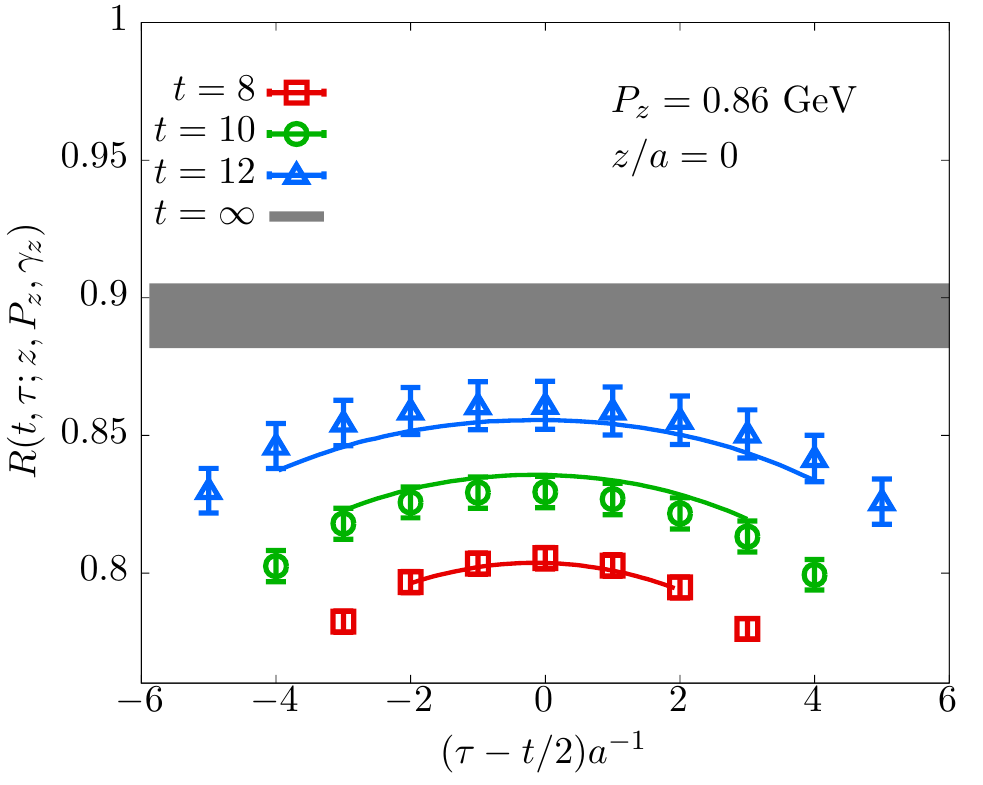}
\includegraphics[scale=0.48]{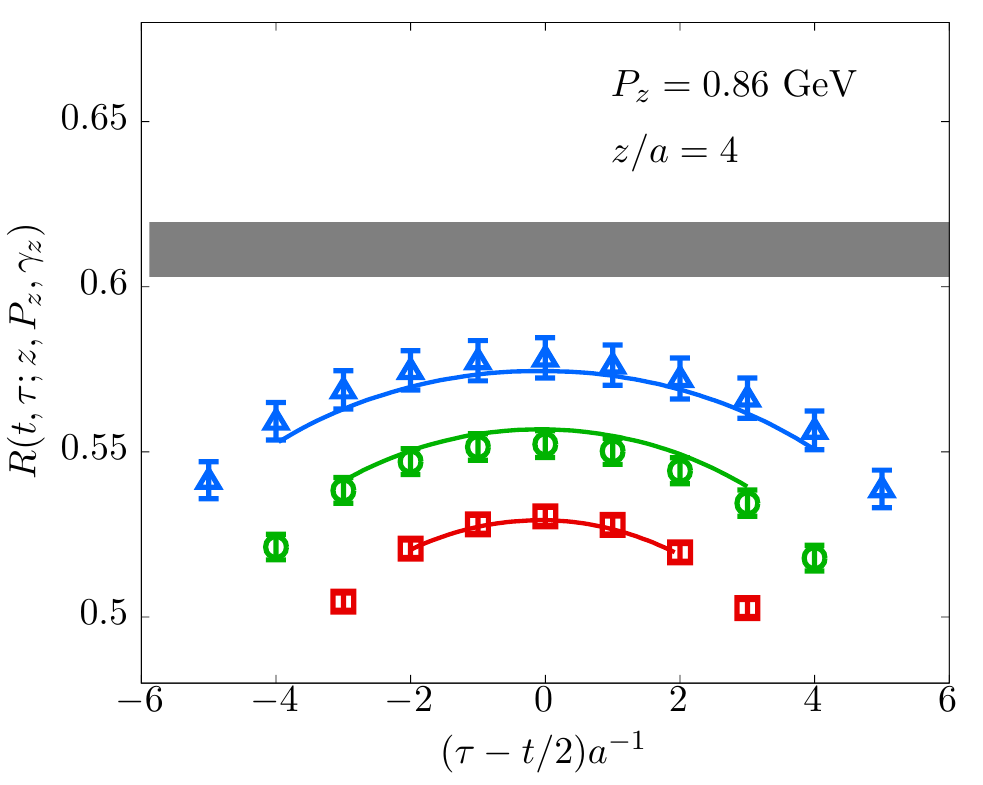}
\includegraphics[scale=0.48]{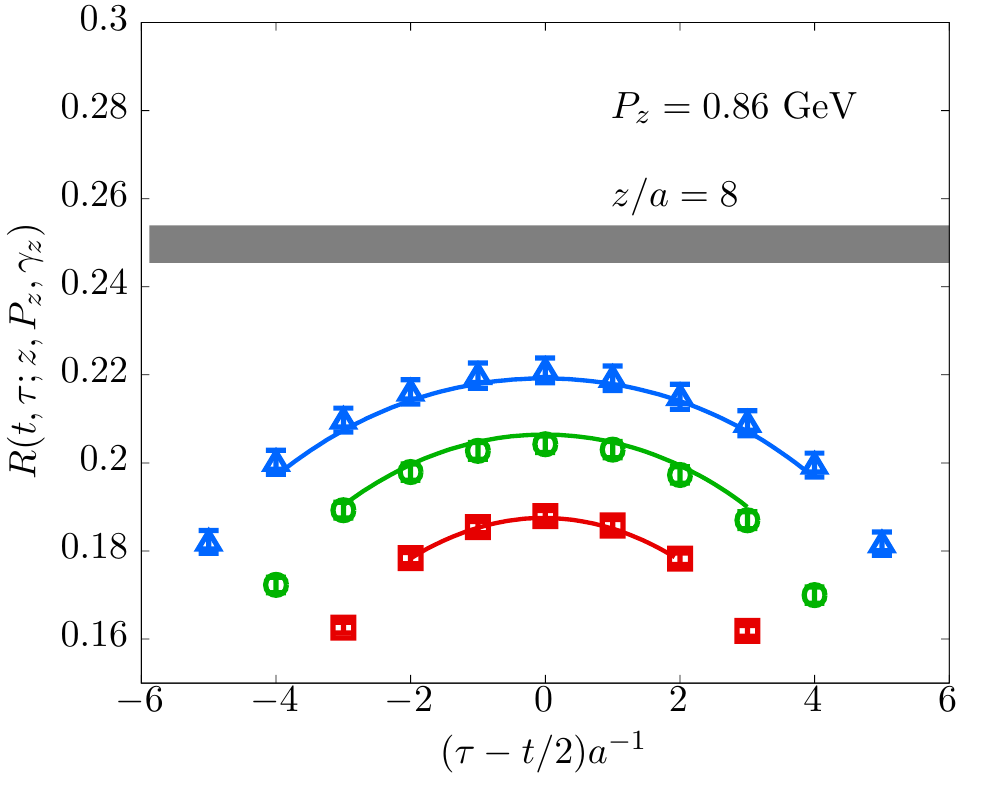}

\includegraphics[scale=0.48]{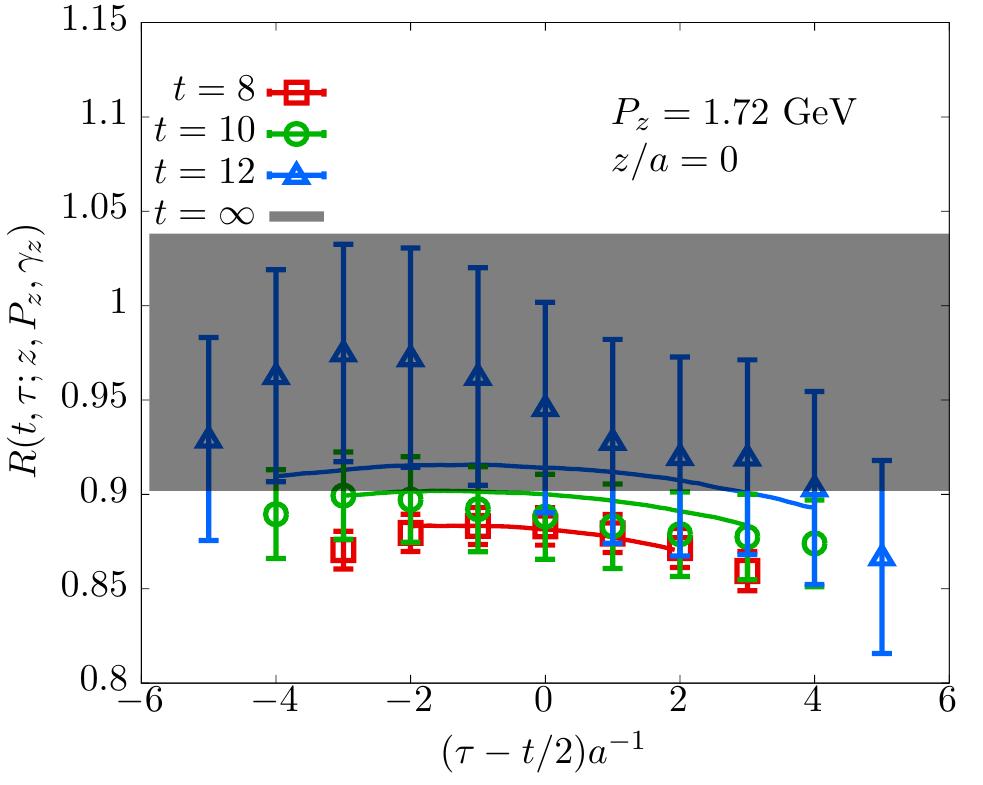}
\includegraphics[scale=0.48]{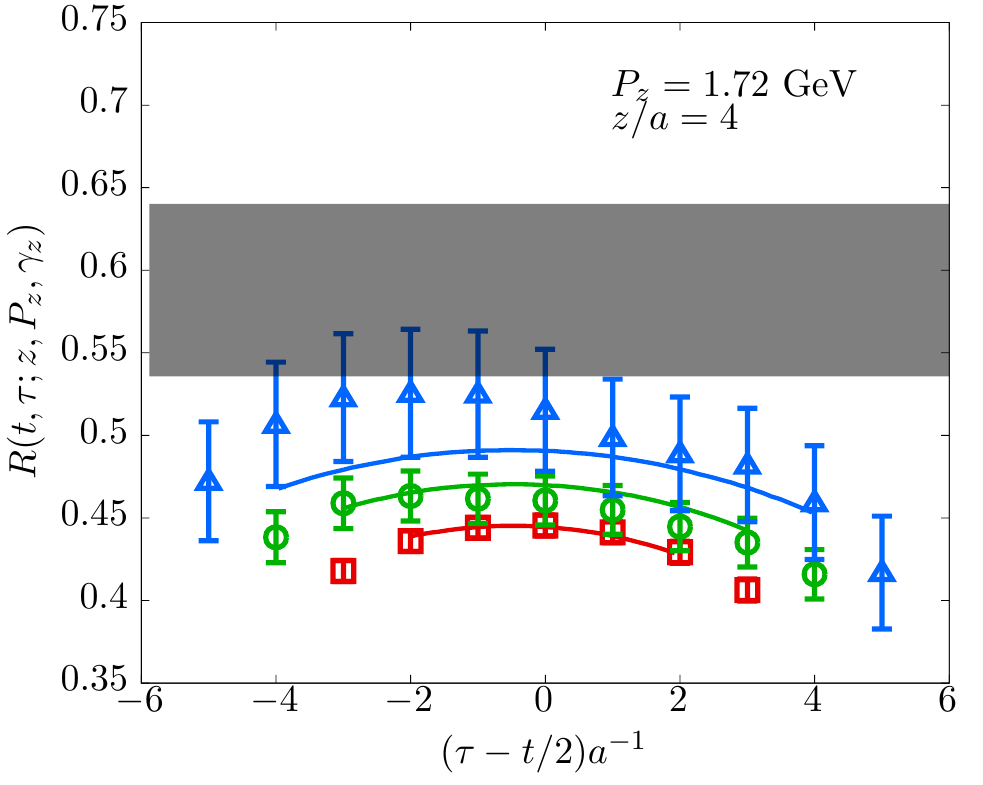}
\includegraphics[scale=0.48]{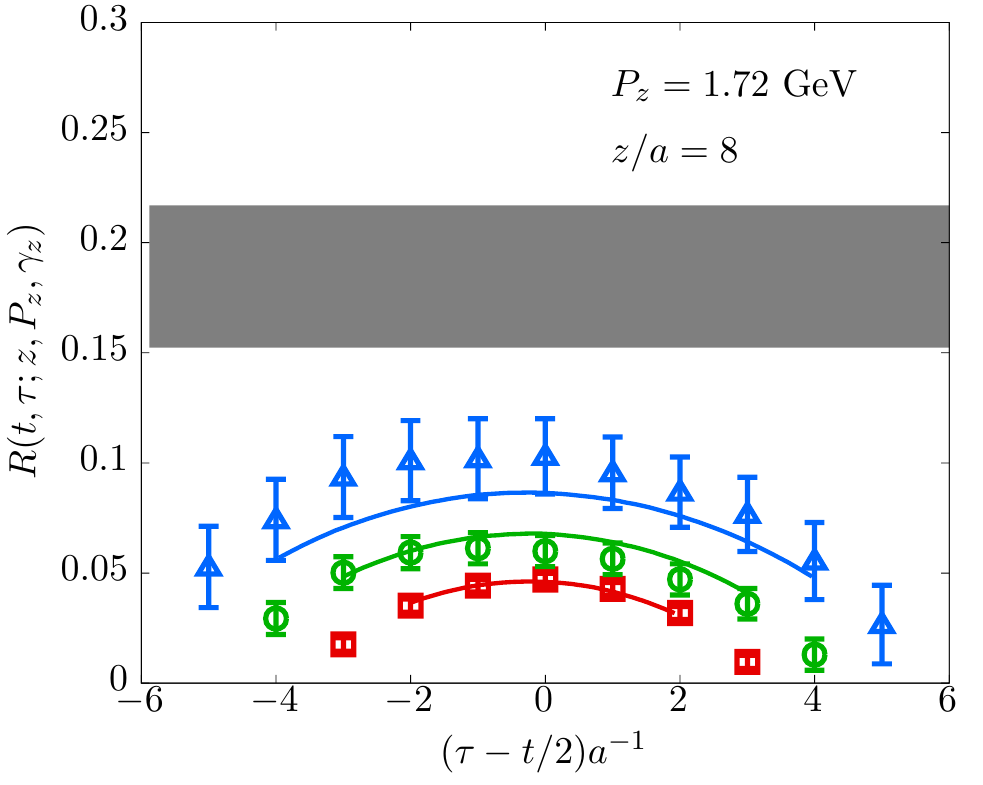}
\caption{
The ratio of the three point function to the two-point function,
$R(t, \tau; z, P_z, \gamma_z)$ for $\Gamma=\gamma_z$ is shown as function of $\tau-t/2$
for $z/a=0$, 4 and 8 (from left to right) and $P_z=0,0.483,0.86,$
and 1.72 GeV (top to bottom). The corresponding plots for $P_z=1.29$ GeV are shown in \fgn{3pt_fit} in the main text. 
The central values of the two-state
fits to the lattice results for different source-sink separations
are shown as the curves. The horizontal band corresponds to the
extrapolated result for infinite source-sink separation.
}
\label{fig:3pt_fit_gammaz_all}
\end{figure}

\begin{figure}

\centering
\includegraphics[scale=0.48]{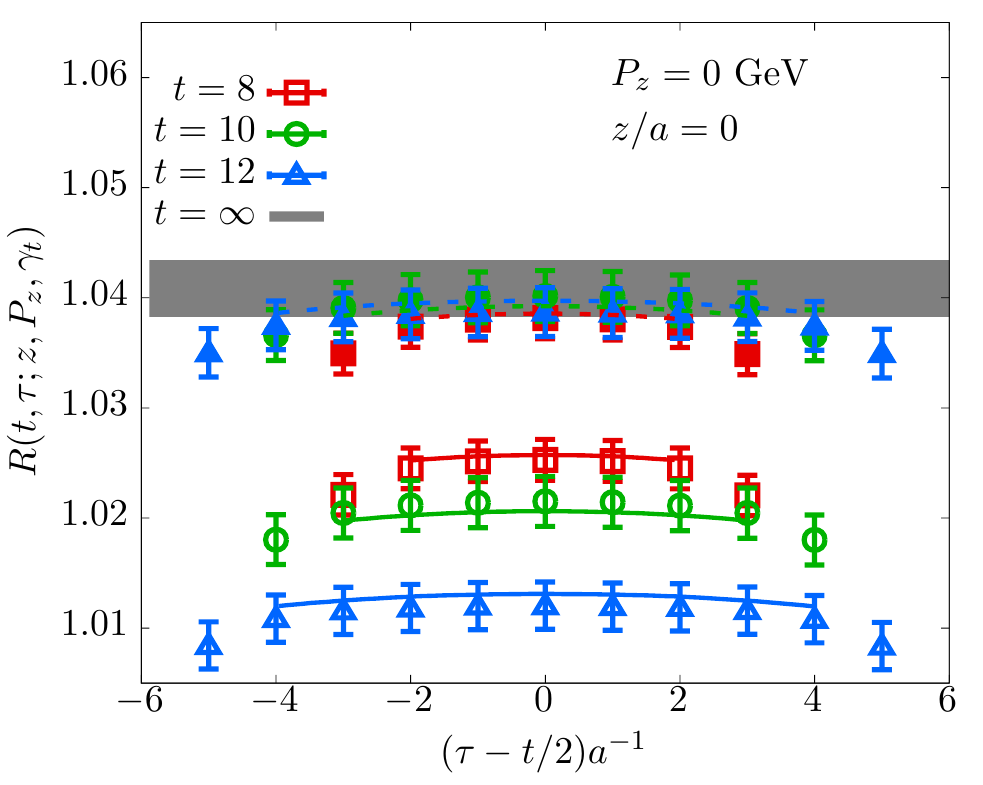}
\includegraphics[scale=0.48]{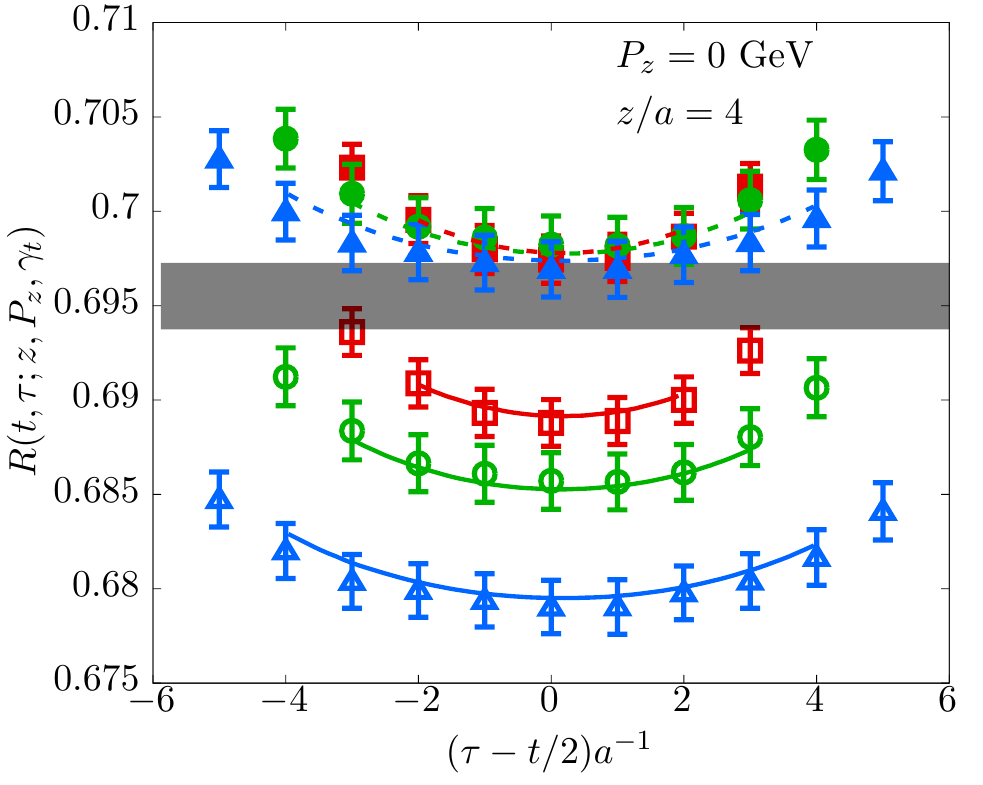}
\includegraphics[scale=0.48]{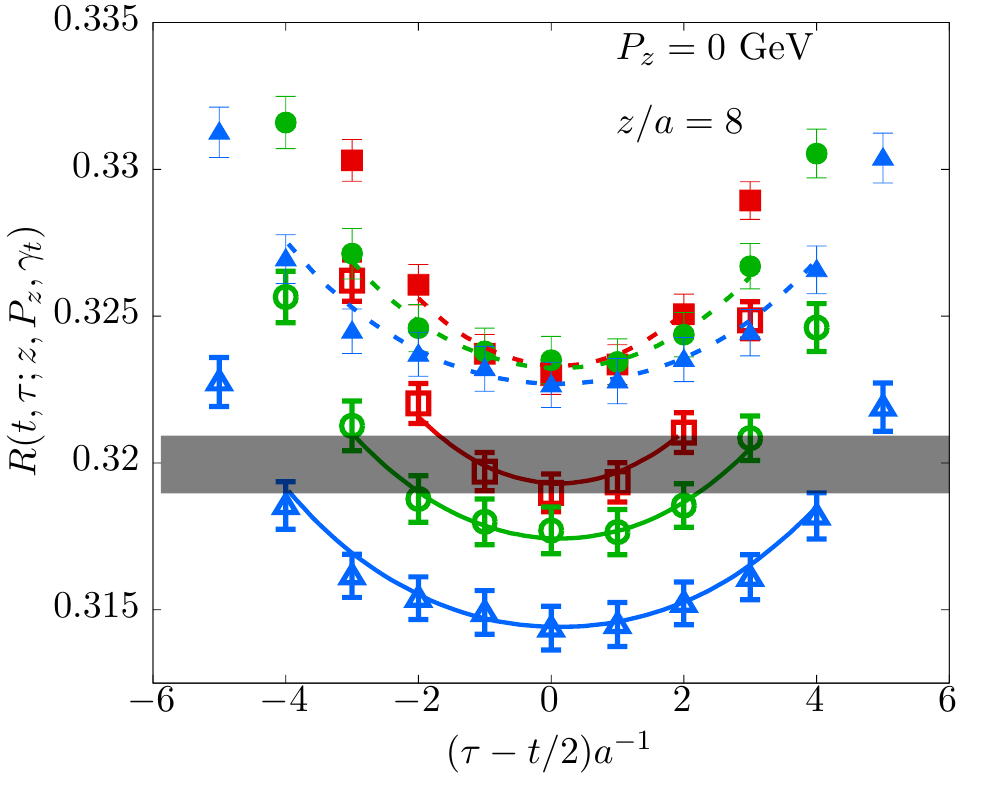}

\includegraphics[scale=0.48]{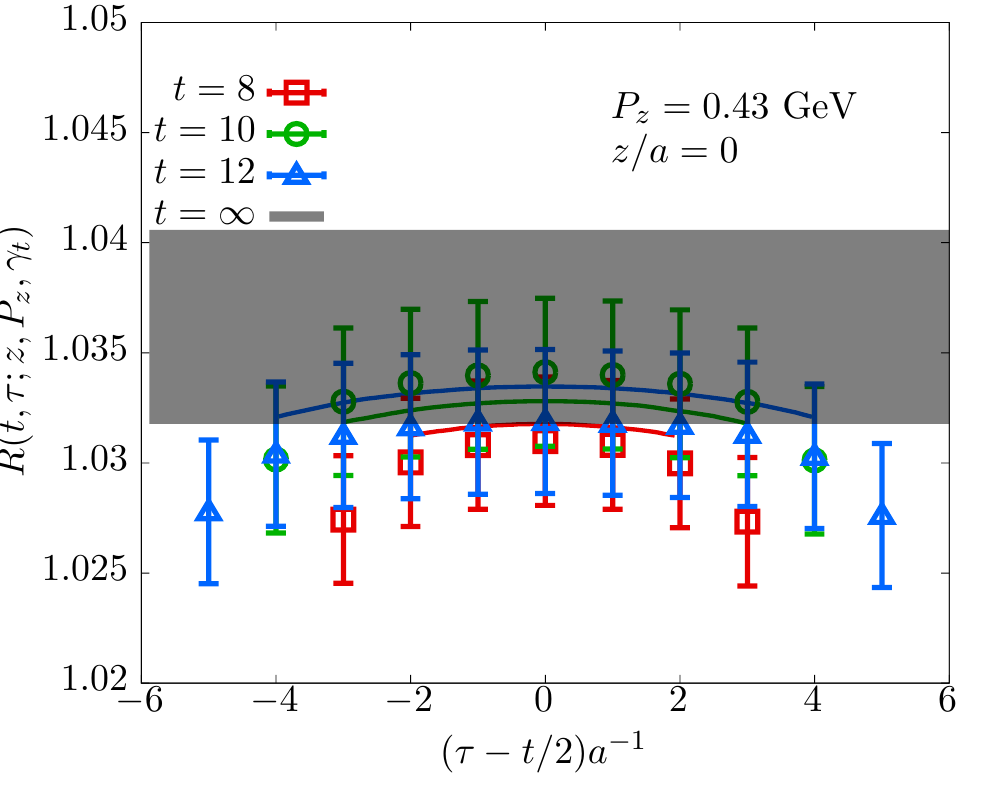}
\includegraphics[scale=0.48]{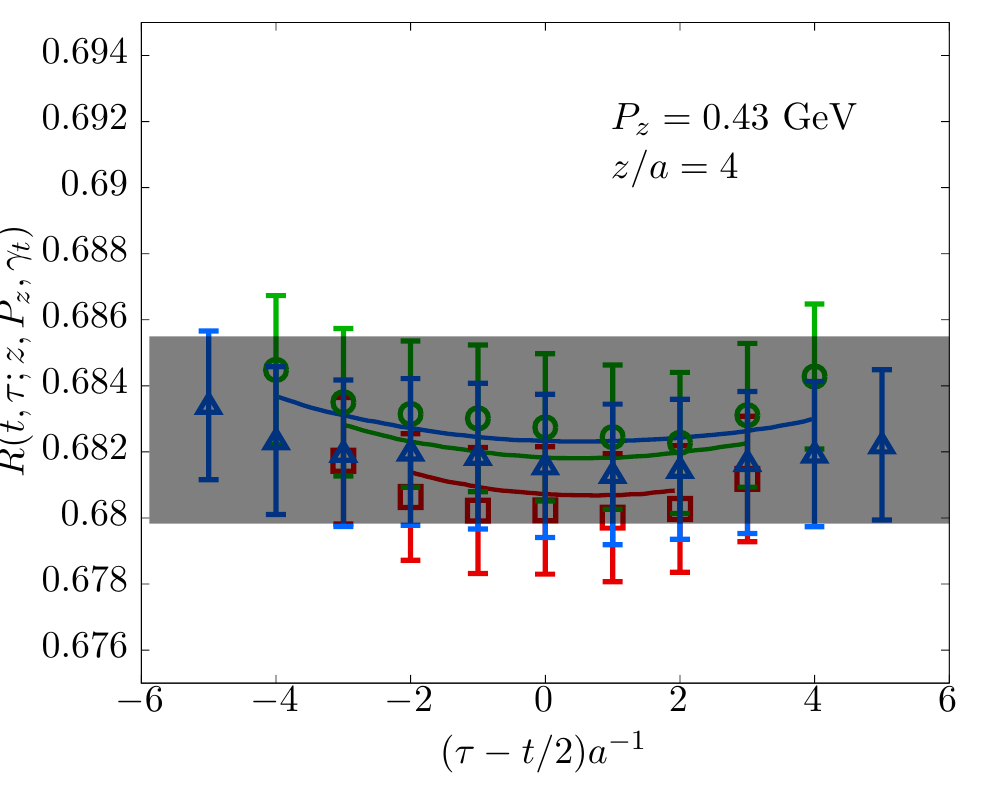}
\includegraphics[scale=0.48]{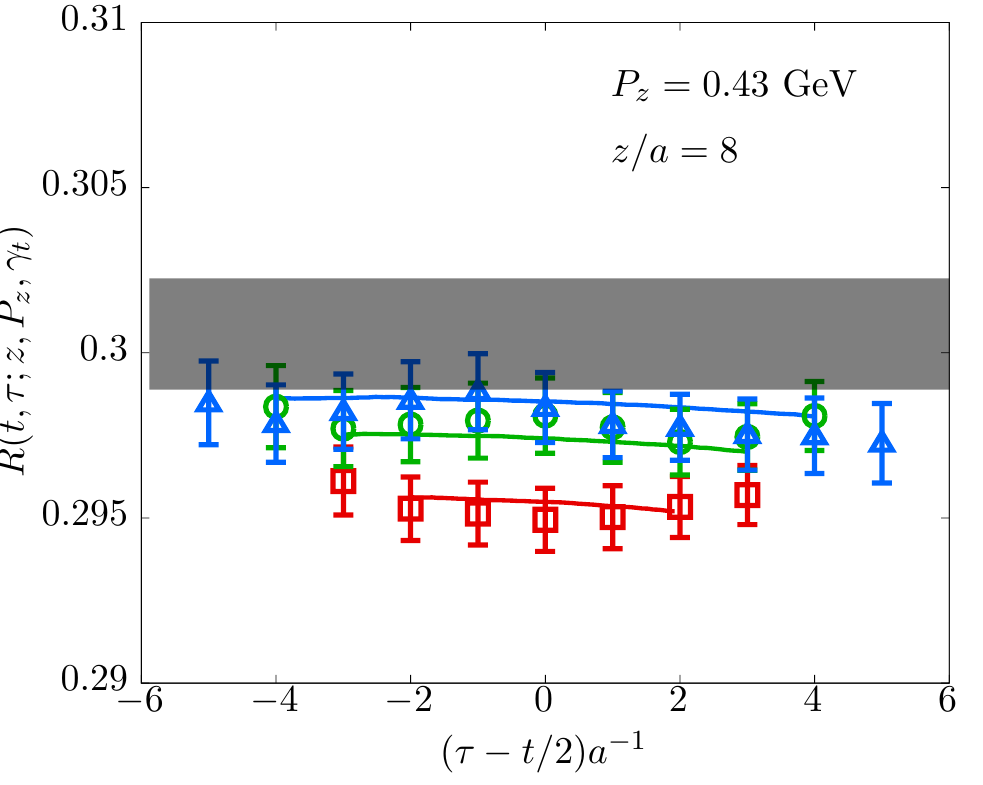}

\includegraphics[scale=0.48]{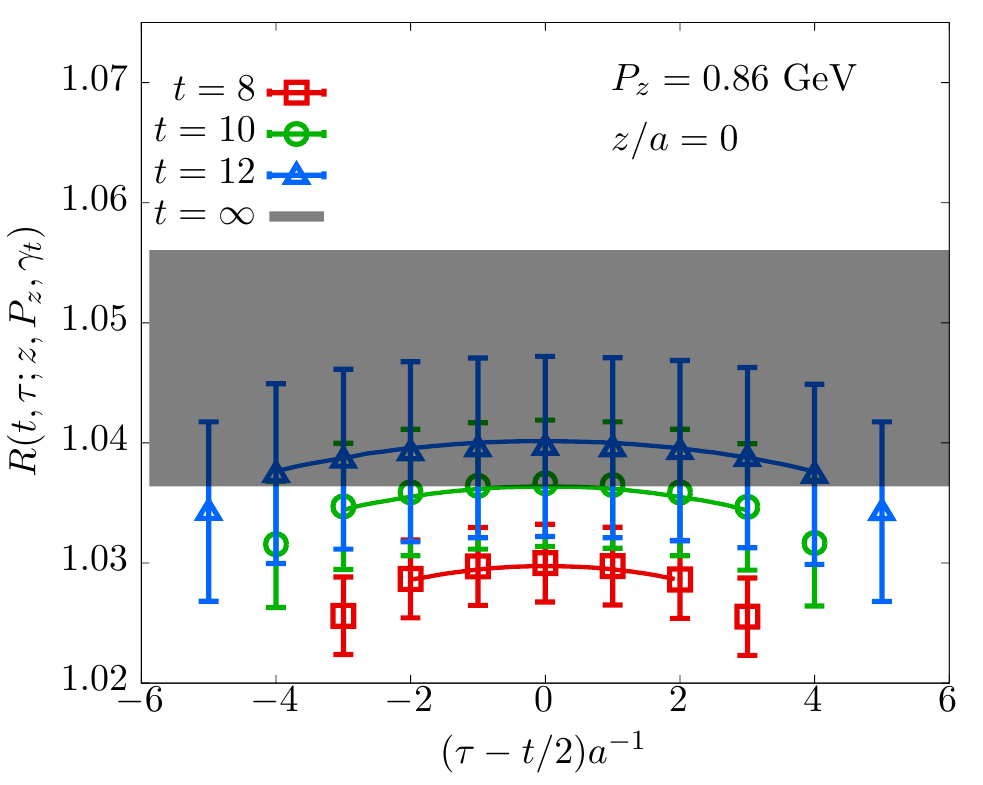}
\includegraphics[scale=0.48]{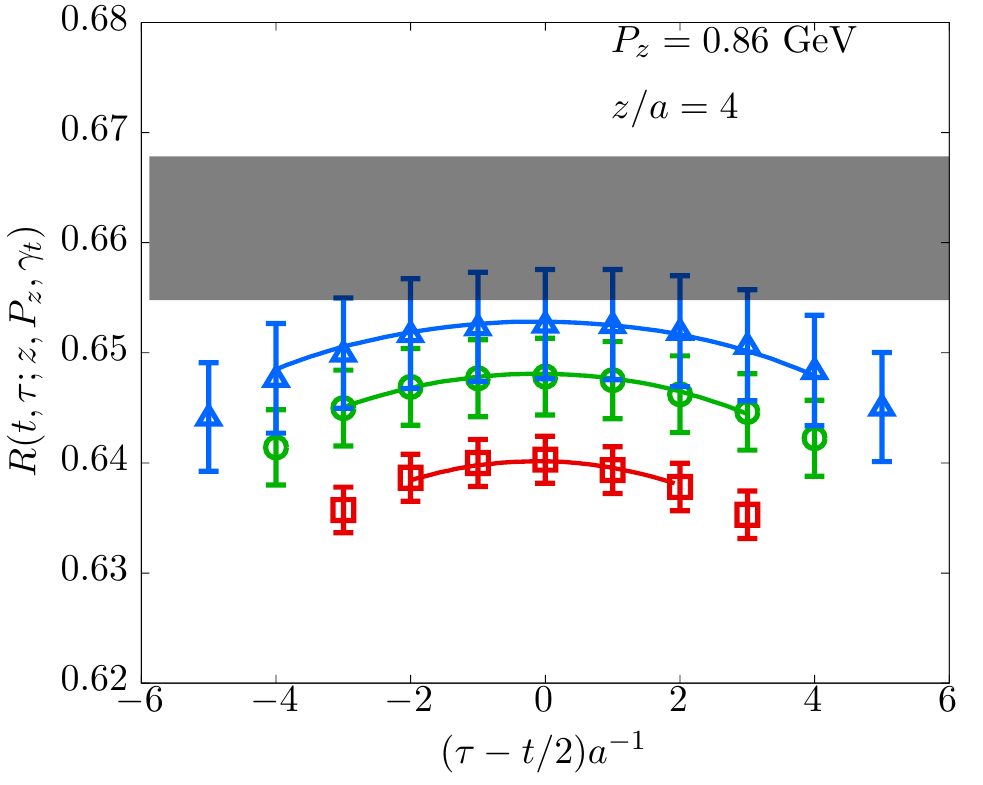}
\includegraphics[scale=0.48]{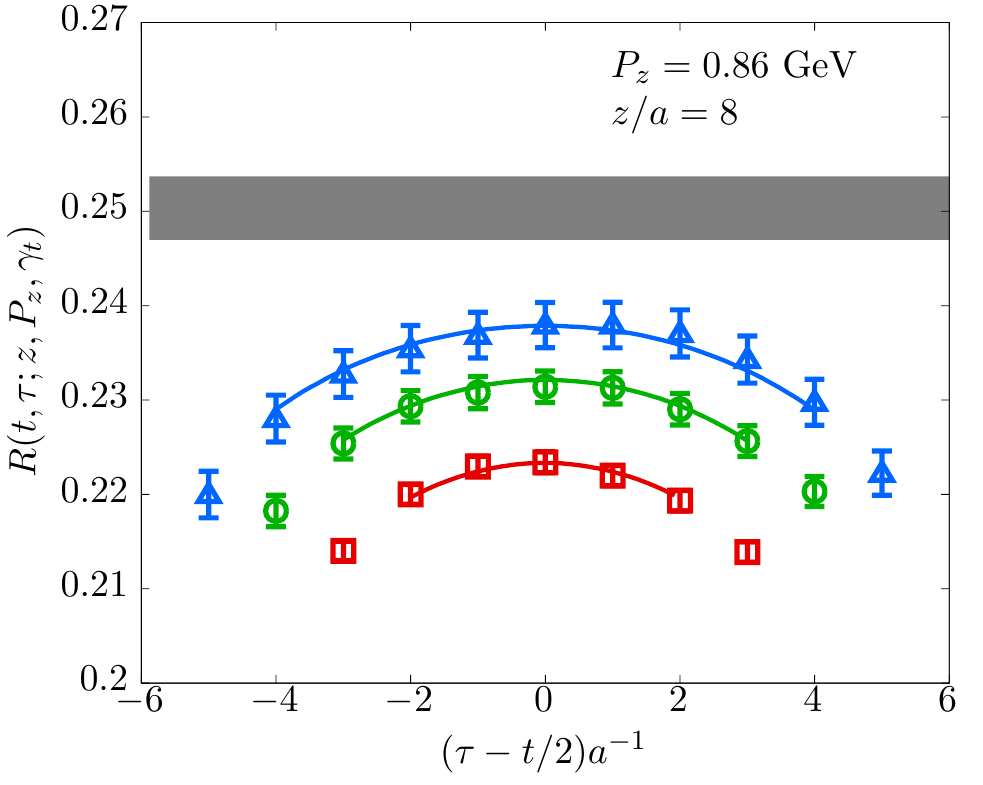}

\includegraphics[scale=0.48]{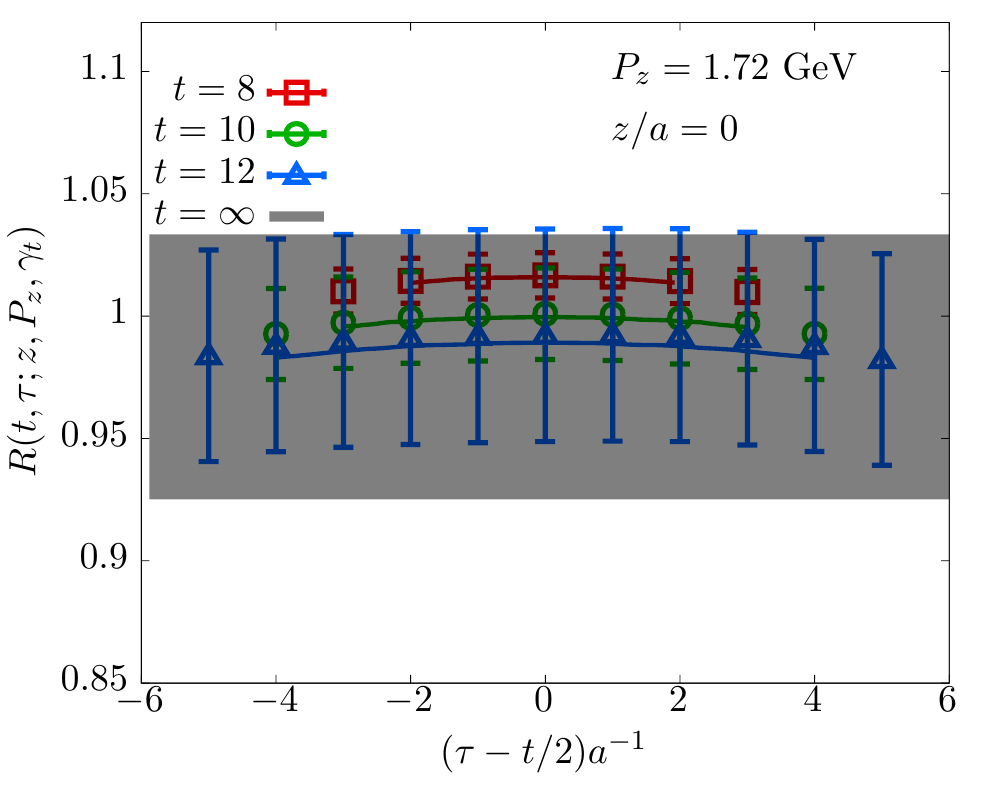}
\includegraphics[scale=0.48]{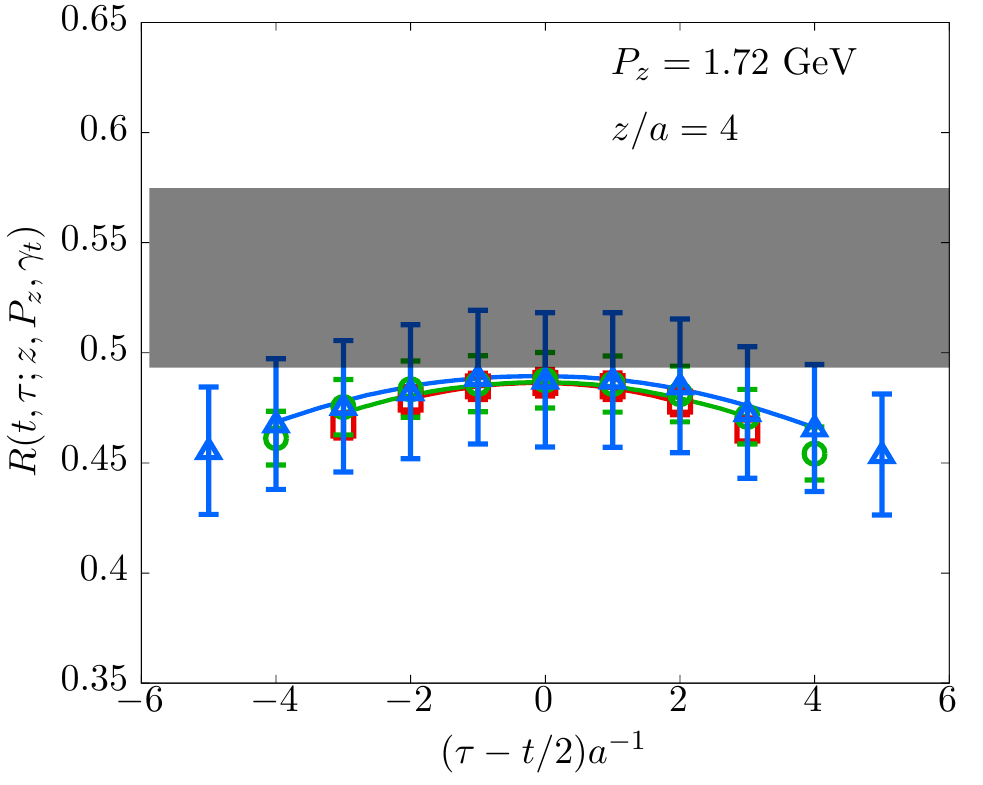}
\includegraphics[scale=0.48]{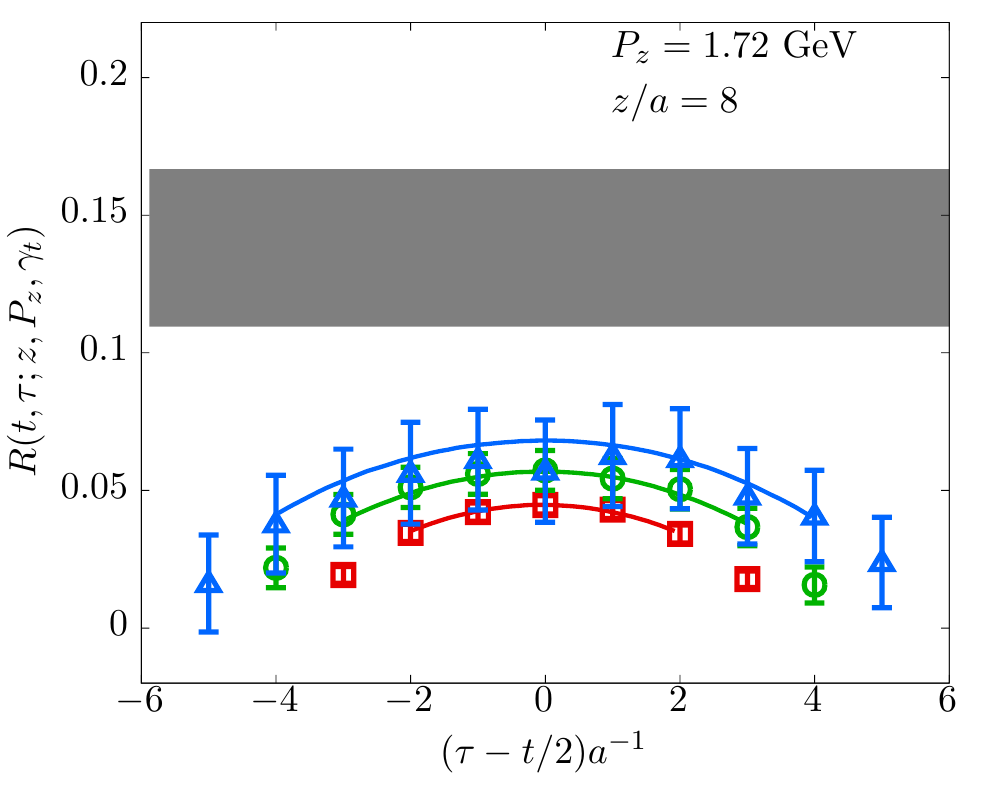}
\caption{
The ratio of the three point function to the two-point function,
$R(t, \tau; z, P_z, \gamma_t)$ for $\Gamma=\gamma_t$ is shown as
function of $\tau-t/2$ for $z/a=0$, 4 and 8 (from left to right)
and $P_z=0,0.483,0.86,$ and 1.72 GeV (top to bottom).  The corresponding
plots for $P_z=1.29$ GeV are shown in \fgn{3pt_fit} in the main
text.  The central values of the two-state fits to the lattice
results for different source-sink separations are shown as the
curves. The horizontal band corresponds to the extrapolated result
for infinite source-sink separation.  The case of $P_z=0$, in the
top-most panels, is special due to the presence of the effect of
lattice periodicity, and hence, the various symbols and curves for
the top-most panels are explained in detail in the text of
Appendix~\ref{matrixelements}.
}
\label{fig:3pt_fit_gammat_all}
\end{figure}

\twocolumngrid
\clearpage
\newpage

\bibliography{ref.bib}

\end{document}